\documentclass[namedreferences]{SolarPhysics}
\usepackage[optionalrh]{spr-sola-addons} 
\usepackage{graphicx}        
\usepackage{color}           
\usepackage{url}             

\usepackage{solaheader}	
\newcommand{\doi}{010-9592-6}
\newcommand{\algname}{Bayesian probability--based oscillation detection}

\newcommand{\md}{\mathbf{d}}
\newcommand{\mG}{\mathbf{G}}
\newcommand{\mb}{\mathbf{b}}
\newcommand{\mx}{\mathbf{x}}
\newcommand{\mf}{\mathbf{f}}
\newcommand{\ww}{\{\omega\}}

\begin{document}
\begin{article}
\begin{opening}

  \title{Automated Detection of Oscillating Regions in the Solar Atmosphere}

\author{J.     \surname{Ireland}$^{1}$\sep
        M.S.   \surname{Marsh}$^{2}$\sep
        T.A.   \surname{Kucera}$^{3}$]sep
        C.A.   \surname{Young}$^{1}$}

      \runningauthor{Ireland {\it et al.}}  \runningtitle{Automated
        Detection of Oscillating Regions in the Solar Atmosphere}

      \institute{ $^{1}$ ADNET Systems, Inc., NASA's Goddard
        Spaceflight
        Center, Mail Code 671.1, Greenbelt, MD 20771, USA.\email{Jack.Ireland@nasa.gov}\\
        $^{2}$ Jeremiah Horrocks Institute, University of Central
        Lancashire, Preston, PR1 2HE, UK.\\
        $^{3}$ NASA's Goddard Spaceflight
        Center, Mail Code 671.1, Greenbelt, MD 20771, USA.\\
      }

\begin{abstract}
  Recently observed oscillations in the solar atmosphere have been
  interpreted and modeled as magnetohydrodynamic wave modes.  This has
  allowed the estimation of parameters that are otherwise hard to
  derive, such as the coronal magnetic-field strength.  This work
  crucially relies on the initial detection of the oscillations, which
  is commonly done manually.  The volume of {\it Solar Dynamics
    Observatory} (SDO) data will make manual detection inefficient for
  detecting all of the oscillating regions.  An algorithm is presented
  which automates the detection of areas of the solar atmosphere that
  support spatially extended oscillations. The algorithm identifies
  areas in the solar atmosphere whose oscillation content is described
  by a single, dominant oscillation within a user-defined frequency
  range. The method is based on Bayesian spectral analysis of
  time-series and image filtering.  A Bayesian approach sidesteps the
  need for an {\it a-priori} noise estimate to calculate rejection
  criteria for the observed signal, and it also provides estimates of
  oscillation frequency, amplitude and noise, and the error in all
  these quantities, in a self-consistent way.  The algorithm also
  introduces the notion of {\it quality measures} to those regions for
  which a positive detection is claimed, allowing simple
  post-detection discrimination by the user.  The algorithm is
  demonstrated on two {\it Transition Region and Coronal Explorer}
  (TRACE) datasets, and comments regarding its suitability for
  oscillation detection in SDO are made.
\end{abstract}
\end{opening}

\section{Introduction}
\label{sec:Introduction} 

The {\it Solar and Heliospheric Observatory} (SOHO) and TRACE missions
established that the solar corona supports observable oscillations.
The extensive literature on the theory of oscillations in coronal flux
tubes has been used in conjunction with observationally derived
parameters to deduce conditions in the corona. The study of these
oscillations is called coronal seismology (in analogy to terrestrial
seismology).  Their general features, as determined by current
analyses, and their implications for the physics of the corona are
summarized briefly below; more detailed reviews can be found in
\inlinecite{dem2005a} and \inlinecite{nak2005a}.  Quasi-periodic
oscillations, interpreted as propagating slow waves have been detected
in coronal plumes SOHO/Ultraviolet Coronal Spectrograph (UVCS:
\opencite{ofman1997a}); SOHO/Extreme Ultraviolet Imaging Telescope
(EIT: \opencite{deforest1998a}).  A similar phenomenon (at higher
frequency) has also been detected at the base of coronal loops
(\opencite{berghmans1999a}; \opencite{nightingale1999a};
\opencite{dem2000a}; \opencite{robbrecht2001a}).  They are
interpreted as propagating slow waves since they travel at
approximately the sound speed and are seen as intensity (and therefore
density) variations propagating and decaying away from the base of the
flux tube through the corona.  The observed periodicities fall into
distinct ranges around three and five minutes
(\opencite{dem2002a}), suggesting that the oscillations are due to
different connectivity to either sunspot or transition-region moss
magnetic field respectively.  Coronal seismological applications for
these oscillations include determination of the connectivity of the
photosphere to the corona (\opencite{2005ApJ...624L..61D}) as well
as deriving information on the coronal heating function since the wave
observables (e.g., period and damping length) are strongly dependent
on the thermal conditions of the corona (\opencite{dem2003a};
\opencite{dem2004a}).

All of these studies have in common that the oscillating feature was
discovered by a manual examination of data in regions known (or
suspected) to contain oscillating material.  Although a successful
strategy for detecting oscillations, it is clearly impractical for
ever-increasing rates of data acquisition.  For example, the SDO
mission is acquiring >1.4 Tb of science data per day, around 1000
times the data rate of SOHO.  Much of this data will be taken at
cadences and spatial resolutions comparable to the best that TRACE can
produce, the key difference from TRACE being that SDO will provide a
near continuous, full disk coverage of the Sun in multiple wavelengths
simultaneously. Given the diagnostic possibilities of these waves, and
the large amount of data that will become available, an automated
detection algorithm is necessary.  There are at least five published
algorithms, which are briefly described below.

\inlinecite{nak2007permap} use a thresholded fast Fourier Transform to
find locations in TRACE data that may support an oscillatory signal.
The threshold level is defined as three\,--\,four times the average
FFT power; if the maximum FFT power is above this level then the
frequency at which that power occurs is assumed to be real. Since this
method relies on the FFT, it is very fast.  The definition of the
threshold is chosen to speed up the algorithm in comparison to
calculating a threshold; for example, implementing the randomization
method of \inlinecite{linnellnemec} requires at least an estimated 150
times more FFT calculations.  \inlinecite{linnellnemec} establish
frequency acceptance/rejection criteria based on probabilistic
arguments whereas \inlinecite{nak2007permap} use essentially
empirically derived arguments to establish the threshold.  Further,
identification of contiguous groups of pixels that might form an
oscillatory region is carried out by manual inspection.  To aid
identification, the authors assume a null hypothesis that the entire
time-series consists solely of Gaussian distributed noise, and so the
chance that any two neigboring pixels contain the same frequency is
small.  Hence, if two or more neighboring pixels do satsisfy the
selection criteria, then it is likely they are physically connected.
These assumptions ignore two effects: firstly, neighboring pixels are
not statistically independent due to the point spread function of the
instrument.  Secondly, the Sun is observed to have physically
connected structures, such as loops, that exist over many pixels in at
least one direction, and so by continuity, one can expect that
neighboring pixels do influence each other.  Hence the assumption that
neighboring pixels are independent of each other is an
over-simplification of the nature of the image.  The authors suggest
that the algorithm be used to identify regions in the data worthy of
further study, although there is no quoted method of automated region
identification.

\inlinecite{dem2004wavdet} describe an automated oscillation detection
algorithm based on the wavelet analysis routines of
\inlinecite{1998BAMS...79...61T} and the analysis procedure of
\inlinecite{1999AA...347..355I}.  The algorithm finds significant wave
packets ranging from single to multiple wave cycles in duration, by a
wavelet power/confidence level comparison against the null hypothesis
that a given time-series is Gaussian distributed noise.  Shorter
duration detections are rejected.  Contiguous regions of multi-cycle
duration wave packets are found in the data, but are identified and
isolated manually by inspection.  

\inlinecite{2008SoPh..248..395S} base their detection algorithm on
pixelized wavelet filtering (PWF) of a three-dimensional data cube
$(x,y,t)$. This too is based on the wavelet analysis routines of
\inlinecite{1998BAMS...79...61T}, but has a more complex treatment of
the resultant wavelet spectrum.  Regions of interest are found by
first calculating a variance map (\opencite{2003SoPh..213..103G})
of the signal; regions with high variance are candidate oscillatory
regions (note that this must also imply the removal of a background
trend in order for the variance to measure an oscillatory signal, and
not the trend).  Wavelet spectra are calculated for those pixels, and
only the ``significant'' pixels are retained (what constitutes a
significant signal is not stated explicitly). The routine analyzes
further (than \opencite{dem2004wavdet}) the temporal evolution of
the oscillation and so can differentiate between standing and
traveling waves.

The algorithm presented by \inlinecite{2008SoPh..252..321M} can also
differentiate between standing and traveling waves.  The algorithm
begins by Fourier transforming the entire data cube, performing cross
correlations with neighboring pixels in narrow frequency bands, and
filtering the results (by thresholding on various quantities, such as
eliminating areas where the relative error in the calculated phase
speed is large) to determine groups of pixels that are highly
correlated in both time and space.  This correlation technique allows
the discrimination of standing and traveling waves, and the
calculation of other parameters such as the phase speed and the
propagation angle.  There is an implicit null hypothesis in the
algorithm: at one stage, only time-series with a high coherence are
accepted for further analysis.  The null hypothesis here is that the
candidate time-series is pure noise, with the additional assumption
that pairs of noisy time-series have low coherence, and so can be
rejected.  However, this is not strictly true;
\inlinecite{chatfield1996} shows an example of two different noisy
time-series that have a perfect coherence over all spectral frequencies
because both time-series are generated from the same noise process.
Highly coherent patches of non-oscillatory material are probably
filtered out in the next stage of the algorithm, where regions with a
poorly determined phase speed (large error) are discarded; this,
however, has not been explicitly tested (S. W. McIntosh, private
communication, 2008).

It is clear that there is an increasing amount of effort in finding
oscillatory regions and identifying waves in the solar atmosphere.
All the above algorithms show promising results and avenues for
further work.  The algorithm presented here seeks to find regions in
the data which support oscillations {\it via} Bayesian time-series
analysis.  This involves calculating explicitly the probability that a
time series supports an oscillation of a given frequency.  This is in
distinction to the methods above, which rely on statements about null
hypotheses in order to determine if an oscillation is present.
Section \ref{sec:btsa} introduces Bayes' Theorem and an application of
it to time-series analysis. Section \ref{sec:detectalg} describes a
detection algorithm based on the results of Section \ref{sec:btsa},
whilst Section \ref{sec:results} describes the application of this
algorithm to some example datasets from the TRACE mission.  Finally,
Section \ref{sec:conc} discussions some further applications of
Bayesian time-series analysis and automated EUV detection algorithms.

\section{Bayesian Time-Series Analysis}\label{sec:btsa}

Denoting by $p(a|b)$ the conditional probability that proposition $a$ is
true, given that proposition $b$ is true, Bayes' theorem is
\begin{equation}
p(H|D,I) = \frac{p(H|I)p(D|H,I)}{p(D|I)}
\label{eqn:bayes}
\end{equation}
where $H$ is the hypothesis to be tested, $D$ is the observation, and
$I$ is any applicable prior information that we have before making the
observation (\opencite{bayes1763}). The left hand side $p(H|D,I)$
is called the {\it posterior} probability of the hypothesis, given the
data and the prior information, and it encapsulates the available
knowledge about the hypothesis.  The quantity $p(H|I)$ is called the
prior distribution and represents what we know about $H$ prior to any
data collection. Often a prior describes a probability distribution of
likely parameter values.  The sampling distribution or likelihood
($p(D|H,I)$) represents the likelihood of the data given the
hypothesis (as well as any prior information).  The quantity $p(D|I)$
is the prior probability of the data; it is absorbed into a
normalization constant, and does not affect the following analysis for
a given model.  Equation (\ref{eqn:bayes}) is very general and is not
restricted to the mathematical equations: any logical proposition can
be treated in a Bayesian context (\opencite{jaynes2003};
\opencite{gregory2005book}).

\subsection{Signal Containing a Single Frequency}\label{sec:singfreq}
For the purposes of detecting oscillations in solar time-series
$d(t_{i})$, the linear model
\begin{equation}
d(t_{i}) = b_{1}\cos{\omega t_{i}} + b_{2}\sin(\omega t_{i}) + x_{i}
\label{eqn:model}
\end{equation}
is used.  This assumes that all time-series are modeled as a single
oscillation plus Gaussian distributed noise.  Under some simplifying
assumptions ($N>>1$ and that there are no low-frequency oscillations
present), \inlinecite{jaynes1987a}, \inlinecite{bretthorst1988book},
and \inlinecite{ruan} derive an expression for the probability that
the time-series contains an oscillation of frequency $\omega$ for the
model oscillation above, that is,
\begin{equation}
  p(\omega|\md, I) \propto \left[ 1 - \frac{2C(\omega)}{n\rho^{2}} \right]^{(2-n)/2}
\label{eqn:prob}
\end{equation}
where $\rho = 1/n\sum_{j=1}^{n}d_{j}^{2}$ and
\begin{equation}
C(\omega) = \frac{1}{N}\left|\sum_{j}^{n}d_{j}e^{i\omega t_{j}}\right|^{2}
\label{eqn:schuster}
\end{equation}
is the Schuster periodogram (\opencite{schuster1898}).  The full
details of the derivation of Equation (\ref{eqn:prob}) are contained in
the appendix. This is a probability measure that a given frequency is
present in the data that does not require explicit knowledge of the
Gaussian noise present. Equation (\ref{eqn:prob}) can analyze unevenly
sampled data due to its use of the Schuster periodogram.  If the
analysis frequencies are
\begin{equation}
\label{eqn:freq}
\omega_{p} = 2\pi p/n, 0\le p\le n/2-1,
\end{equation}
Equation (\ref{eqn:schuster}) is proportional to the power of the fast
Fourier transform (FFT) of the data (\opencite{cooley1965};
\opencite{chatfield1996}).  The speed of the FFT transform can be
exploited for data which are evenly or close to evenly spaced.
\begin{figure}
  \centerline{
    \includegraphics[width=1.0\textwidth,clip=]{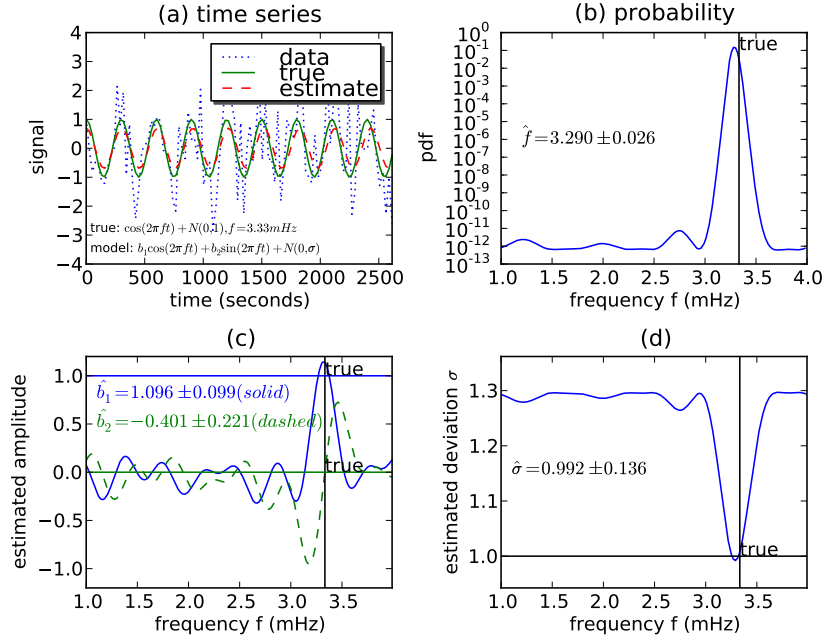}}
  \caption{An example analysis using the Bayesian approach of Section
    \ref{sec:btsa}.  Panel (a) shows the true oscillation, the
    observed (noisy) data, and the estimate.  Panel (b) shows the
    probability-density function arising from the observed data.  Panel
    (c) shows the estimated amplitudes as a function of frequency [$f$],
    and panel (d) shows the estimated noise as a function of
    frequency.  Final quantity estimates and an estimated error are
    shown also.  See Section \ref{sec:btsa} for more detail.}\label{fig:genlinex}
\end{figure}

\subsection{Example Analysis}\label{sec:exan}
Figure \ref{fig:genlinex} demonstrates the application of the above
equations to the analysis of a single time-series.  The true
oscillation is an evenly sampled simple cosine at frequency $f = 3.33$
mHz (100 data points at a cadence of 26.4 seconds) with unit
amplitude.  The fake data is this oscillation corrupted by Gaussian
distributed noise with deviation of $\sigma=1$, a signal-to-noise
ratio of 1 (Figure \ref{fig:genlinex}(a)).  The probability function,
Equation \ref{eqn:prob}, is highly peaked close to the true frequency
and the probability that the signal contains a frequency within the
range 3.0\,--\,3.5 mHz exceeds 99\% (Figure \ref{fig:genlinex}(b)).
Figures \ref{fig:genlinex}(c,d) show the measured basis function
amplitudes and Gaussian deviation estimate respectively as a function
of frequency. Both lie close to their true values at the true
frequency, as indicated by the solid vertical line.  The quoted values
shown in each plot (Figure \ref{fig:genlinex}(b,c,d)) are found at
$\omega = \hat{\omega}$, the most probable frequency.  An error
estimate is given by the standard deviation
\begin{equation}\label{eqn:expect:sd}
\left[\mbox{sd}\left(\hat{\phi}\right)\right]^{2} =
\sum_{i=1}^{N_{\omega}}p(\omega_{i})\left[\phi(\omega_{i})-\phi(\hat{\omega})\right]^{2}
\end{equation}
where $\phi(\omega)$ stands for the analyzed frequency range, the
distribution of amplitudes (found from Equation (\ref{eqn:amp}) -- see
appendix) and the distribution of $\sigma$ (found from Equation
(\ref{eqn:sigma}) -- see appendix) as a function of the analysis
frequency set $\omega_{i}, 1\le i \le N_{\omega}$ (such as the Fourier
set, Equation (\ref{eqn:freq})).

These estimates of frequency probability, oscillation amplitude and
Gaussian noise deviation have all been obtained without explicitly
performing a least-squares fit, or with no special {\it a-priori}
knowledge of the noise in the signal other than the assumption of a
Gaussian distribution.

The peakedness of the Bayesian probability-density function
immediately suggests that a search for portions of the $\ww$ parameter
space that contain most of the probability are the values that are of
greatest interest.  In addition, prior knowledge from previous studies
of solar atmospheric oscillations suggests which regions of the
parameter space are of interest.  These two observations are combined
below to take the first steps towards creating an automated
oscillation detection algorithm which can identify oscillation regions
and return useful information, such as their amplitude.

\section{Detection using Bayesian Spectral Analysis}
\label{sec:detectalg}

\subsection{Data}\label{sec:data}
The data that we will use are three dimensional datacubes $(x,y,t)$,
with the $x$ and $y$ locations referring to spatial locations (pixels)
on the Sun, and $t$, time.  time-series are formed by choosing a
particular location on the Sun and extracting a one-dimensional
time-series (occasionally super-pixels formed by the sum of
neighbouring pixels are used to increase the signal-to-
noise ratio).

\subsection{Probability and Frequency Bands}\label{sec:int}
In order to use Equation (\ref{eqn:prob}) to detect oscillations,
further assumptions must be made.  The algorithm assumes that the
range of possible frequencies is limited to that spanned by the Fast
Fourier Transform applied to a time-series of similar length.  This
permits us to normalize the probablity density function over a fixed
range of frequencies.  We are commonly interested in detecting
oscillations in given frequency bands (say the three or five minute
oscillation frequency bands), and so we calculate the probability that
the oscillation in a given pixel lies within the range
[$\omega_{1},\omega_{2}$], that is,
\begin{equation}
p_{\omega_{1},\omega_{2}}=\int_{\omega_{1}}^{\omega_{2}}p(\omega|D,I)\mbox{d}\omega
\label{eqn:int}
\end{equation}
at every point in the image.  Large values of
$p_{\omega_{1},\omega_{2}}$ indicate that the true oscillation
frequency is very likely to be within the range
[$\omega_{1},\omega_{2}$].

The Bayesian formulation handily yields both computational and logical
advantage over other frequentist approaches for oscillating-pixel
detection.  The prime derived data product for each pixel is a
probability distribution describing the probability of a given
frequency in the data and so there is no need to perform secondary,
often computationally expensive calculations, such as randomization
tests (\opencite{linnellnemec}; \opencite{oshea2001}) to
assess the probability that the frequency is present or not.
Frequentist-based approaches to detecting oscillations rely on
calculating the expectated value of the Fourier transform power at a
given confidence level, assuming that the observed data arises from
the null hypothesis that the time-series is pure noise.  If the
Fourier transform power of the observed time-series exceeds the
expectation value, then the null hypothesis is rejected at that
confidence level.  This, however, strictly cannot be used to imply the
presence of an oscillation; we have merely rejected the null
hypothesis.  In comparison, Equation (\ref{eqn:int}) {\it directly}
calculates the probability that the frequency of oscillation lies in
the range [$\omega_{1},\omega_{2}$], under the modeling assumptions of
Section \ref{sec:btsa}.

\subsection{An Algorithm} \label{sec:algorithm} Figure
\ref{fig:algorithm} describes the general-purpose algorithm used to
find oscillating spatial locations in the data cube.  Step 2 relies on
instrument calibration routines provided by instrument teams, and on
standard solar de-rotation routines, both provided in the {\it
  IDL/Solarsoft} package.  Time-series de-trending (Step 3) is
necessary for two reasons.  The first is dictated by our choice of
oscillation model, Equation (\ref{eqn:model}). This model assumes that
the time series contains a single oscillation only, and no other
features. (It is certainly possible to define other models that do
contain a background trend, or multiple frequencies, and calculate
probability density functions for those time-series.  However, these
more sophisticated analyses are not necessary for us to make progress
in the current application of locating oscillating material).
Secondly, strong background trends can pollute the Fourier power
spectrum with spectral power unrelated to the oscillation.  This can
lead to the mis-identification of peaks in the power spectrum as
oscillatory when they are not.  It should be noted that the influence
of background trend on locating oscillations of a given frequency will
influence all proposed automation algorithms.

\begin{figure}
\begin{tabbing}
\bf{Algorithm: \algname} \\
1 be\=gin   \\
2   \> Prepare data: apply instrument calibrations and de-rotate data cube.\\
3   \> Detrend data and ensure data has zero mean. \\
4   \> for \= each time-series $d_{x,y}(t)$ in the data cube $(x,y,t)$ \\
5   \>       \> Calculate $p_{\omega_{1},\omega_{2}}$ for given $\omega_{1},\omega_{2}$ at $(x,y)$.\\
6   \> end \\
7   \> Generate map $M$ of the spatial distribution of probability $p_{\omega_{1},\omega_{2}}$ \\
8   \> Filter map to find the ``highly probable'' oscillation areas. \\
9   \> Report these areas and derive useful parameters. \\
10 end \\
\end{tabbing}
\caption{Pseudo-code algorithm to find oscillating locations in
  $(x,y,t)$ data cubes.}
\label{fig:algorithm}
\end{figure}

De-trending in Step 3 is accomplished by subtracting a running average
of the time-series (a window of size $R$ seconds is slid across the
time-series and the running average of the data lying entirely within
that window is calculated). Since previous experience has told us that
there are oscillations of interest that have periods less than 500
seconds, it seems to us that 500 seconds is a reasonable to chose for
these data (\opencite{2002SoPh..206...99A}). Timescales longer than
$R$ can be considered to be associated with the background
trend. Timescales shorter than $R$ potentially support an
oscillation. Fourier power spectra of smooth time-series that have had
their background trend subtracted have extremely low power at
frequencies less than that corresponding to $R$. Therefore, these low
frequencies corresponding to the background trend have an extremely
low probability and are not selected.

Step 5 finds the probability that the time-series has a single
frequency in the range $[\omega_{1},\omega_{2}]$. There are several ways
of defining $[\omega_{1},\omega_{2}]$. In the work below the algorithm
begins by first finding the location $\omega_{\mbox{max}}$ of the highest
peak in the probability distribution function (PDF) $p(\omega|D, I)$.
If $[\Omega_{1}\le \omega_{\mbox{max}} \le \Omega_{2}]$, where
$[\Omega_{1},\Omega_{2}]$ are defined by the user (a frequency filter),
then the algorithm proceeds by stepping away from the peak to find the
nearest turning points in the PDF, located at $[\omega_{1},\omega_{2}]$.
The probability [$p_{\omega_{1},\omega_{2}}$] is then calculated. This
creates the probability map $M$ of Step 7.

%
%
\begin{figure}
  \centerline{\hspace*{0.015\textwidth}
    \includegraphics[width=0.515\textwidth,clip=]{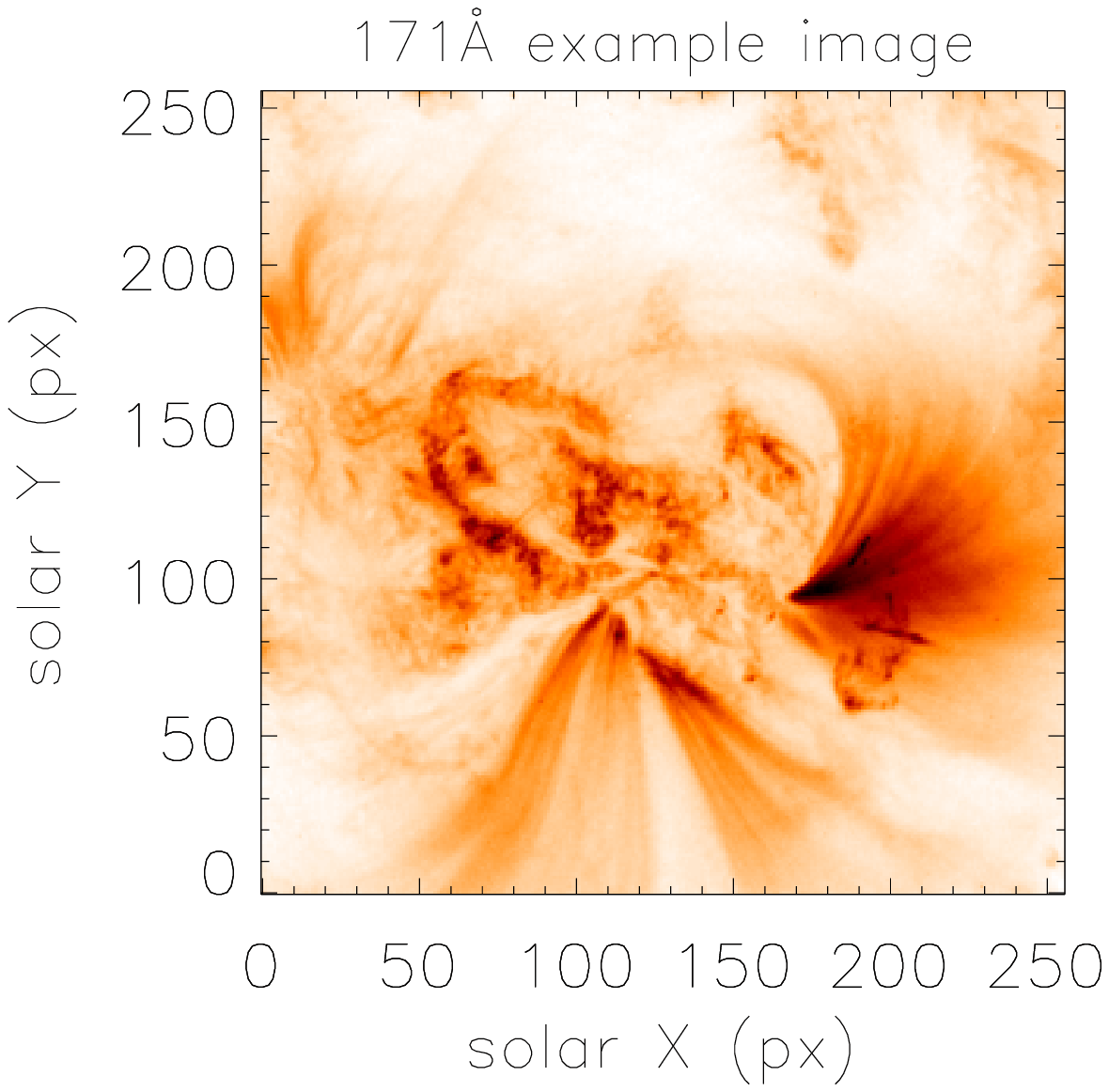}
    \hspace*{-0.03\textwidth}
    \includegraphics[width=0.515\textwidth,clip=]{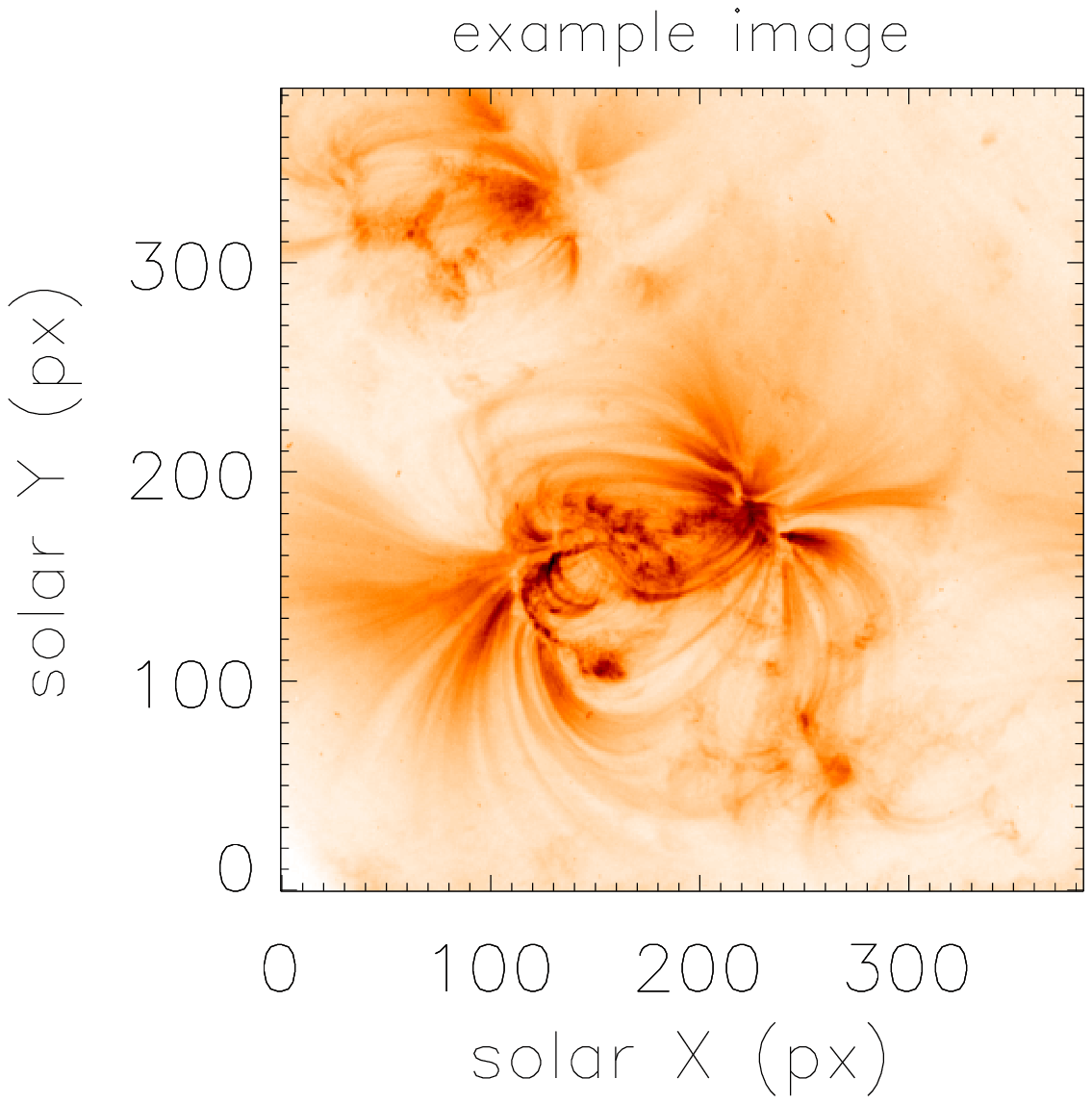}
    \vspace{0.0 \textwidth}}
  \vspace{-0.38\textwidth}   
  \centerline{
    \hspace{0.0 \textwidth}  \color{black}{Jul. 1}
    \hspace{0.415\textwidth}  \color{black}{Jul. 14}
    \hfill}
  \vspace{0.02\textwidth}   
  \centerline{
    \hspace{0.0 \textwidth}  \color{black}{(a)}
    \hspace{0.435\textwidth}  \color{black}{(d)}
    \hfill}
  \vspace{0.30\textwidth}    
  \centerline{\hspace*{0.015\textwidth}
    \includegraphics[width=0.515\textwidth,clip=]{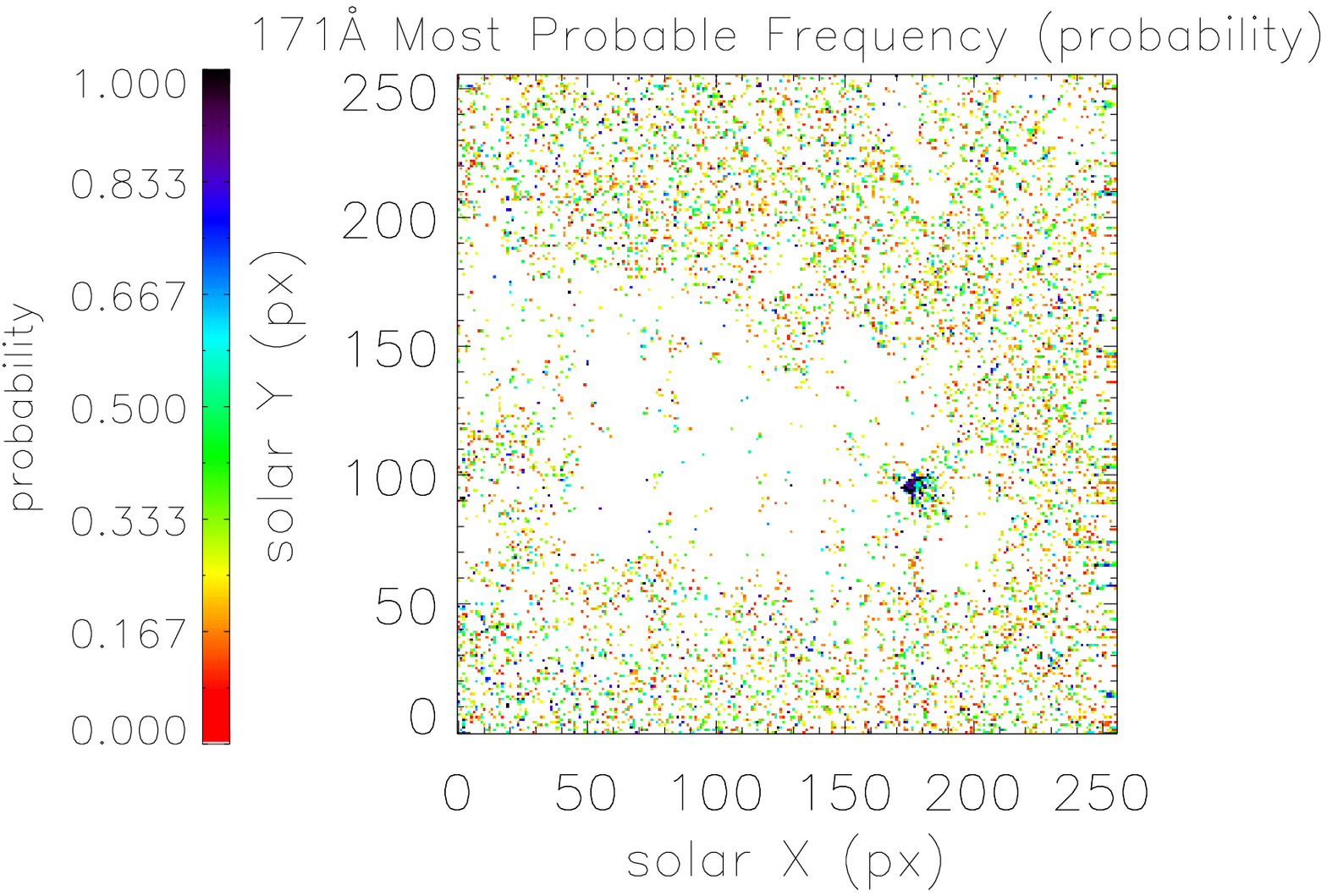}
   \hspace*{-0.03\textwidth}
    \includegraphics[width=0.515\textwidth,clip=]{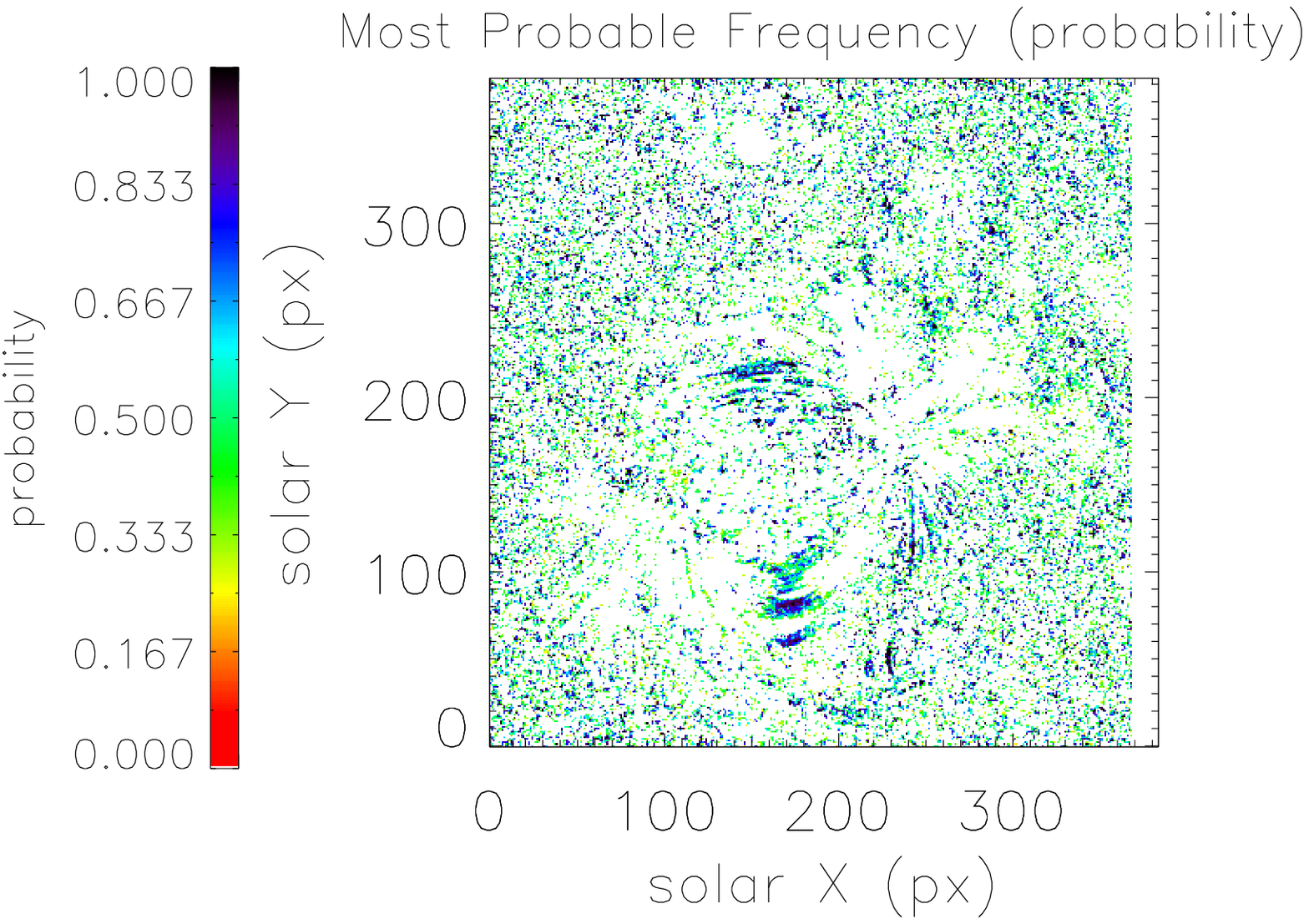}
   \vspace{0.0 \textwidth}}
  \vspace{-0.38\textwidth}   
  \centerline{
    \hspace{0.0 \textwidth}  \color{black}{(b)}
    \hspace{0.435\textwidth}  \color{black}{(e)}
    \hfill}
  \vspace{0.35\textwidth}    
  \centerline{\hspace*{0.015\textwidth}
    \includegraphics[width=0.515\textwidth,clip=]{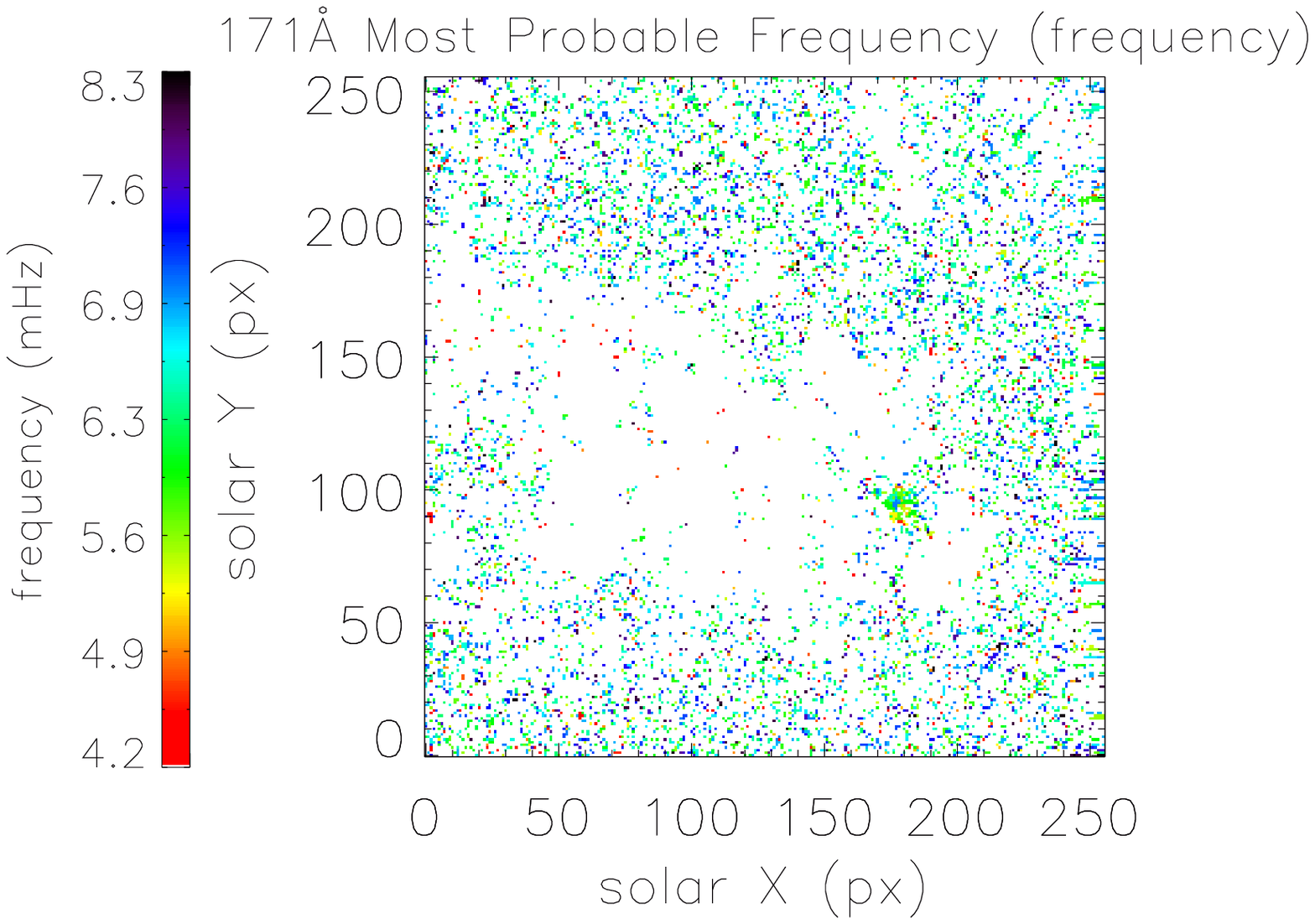}
   \hspace*{-0.03\textwidth}
    \includegraphics[width=0.515\textwidth,clip=]{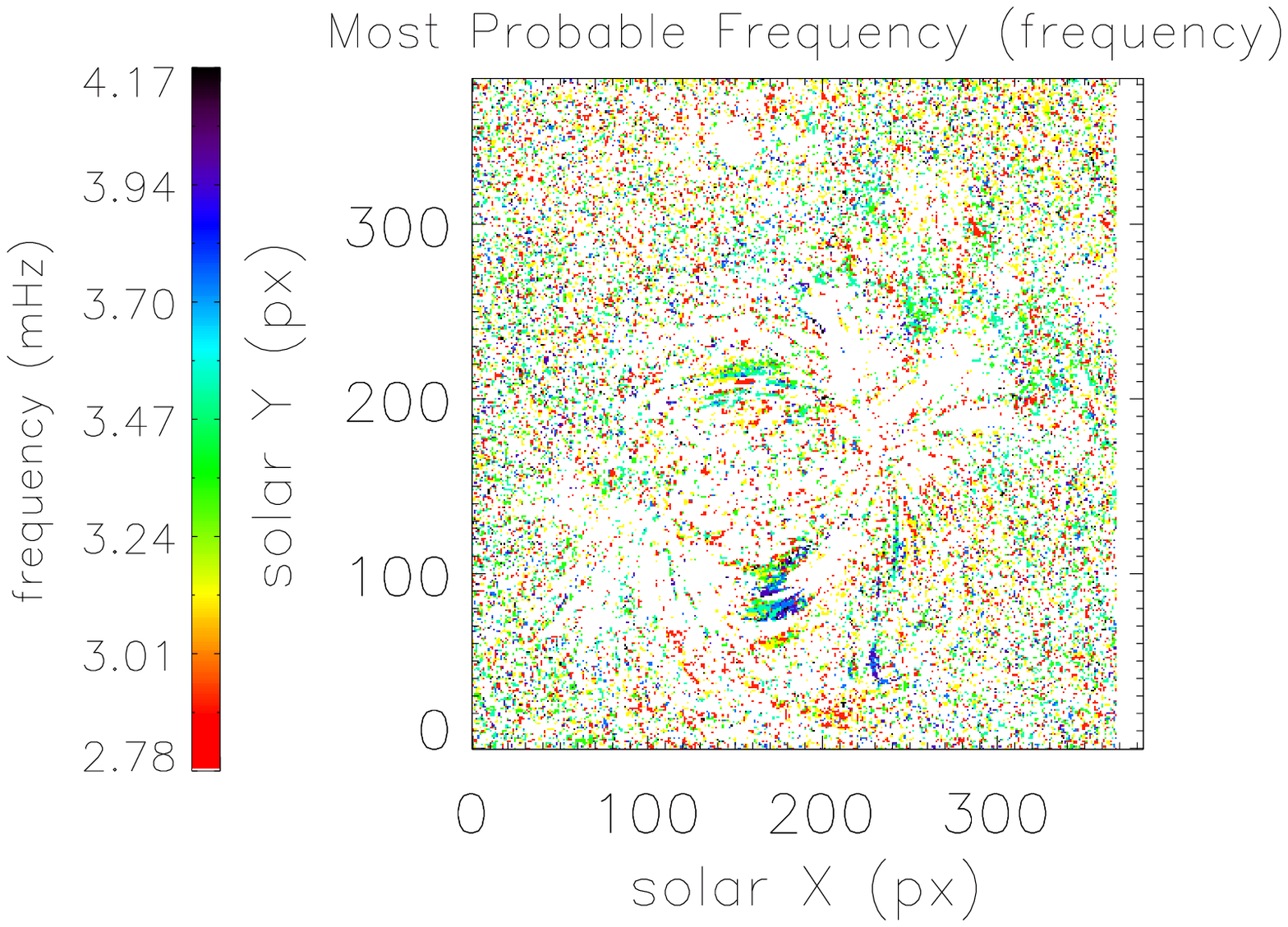}
   \vspace{0.0 \textwidth}}
  \vspace{-0.38\textwidth}   
  \centerline{
    \hspace{0.0 \textwidth}  \color{black}{(c)}
    \hspace{0.435\textwidth}  \color{black}{(f)}
    \hfill}
  \vspace{0.35\textwidth}    
  \caption{Example image data and probability and frequency maps.
    Panels a, b, and c refer to TRACE 171\AA\ data taken on 1st July
    1998 and analyzed for oscillations in the three-minute frequency band
    $4.17\le f_{\mbox{max}}\le 8.33$ mHz), and
    panels d, e and f refer to TRACE 171\AA\ data taken on 14 July
    1998 and analyzed for oscillations in the five-minute
    frequency band ($2.78\le f_{\mbox{max}}\le 4.17$ mHz) .}
  \label{fig:probex}
\end{figure}

Figure \ref{fig:probex}(a,b,c) shows some example data, and its
analysis up to Step 7.  Two datasets are examined, both in the TRACE
171\AA\ wave-band.  The first dataset consists of 391 observations
with a mean 31s cadence taken on 1 July 1998 12:03:10 UT -- 15:28:52
UT.  The resulting probability map and its concomitant frequency map
are shown in Figures \ref{fig:probex}(b,c) respectively. There appears
to be a single large area that appears to contain a significant
oscillation at differing frequencies and probabilities.

Cleanly extracting oscillating features is much more difficult for the
second example dataset shows in Figures \ref{fig:probex}(d,e,f).  Here
there are many groups of pixels having similar frequencies and high
probability close to each other.  In addition, there are many pixels
that have high probability scattered all over the field of view,
further complicating the task of extracting a single group.  The eye
is very good at picking out such groups, and the algorithm attempts to
mimic some of this behavior.  The detection algorithm attempts to
mimic this process by examining the local spatial distribution of
probability.  Firstly, it is assumed that nearby pixels of high
probability are physically related, even if their frequencies may be
different.  However, these groupings may not be contiguous due to
noise, for example.  This may be overcome by smoothing the probability
map at different lengthscales, and flagging those areas which exceed
a certain threshold.  This has the effect of the allowing
discontinuous areas of high probability to merge, in a similar way to
how the eye may integrate physically ``close'' oscillating areas into
one group.  More formally, Step 8 implements
\begin{enumerate}
\item for $m$ an integer in the range $[L_{1},L_{2}]:$
  \begin{enumerate}
  \item Smooth the probability map [$M$] with a Gaussian filter of
    width $m$.
  \item Generate a mask [$Z_{m}$] locating all the areas in the
    smoothed map that have a smoothed probability at or over 0.5
  \end{enumerate}
  The lengthscales $m$ allow the user to examine the average spatial
  probability structure of the data on multiple lengthscales.  This
  procedure is meant to mimic the way the eye examines such
  probability maps, where the eye finds it easy to integrate over
  neighbouring high probability features.  The upper lengthscale
  [$L_{2}$] describes the maximum distance that two pixels are assumed
  to be potentially physically related to each other.  Smaller maximum
  values of $m$ cause the smoothed probability masks to break up more,
  whereas larger values can cause the creation of groups of pixels
  which are essentially unrelated.  The masks [$Z_{m}$] found in step
  1b denote the areas that are more likely than not to support
  contiguous oscillations over the lengthscale $m$.  The net effect of
  step 1 is to implement a simple multi-scale analysis of the spatial
  probability structure.
\item Add all the masks [$Z_{m}$] together.  This is the mask --- $Z$
  --- of all the candidate pixel groups that may support a significant
  oscillations.
\item Remove from $Z$ all the candidate pixel groups that form a group
  smaller than $g$ pixels.  Nominally, we set a minimum group size
  $g=L_{2}^{2}$; that is, we look for groups of pixels that are
  connected over the largest distance set in the multi-scale analysis
  of step 1.
\item For a candidate group of area $A$, remove that group from $Z$ if
  it has more than $hA$ zero-probability pixels, $0\le h \le 1$.  The
  remaining pixel groups are considered to be regions in the solar
  atmosphere that support oscillations in the range
  $\Omega_{1},\Omega_{2}$.
\end{enumerate}
In the results below, $L_{1}=1$ and $L_{2}=4$, $g=16$; these are
chosen in order to allow smaller width strands to be found in the
data.  
The last part of Step 8 acknowledges that in performing the
probability-map smoothing, pixels that contain no oscillation in the
range $\Omega_{1},\Omega_{2}$ can be swept into a candidate pixel
group $A$, and this must have a limit lest the candidate pixel group
have too many lacunae. The final step (Step 9,
Figure\ref{fig:algorithm}) is to report properties of the remaining
pixel groups such as the average frequency [$\overline{f}$] and the
standard deviation [$\sigma_{f}$] of the frequency.  The full table of
reported results is given in Table \ref{tab:measures}.
\begin{table}
\begin{tabular}{clc}
quantity             & definition  & units \\ \hline
$\overline{f}$       & average frequency in a group $G$  & mHz\\
$\sigma_{f}$         & standard deviation of frequency  & mHz\\
$A$                  & area of $G$ with a non-zero probability & area (px) \\
$A_{0.95}$            & area of $G$ with $p_{\omega_{1},\omega_{2}}\ge 0.95$  & area (px)\\
$F$                  & $A/(\mbox{area of group $G$})$ & {\it dimensionless} \\
$\overline{p}$       & average probability of all pixels having a non-zero probability & {\it dimensionless} \\
$Q$                  & $F\overline{p}$ ({\it quality} of the group $G$) & {\it dimensionless} \\
$E$                  & $AQ$ (the {\it equivalent area} of the group $G$) & area (px)\\
\end{tabular}
\caption{Quantities reported for a contiguous group [$G$] of pixels found by the detection algorithm of Figure \ref{fig:algorithm}.}
\label{tab:measures}
\end{table}
In any given analysis, the regions found result from a series of
filters and thresholds made by the user.  However, not all of the
regions that survive this process are necessarily equivalent, they
have merely satisfied some set of criteria.  These quantities attempt
to measure differences between the surviving oscillatory regions, and
allow the user to further discriminate them at their discretion.
Clearly the size [area $A$] of the pixel group is important, and the
number of highly probable pixels [$A_{0.95}$].  The quantity $F$
measures how complete the group is, and $\overline{p}$ measures the
average probability that the average oscillation frequency of $G$ is
indeed between $\Omega_{1},\Omega_{2}$.  The {\it quality} [$Q$] is an
average probability for the entire group, and the {\it equivalent
  area} [$E$] is the area that the group would have if it were
entirely complete (no lacunae) and every oscillation detected in it
had $p_{\omega_{1},\omega_{2}}=1$.  These measures, along with maps of
the detected regions showing frequency, amplitude, and noise estimates
(along with error estimates to each of these quantities), are the
final derived products of this automated detection analysis.

\section{Results}
\label{sec:results}

Data was analyzed from two TRACE observing sequences.  Data was
prepared using standard {\it IDL/Solarsoft} TRACE routines ({\it
  TRACE\_PREP} with the keywords {\it /wave2point, /unspike,
  /destreak, /deripple, /norm, /float} switched on) and derotated
using {\it IDL/Solarsoft SHIFT\_XY} to calculate the motion of the
observed piece of Sun, and cubic two-dimensional interpolation to
shift subsequent images back to the location of the first one. In
addition, a strip of data of width ten pixels at the extreme right of
the image is removed, as this is the location of severe image
distortion due to image derotation. Further, all time-series are
detrended by removing a running average taken over a $R=500$ second
timescale (see Step 3, Figure \ref{fig:algorithm}).  In the following
results, signal-to-noise ratio (SNR) is calculated as the estimated
amplitude $\sqrt{ \hat{b_{1}}^{2} + \hat{b_{2}}^{2}}$ (see Equation
(\ref{eqn:model})) divided by the Gaussian noise estimate
$\hat{\sigma}$.

\subsection{1 July 1998}
\label{sec:results:jul1}
\begin{figure}
  \centerline{\hspace*{0.015\textwidth}
    \includegraphics[width=0.515\textwidth,clip=]{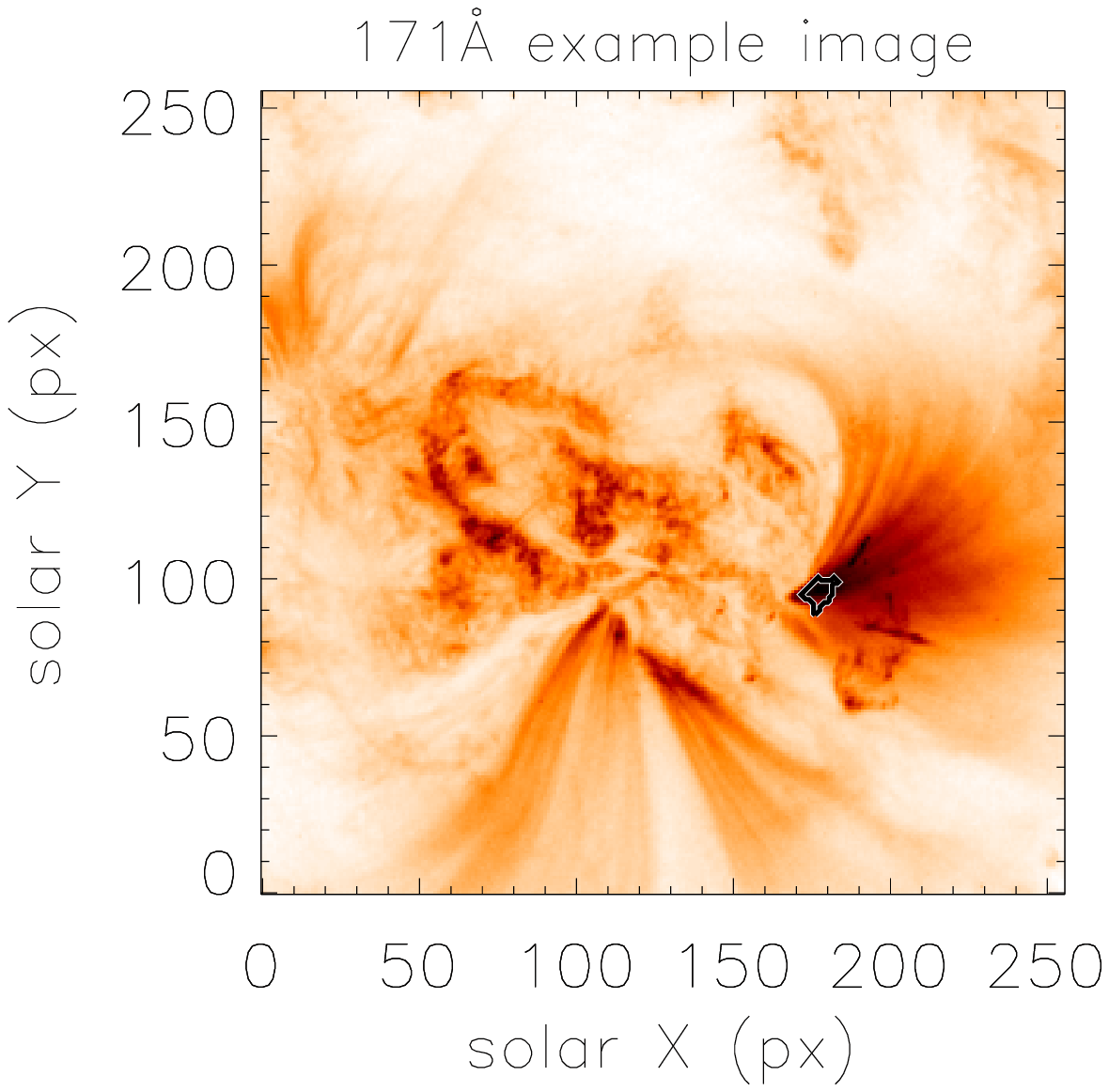}
  \hspace*{-0.03\textwidth}
    \includegraphics[width=0.515\textwidth,clip=]{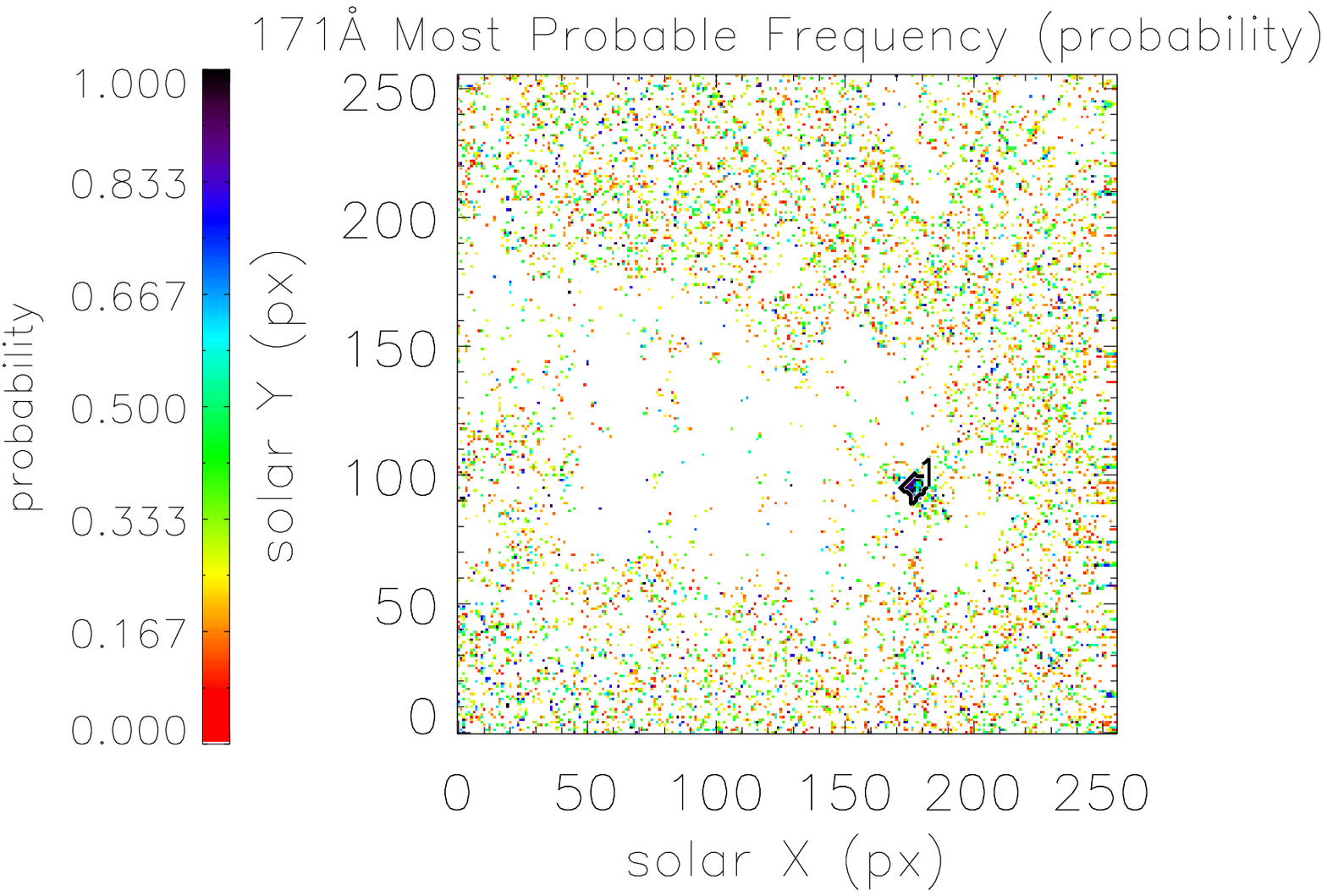}
  \vspace{0.0 \textwidth}}
  \vspace{-0.38\textwidth}   
  \centerline{
    \hspace{0.0 \textwidth}  \color{black}{(a)}
    \hspace{0.435\textwidth}  \color{black}{(b)}
    \hfill}
  \vspace{0.35\textwidth}    
  \centerline{\hspace*{0.015\textwidth}
    \includegraphics[width=0.515\textwidth,clip=]{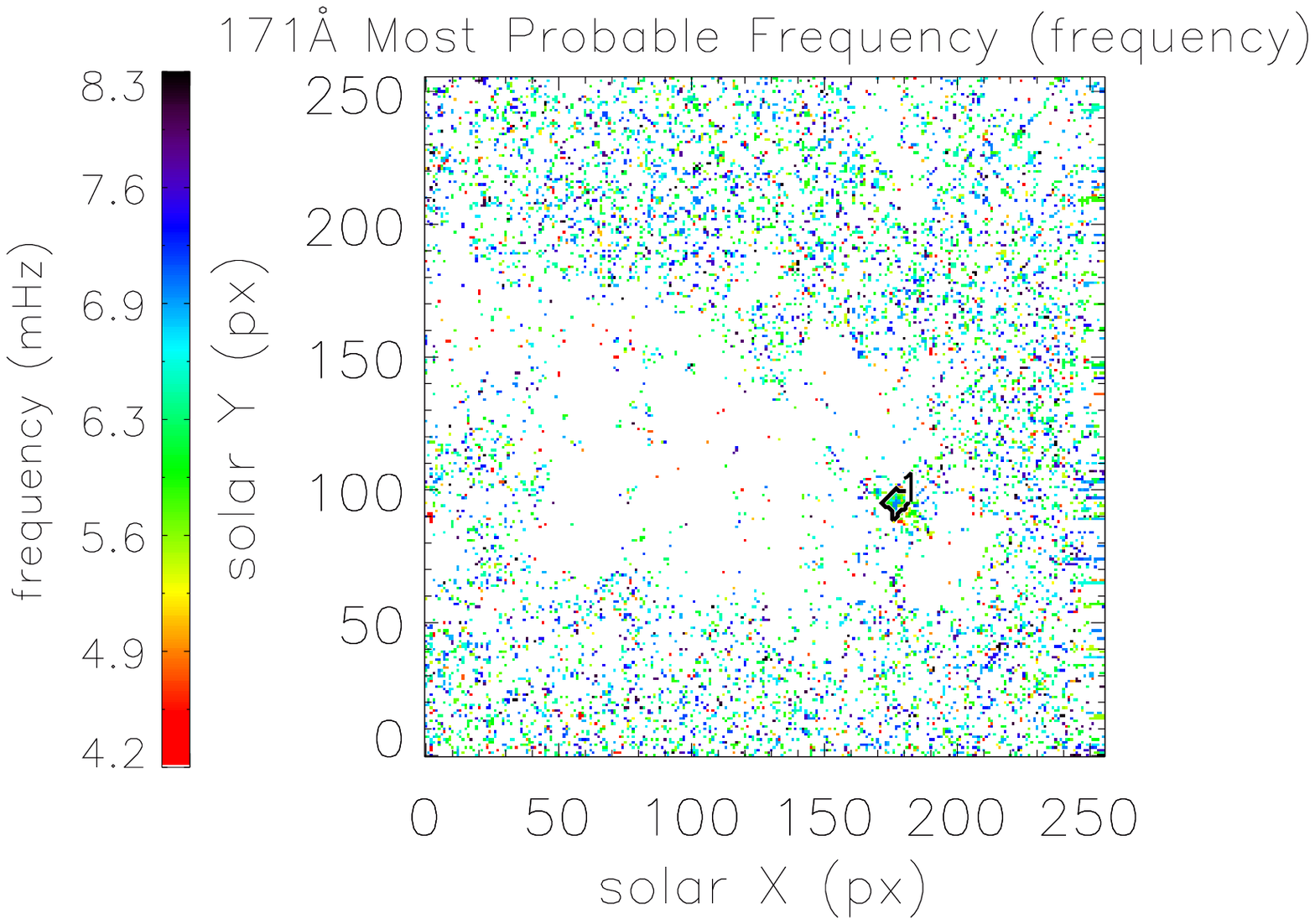}
    \hspace*{-0.03\textwidth}
    \includegraphics[width=0.515\textwidth,clip=]{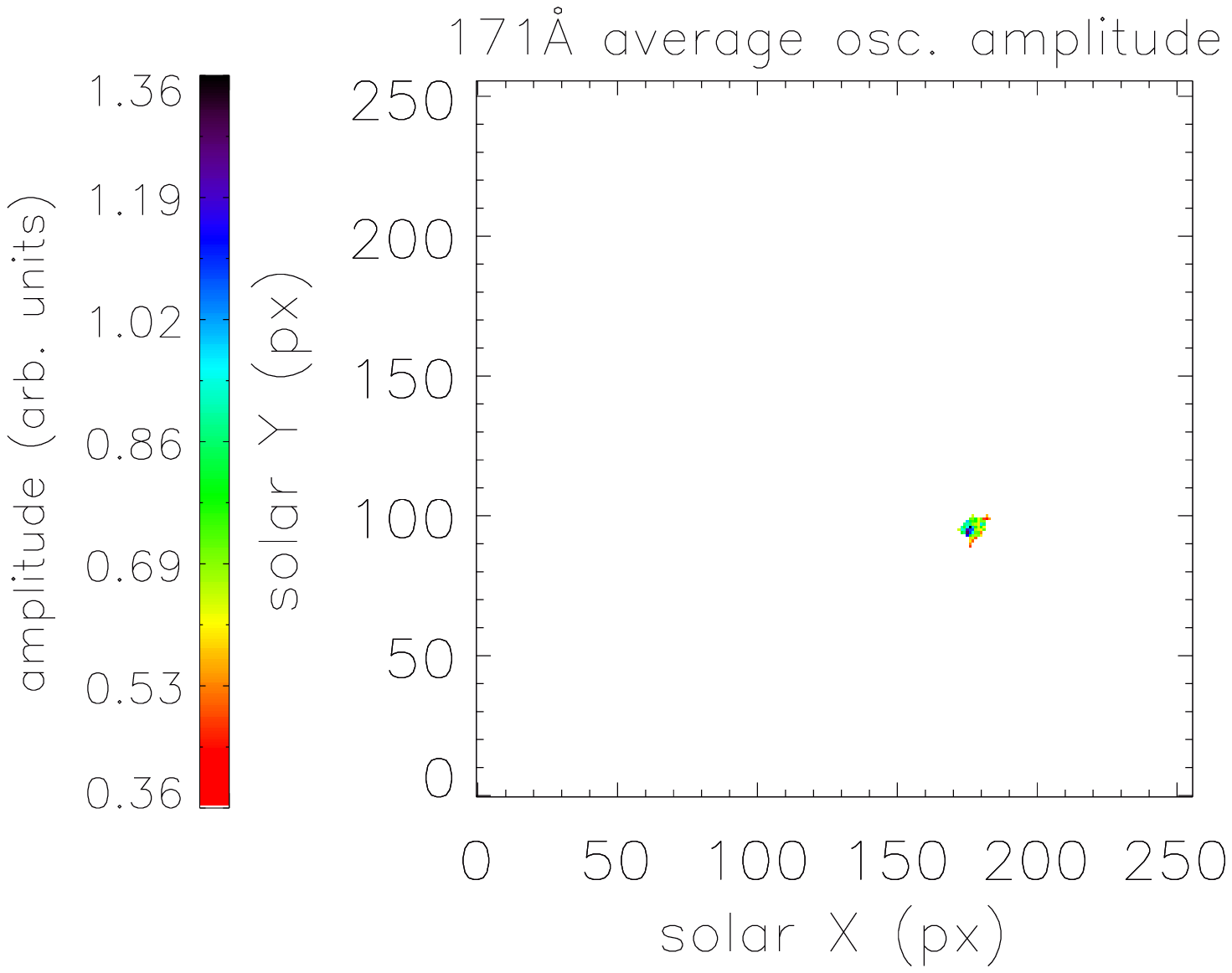}
    \vspace{0.0 \textwidth}}
  \vspace{-0.38\textwidth}   
  \centerline{
    \hspace{0.0 \textwidth}  \color{black}{(c)}
    \hspace{0.435\textwidth}  \color{black}{(d)}
    \hfill}
  \vspace{0.35\textwidth}    
  \centerline{\hspace*{0.015\textwidth}
    \includegraphics[width=0.515\textwidth,clip=]{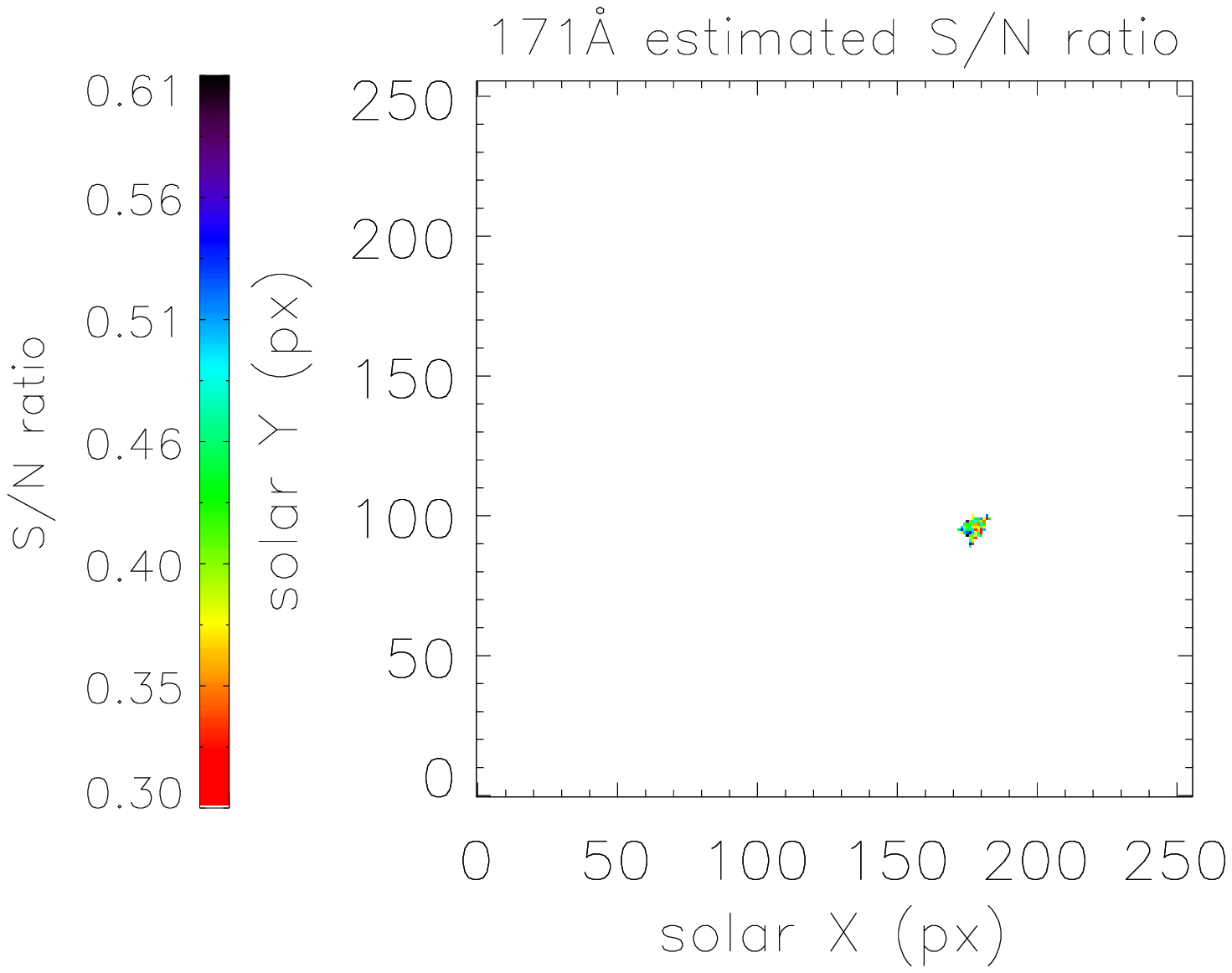}
    \hspace*{-0.03\textwidth}
    \includegraphics[width=0.515\textwidth,clip=]{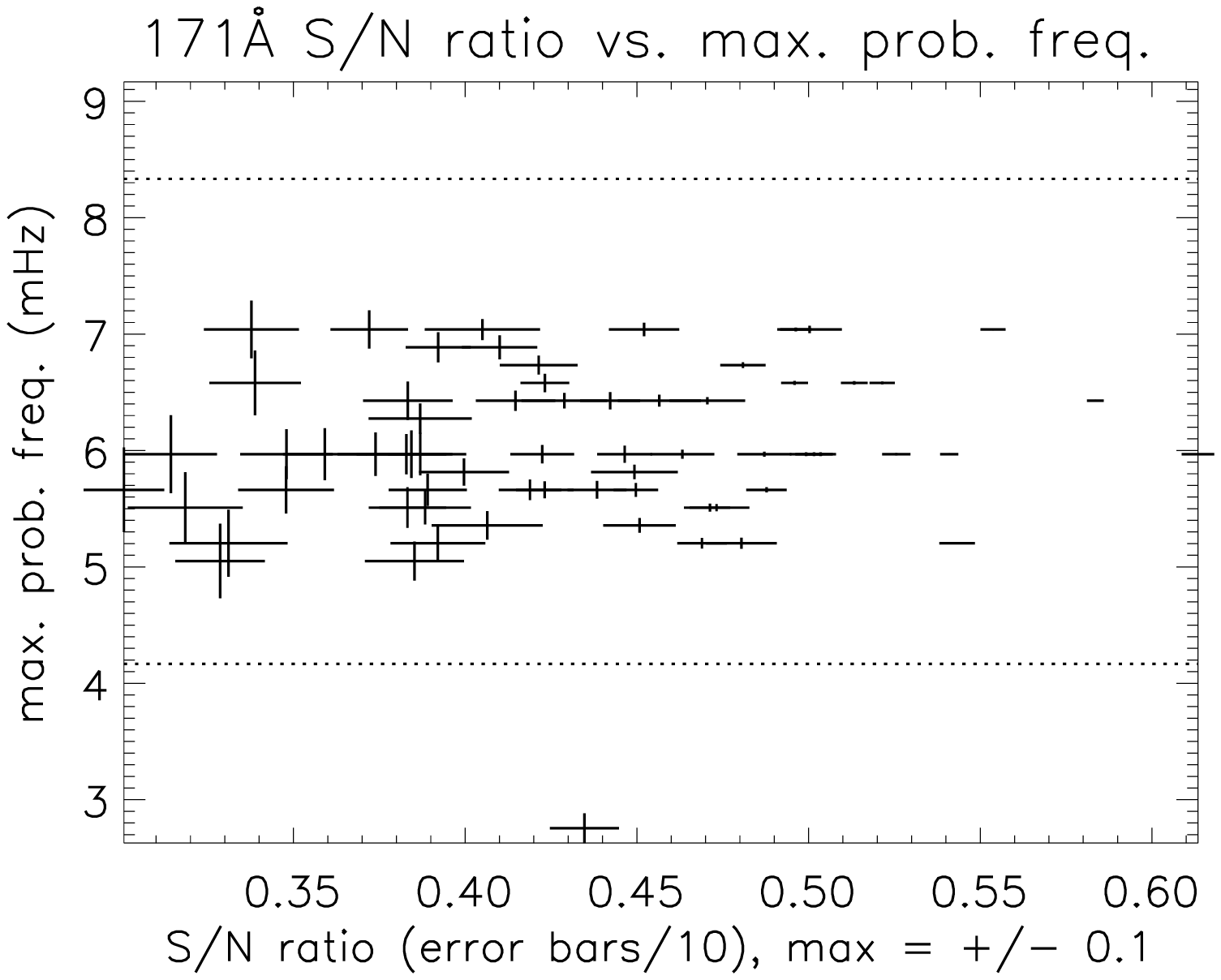}
    \vspace{0.0 \textwidth}}
  \vspace{-0.38\textwidth}   
  \centerline{
    \hspace{0.0 \textwidth}  \color{black}{(e)}
    \hspace{0.435\textwidth}  \color{black}{(f)}
    \hfill}
  \vspace{0.35\textwidth}    
  \caption{Example image data and results from the analysis algorithm
    for TRACE 171\AA\ data taken on 1 July 1998 (panel (a)). Panel (b)
    is the probability that a given pixel supports an oscillation
    within the frequency band indicated in panel (c), in this case, a
    three-minute oscillation frequency band.  Panel (c) shows the
    frquency supported at those pixels.  Also indicated in panels
    (a\,--\,c) are the detected oscillation regions.  Note that the
    region agrees with that found manually by
    \protect\inlinecite{king2003} on the following day.  The
    oscillation region is found at the base of a coronal loop
    structure, as has been found by many authors
    (\protect\opencite{dem2002a}).  Panel (d) shows a map of
    estimated amplitude for the detected regions, whilst panel (e)
    shows a map of the signal to noise ratio.  Finally, panel (f)
    shows a panel of the detected frequency as a function of the
    signal to noise ratio.  The error bars on the abscissa values are
    shown at one-tenth of their actual size in order to better show
    the lower error values at higher signal-to-noise ratio.  Table
    \protect\ref{tab:jul1}(a) shows the values for the detected region
    for the parameters listed in Table \protect\ref{tab:measures}. }
  \label{fig:jul1_171_3min}
\end{figure}

\begin{figure}
  \centerline{\hspace*{0.015\textwidth}
    \includegraphics[width=0.515\textwidth,clip=]{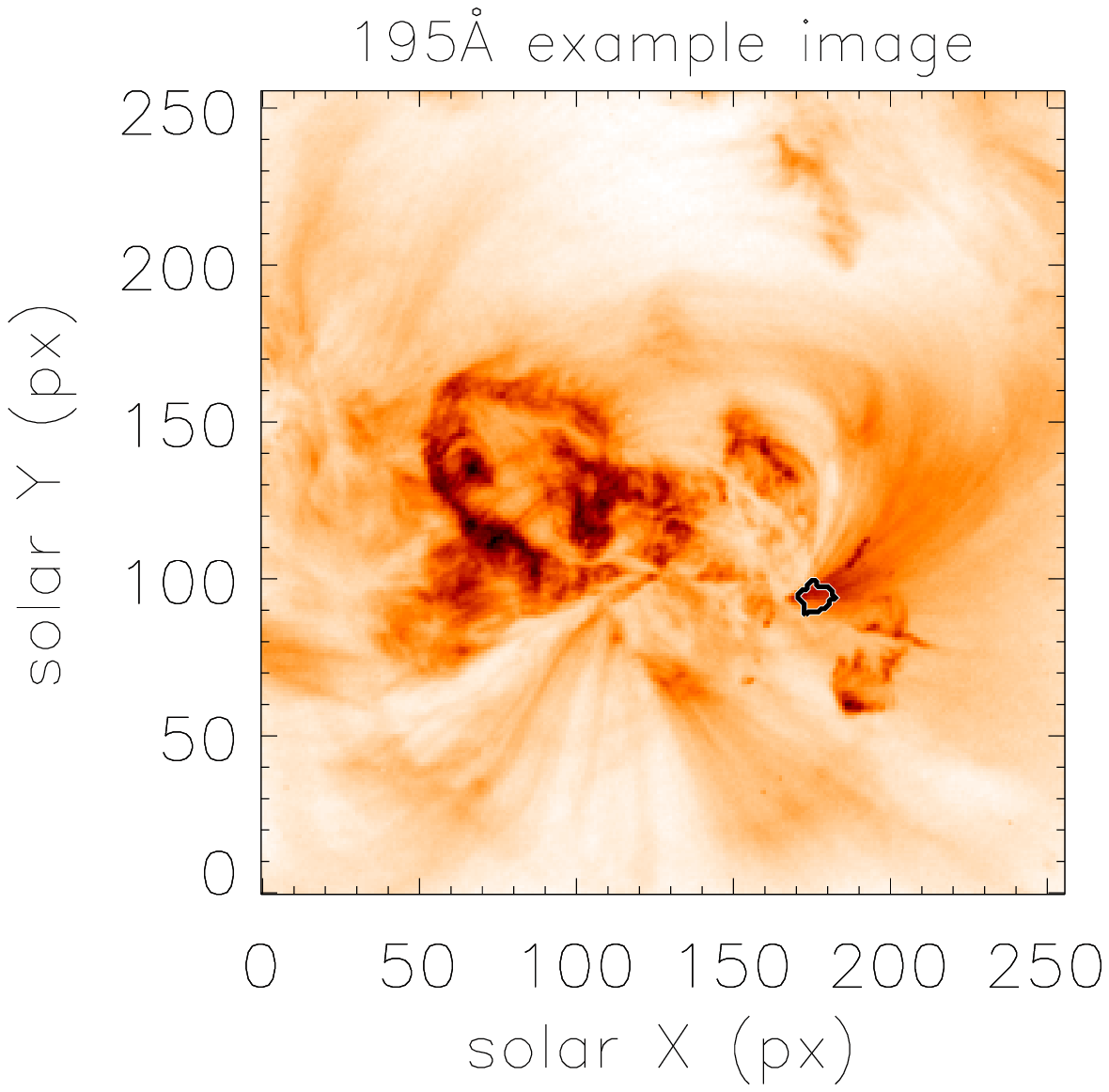}
   \hspace*{-0.03\textwidth}
    \includegraphics[width=0.515\textwidth,clip=]{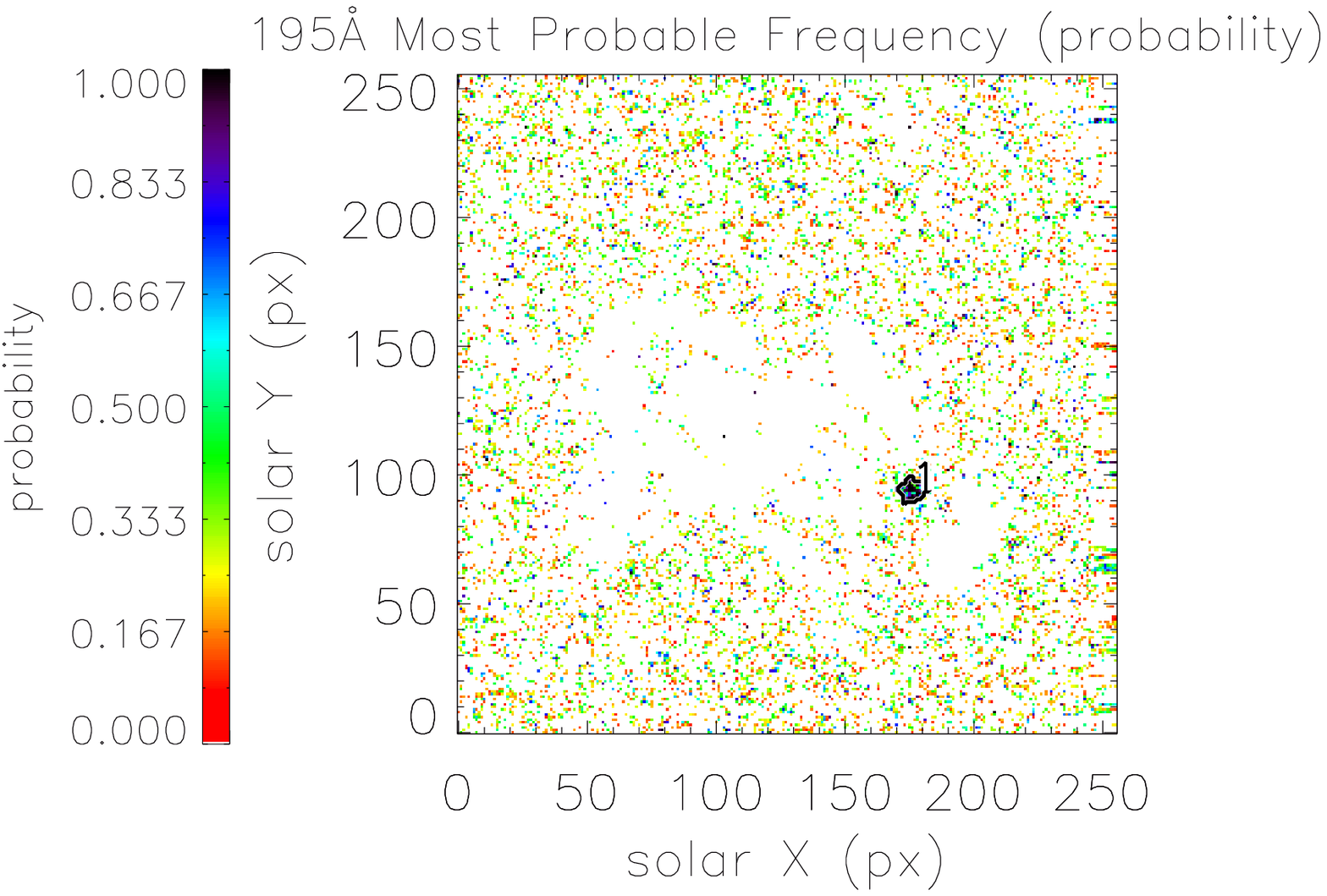}
   \vspace{0.0 \textwidth}}
  \vspace{-0.38\textwidth}   
  \centerline{
    \hspace{0.0 \textwidth}  \color{black}{(a)}
    \hspace{0.435\textwidth}  \color{black}{(b)}
    \hfill}
  \vspace{0.35\textwidth}    
  \centerline{\hspace*{0.015\textwidth}
    \includegraphics[width=0.515\textwidth,clip=]{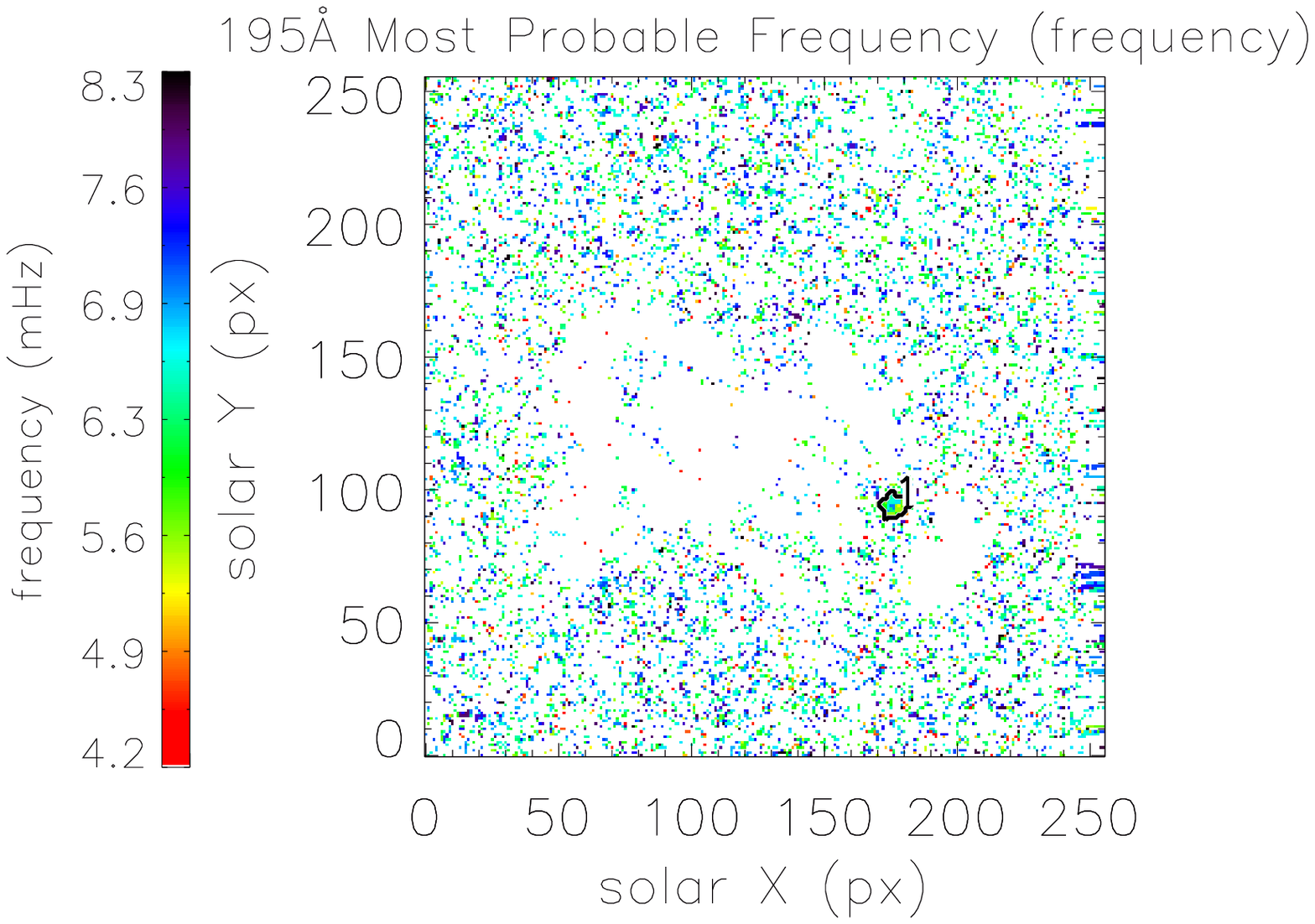}
   \hspace*{-0.03\textwidth}
    \includegraphics[width=0.515\textwidth,clip=]{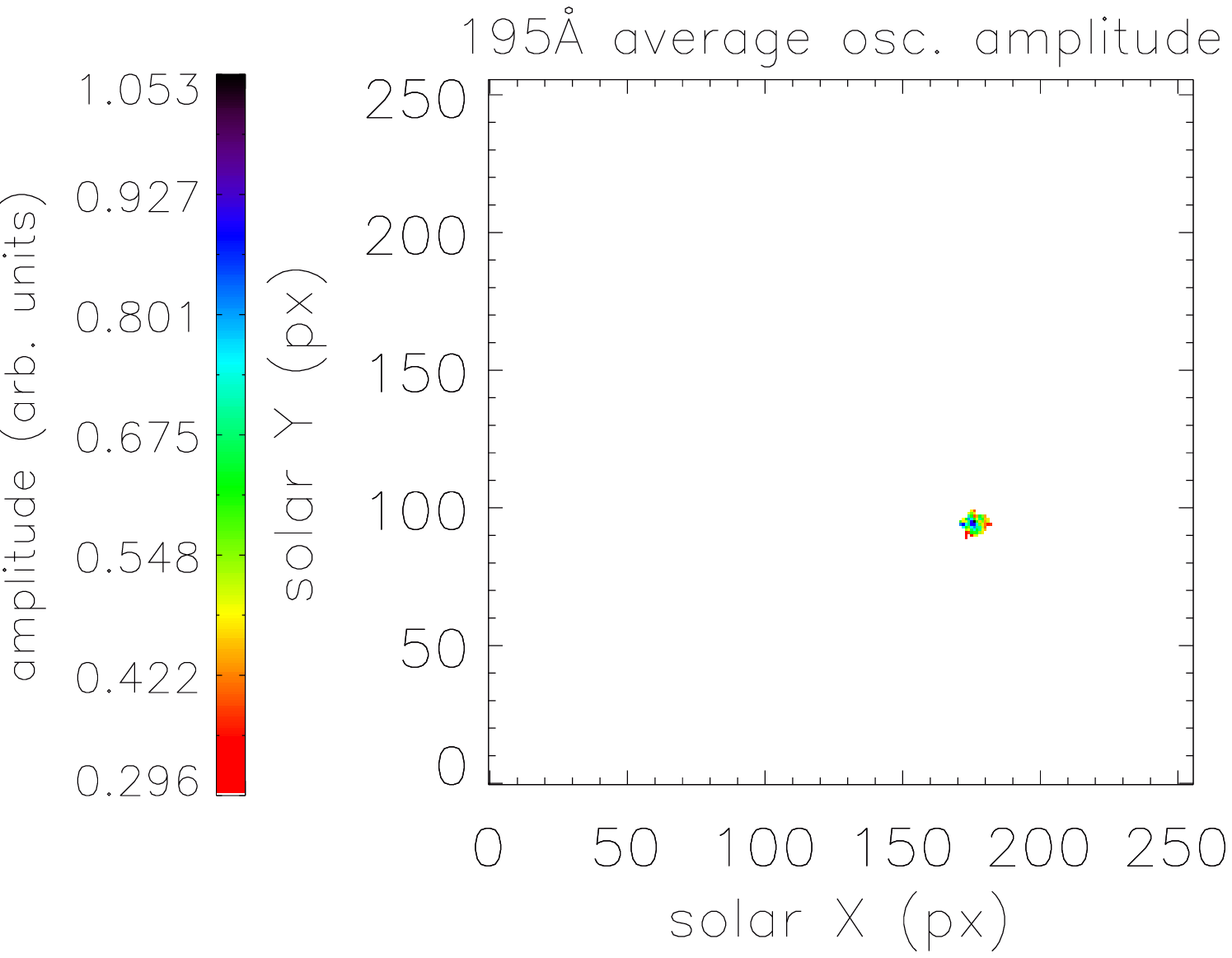}
   \vspace{0.0 \textwidth}}
  \vspace{-0.38\textwidth}   
  \centerline{
    \hspace{0.0 \textwidth}  \color{black}{(c)}
    \hspace{0.435\textwidth}  \color{black}{(d)}
    \hfill}
  \vspace{0.35\textwidth}    
  \centerline{\hspace*{0.015\textwidth}
    \includegraphics[width=0.515\textwidth,clip=]{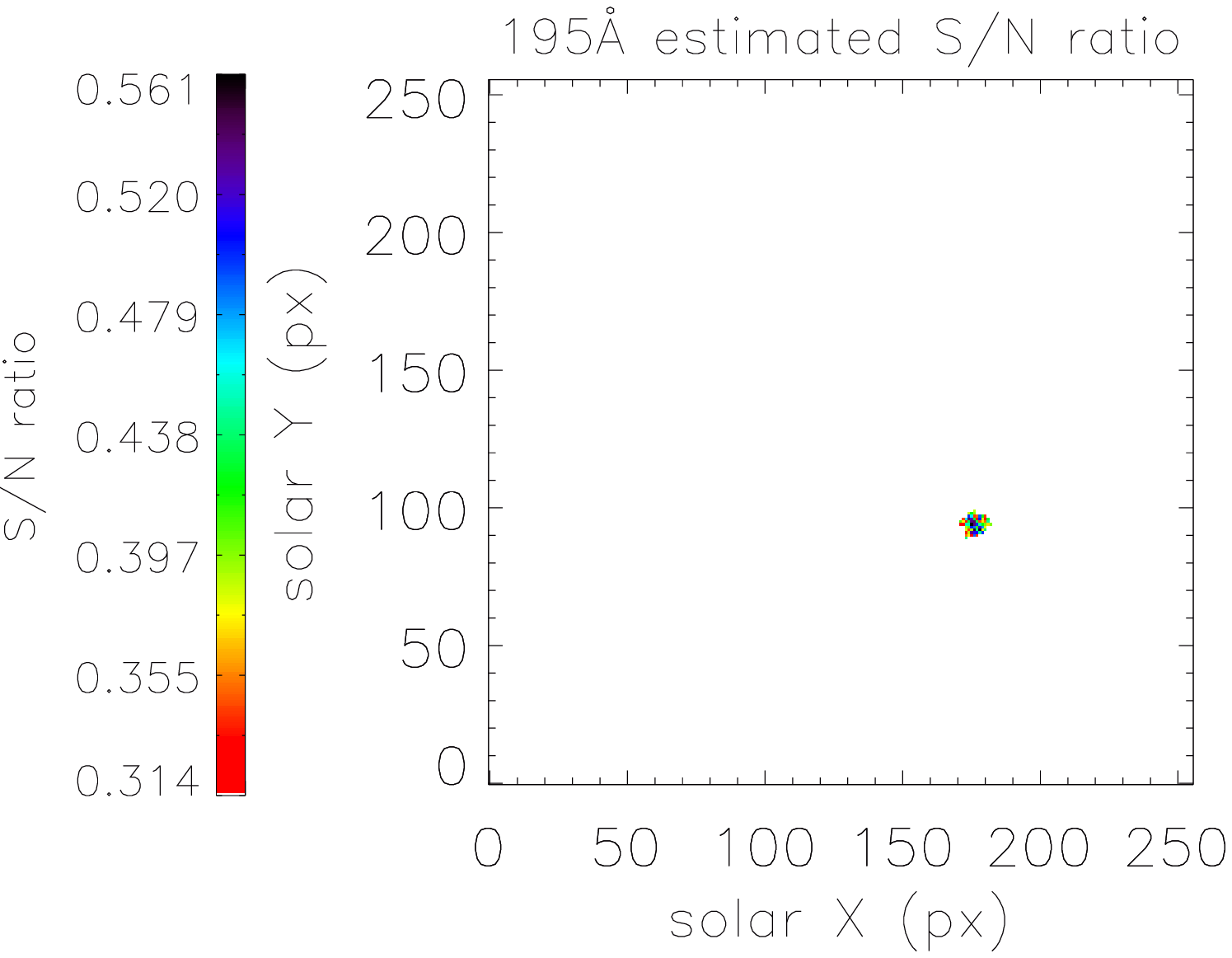}
   \hspace*{-0.03\textwidth}
    \includegraphics[width=0.515\textwidth,clip=]{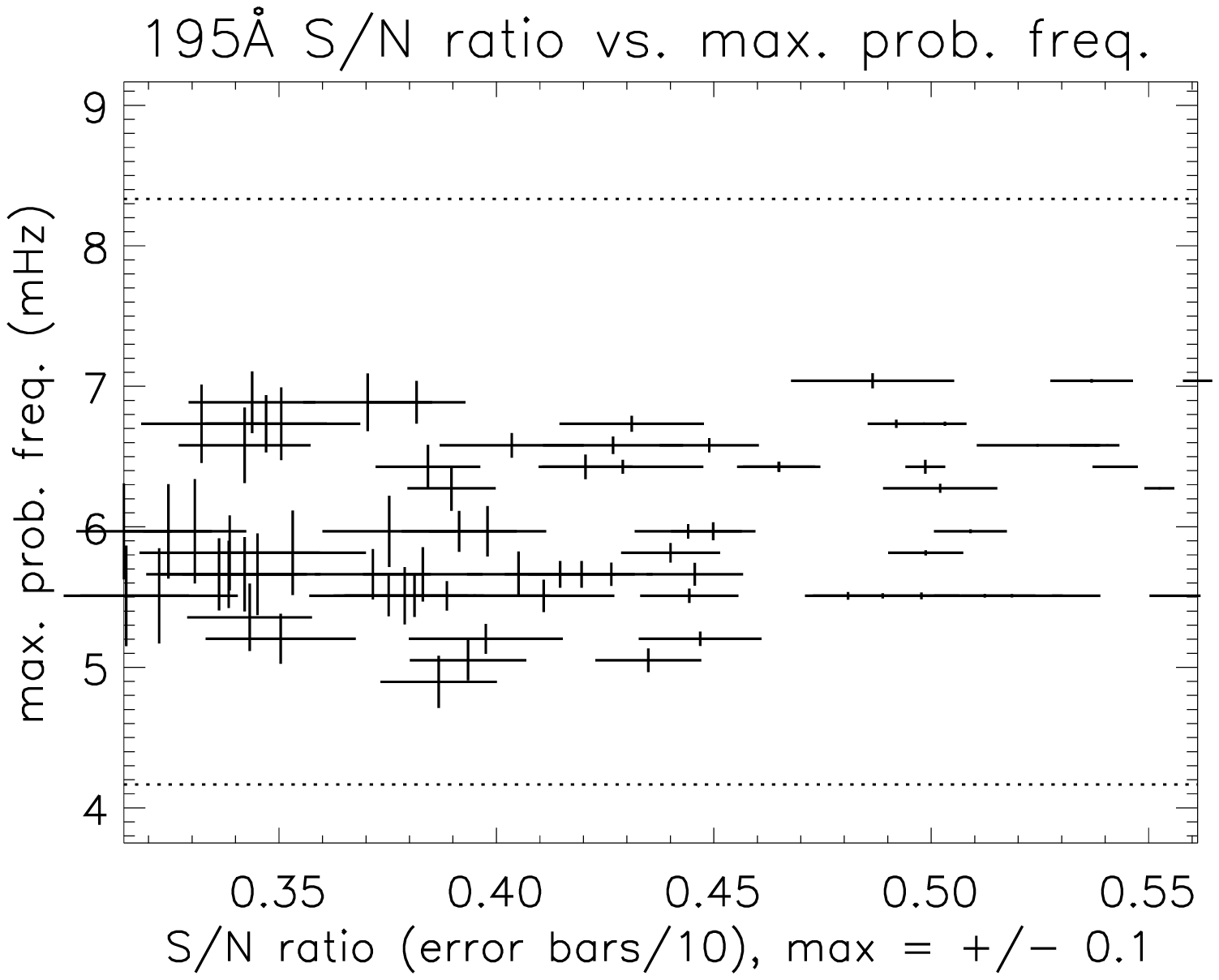}
   \vspace{0.0 \textwidth}}
  \vspace{-0.38\textwidth}   
  \centerline{
    \hspace{0.0 \textwidth}  \color{black}{(e)}
    \hspace{0.435\textwidth}  \color{black}{(f)}
    \hfill}
  \vspace{0.35\textwidth}    
  \caption{Example image data and results from the analysis algorithm
    for TRACE 195\AA\ data taken on 1 July 1998 (panel (a)). Panel (b)
    is the probability that a given pixel supports an oscillation
    within the frequency band indicated in panel (c), in this case, a
    five-minute oscillation frequency band. Note that the region
    agrees with that found manually by \protect\inlinecite{king2003}
    on the following day.  Panels (d\,--\,f) show similar maps to those
    shown and described in Figure \protect\ref{fig:jul1_171_3min}.  Table
    \protect\ref{tab:jul1}(b) shows the values for the detected region
    for the parameters listed in Table \protect\ref{tab:measures}. }
  \label{fig:jul1_195_3min}
\end{figure}

\begin{figure}
  \centerline{\hspace*{0.015\textwidth}
    \includegraphics[width=0.515\textwidth,clip=]{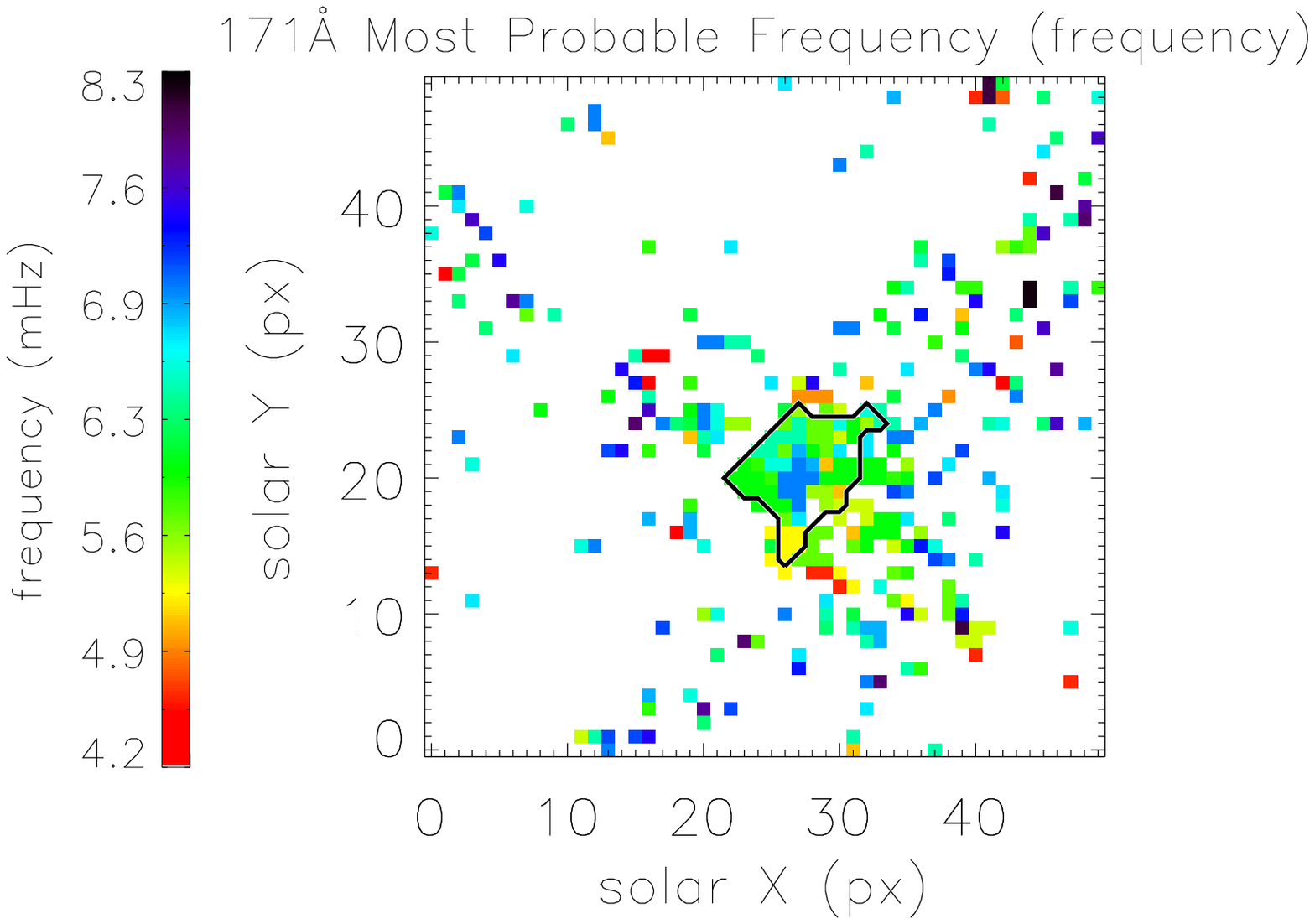}
   \hspace*{-0.03\textwidth}
    \includegraphics[width=0.515\textwidth,clip=]{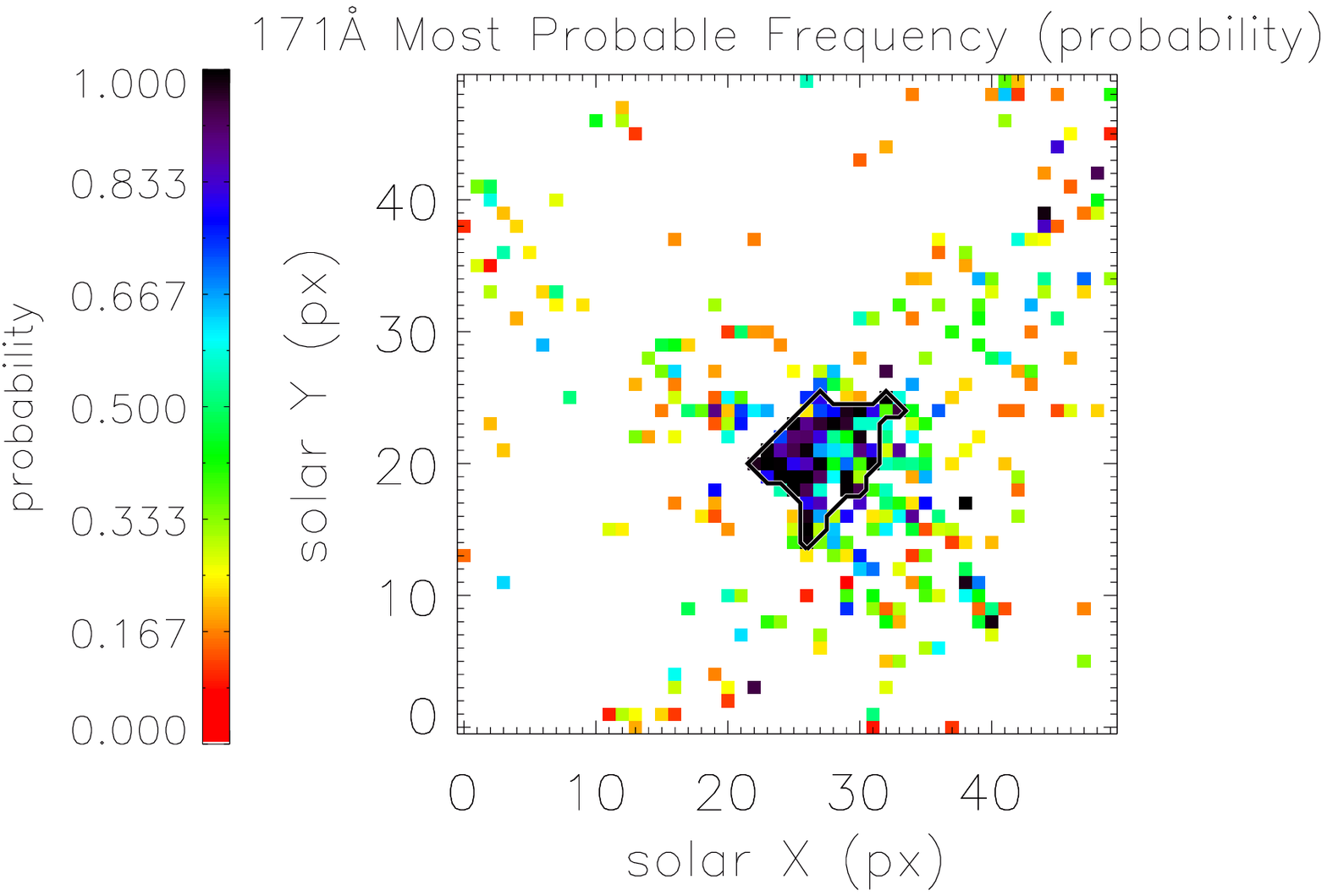}
   \vspace{0.0 \textwidth}}
  \vspace{-0.38\textwidth}   
  \centerline{
    \hspace{0.0 \textwidth}  \color{black}{(a)}
    \hspace{0.435\textwidth}  \color{black}{(b)}
    \hfill}
  \vspace{0.35\textwidth}    
  \centerline{\hspace*{0.015\textwidth}
    \includegraphics[width=0.515\textwidth,clip=]{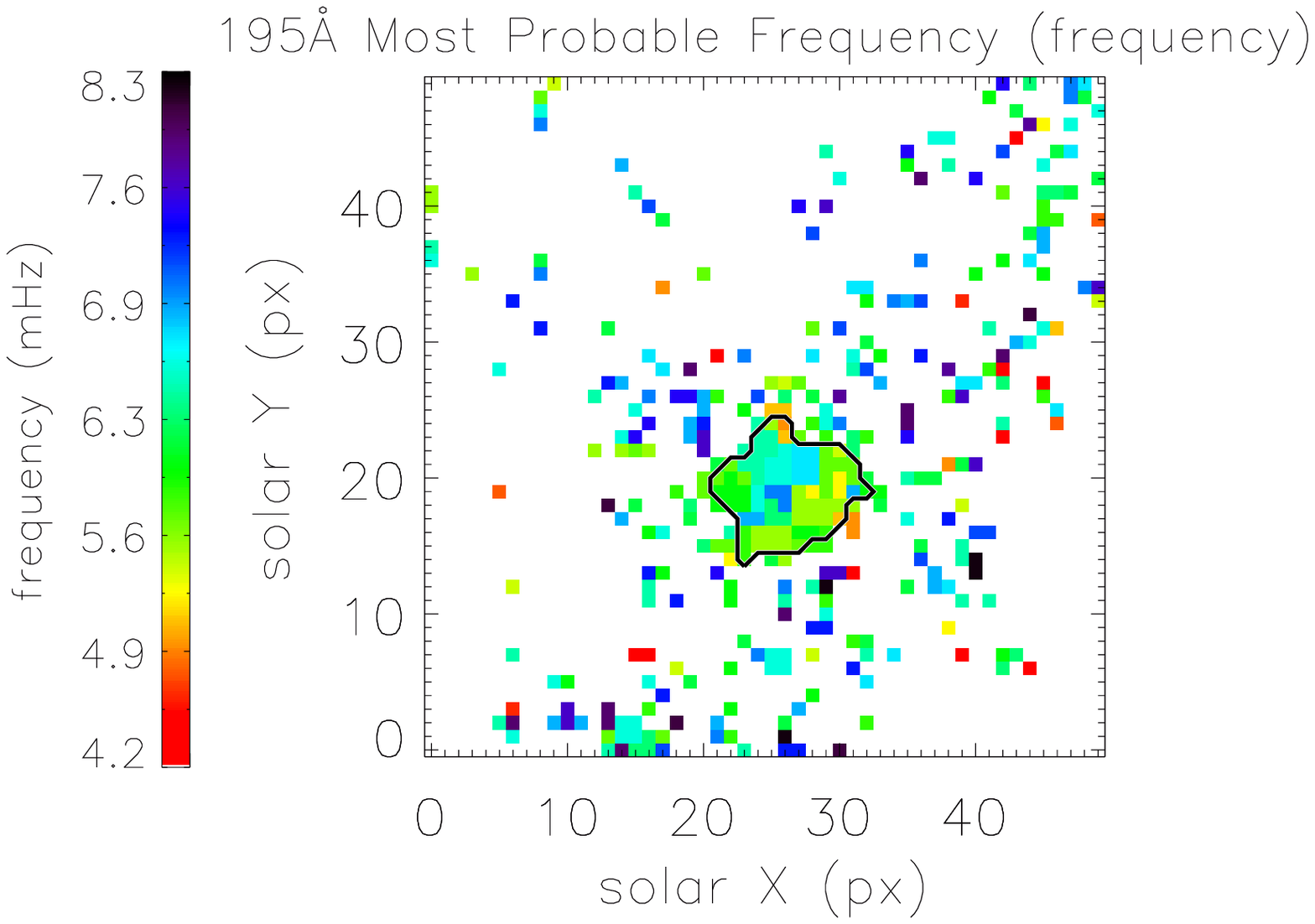}
    \hspace*{-0.03\textwidth}
    \includegraphics[width=0.515\textwidth,clip=]{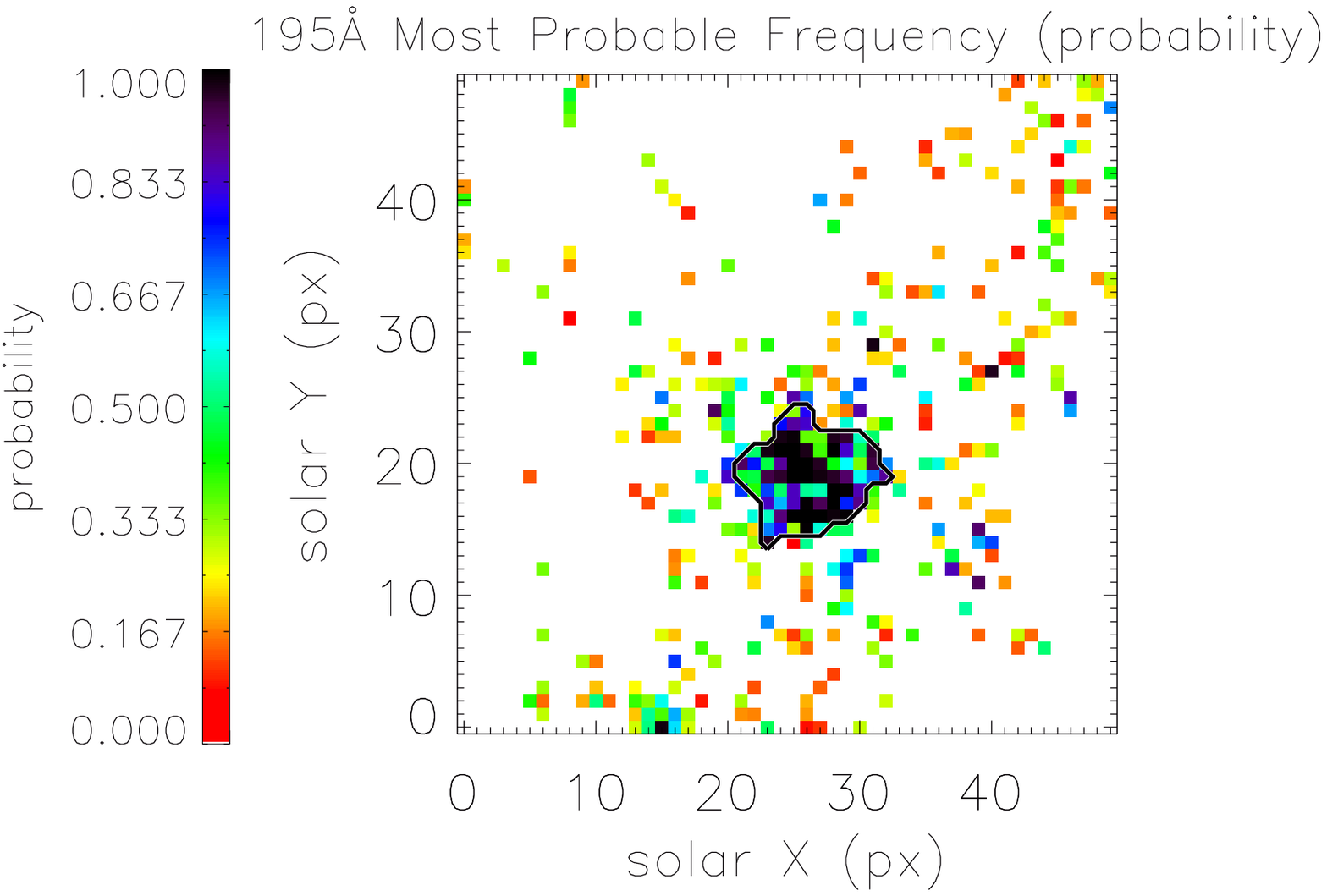}
   \vspace{0.0 \textwidth}}
  \vspace{-0.38\textwidth}   
  \centerline{
    \hspace{0.0 \textwidth}  \color{black}{(c)}
   \hspace{0.435\textwidth}  \color{black}{(d)}
    \hfill}
  \vspace{0.35\textwidth}    
  \caption{Zoomed in view of single large oscillation region found in
    Figures \ref{fig:jul1_171_3min} and \ref{fig:jul1_195_3min}. Panels
    (a) and (b) are from the 171\AA\
  data, and panels (c) and (d) are derived from the 195\AA\ data (1 July
   1998).}
  \label{fig:jul1_3min_zoomin}
\end{figure}

The original data was taken on 1 July 1998 12:30:01 -- 14:24:41 UT
with an average cadence of 31 seconds (220 samples) and an image size
of $512\times 512$ pixels of equivalent size $0.5''$. Data are also
$2\times 2$ summed in space to increase signal-to-noise ratio (this
has the consequent effect of reducing the algorithm run time). The
results of searching in a wideband three-minute range (120 -- 240
seconds, or 4.17\,--\,6.33 mHz) are shown in Figures
\ref{fig:jul1_171_3min} (171 \AA) and \ref{fig:jul1_195_3min} (195
\AA) and in Table \ref{tab:jul1}(a,b). Most of the field of view does
not contain material oscillating in the analysis frequency range. Only
one group of pixels survives the filtering process, at the base of a
coronal-loop fan (Figure \ref{fig:jul1_3min_zoomin}). However, it is
interesting to note that there are areas of the probability maps
Figures \ref{fig:jul1_171_3min}(b), \ref{fig:jul1_195_3min}(b) that
have effectively zero probability of supporting an oscillation in the
frequency band, whilst others have a non-zero and low-probability. We
speculate that with the improved SNR of SDO, many of these low
probability oscillations will become much more likely, leading to the
detection of many more spatially distributed signals in the data.

The single oscillating pixel group found (Figure
\ref{fig:jul1_3min_zoomin}) is in the same location as that identified
manually by \inlinecite{king2003} and automatically by
\inlinecite{2008SoPh..252..321M}. The region is distinctly different
from others in the data. Figures \ref{fig:jul1_171_3min}(f) and
\ref{fig:jul1_195_3min}(f) show that the detected oscillations in this
pixel group has a fairly low estimated signal-to-noise ratio. The
estimated error in the frequency decreases with increasing SNR, as
expected.

\begin{figure}
  \centerline{\hspace*{0.015\textwidth}
    \includegraphics[width=0.515\textwidth,clip=]{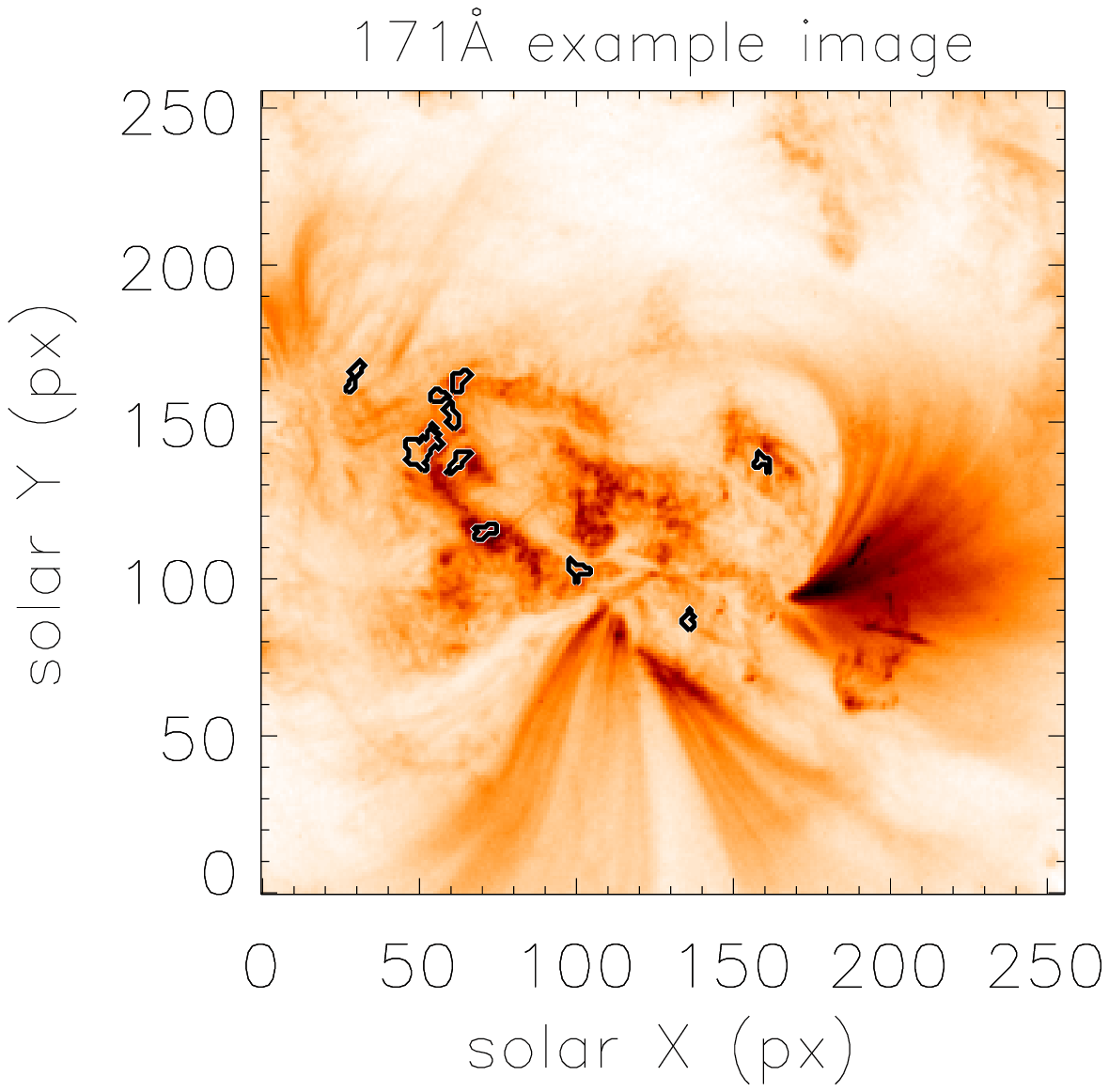}
   \hspace*{-0.03\textwidth}
    \includegraphics[width=0.515\textwidth,clip=]{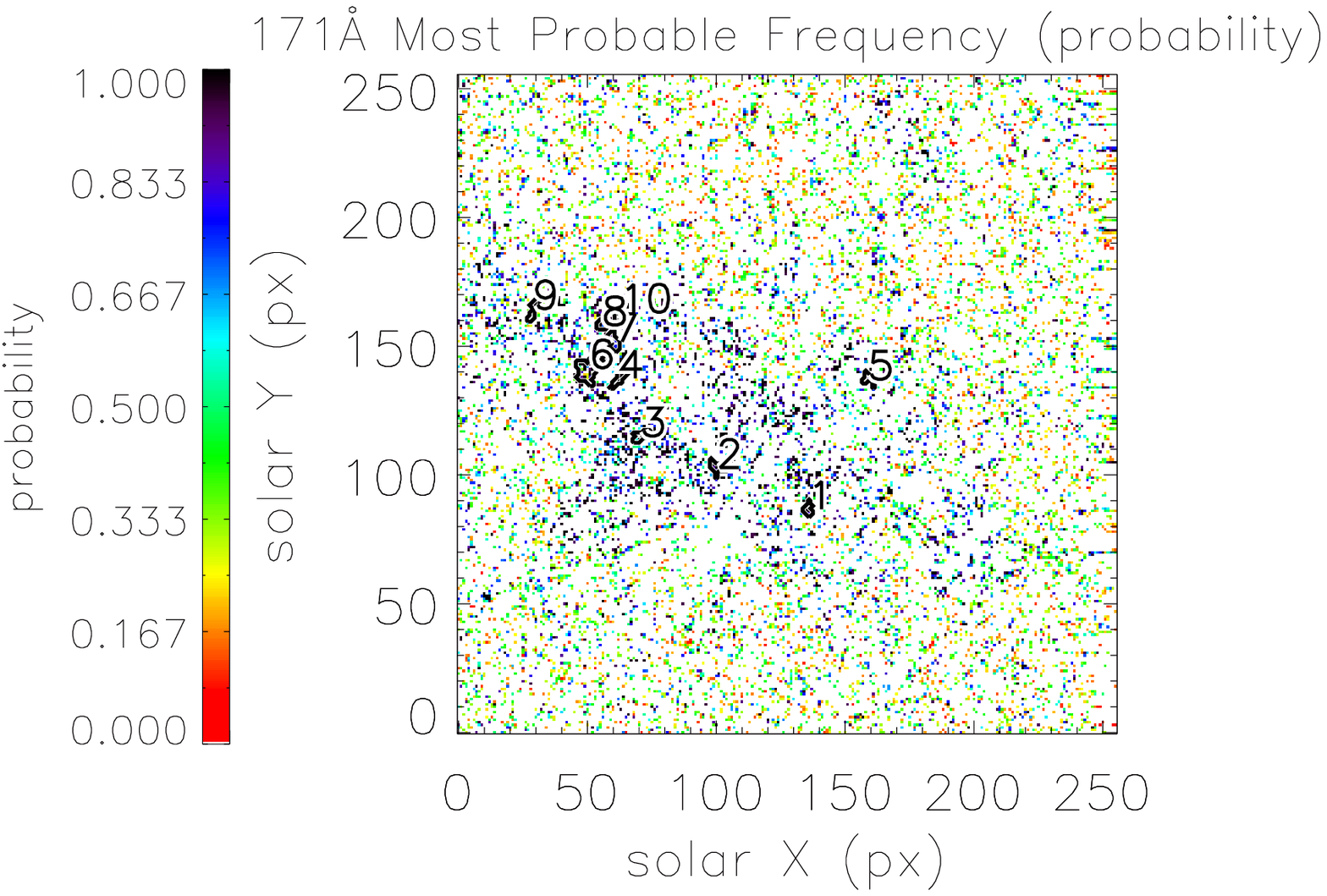}
   \vspace{0.0 \textwidth}}
  \vspace{-0.38\textwidth}   
  \centerline{
    \hspace{0.0 \textwidth}  \color{black}{(a)}
   \hspace{0.435\textwidth}  \color{black}{(b)}
    \hfill}
  \vspace{0.35\textwidth}    
  \centerline{\hspace*{0.015\textwidth}
    \includegraphics[width=0.515\textwidth,clip=]{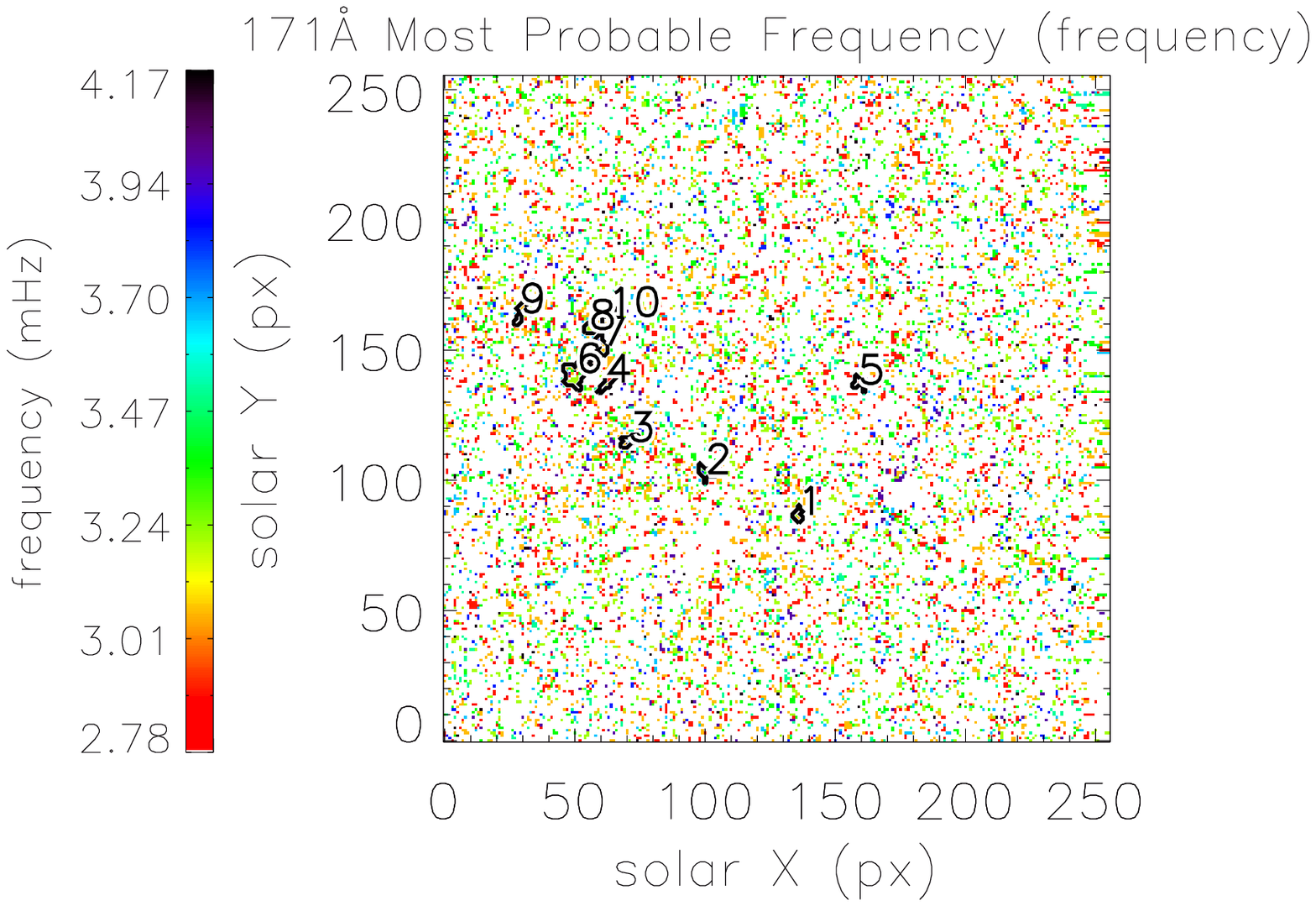}
   \hspace*{-0.03\textwidth}
    \includegraphics[width=0.515\textwidth,clip=]{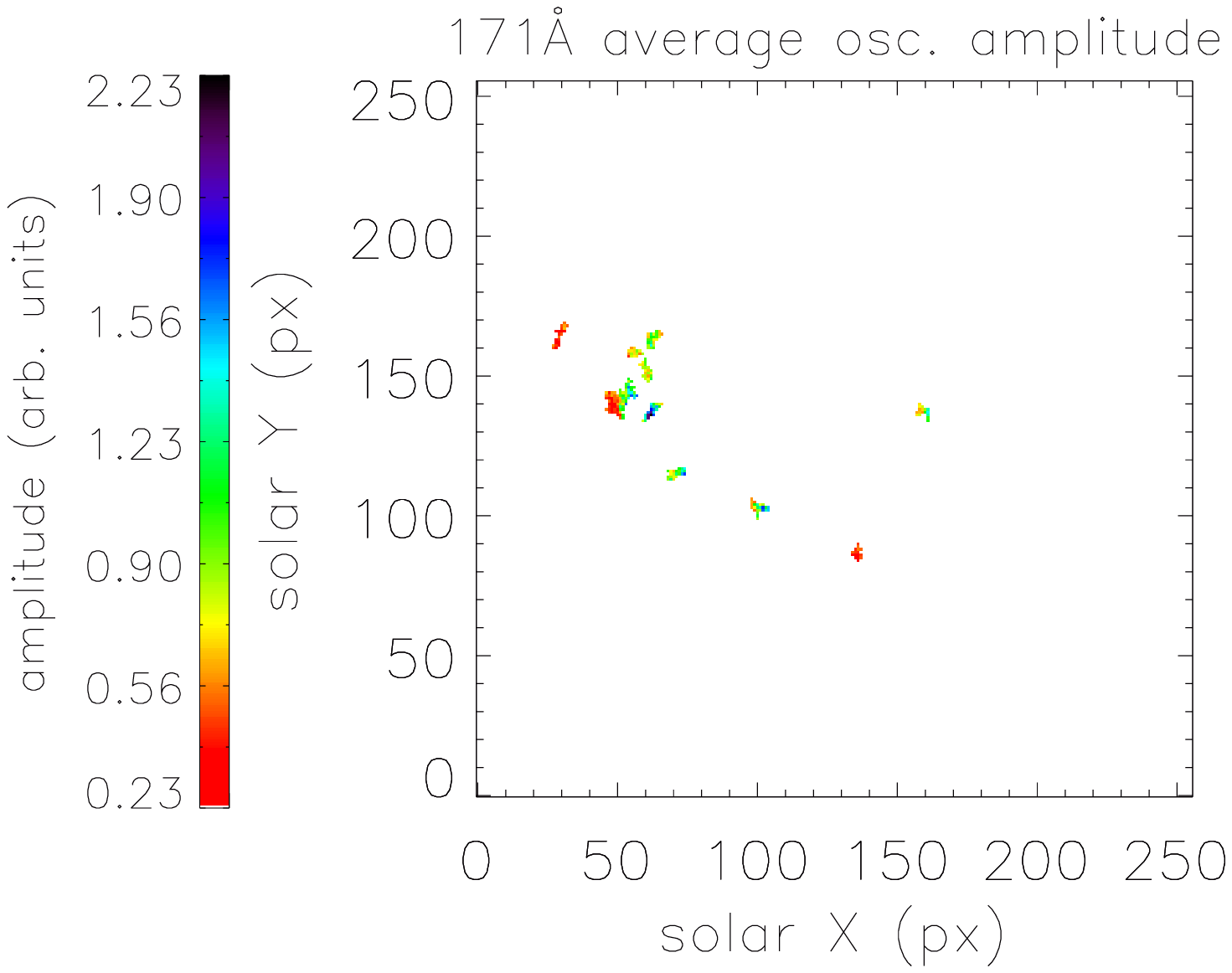}
   \vspace{0.0 \textwidth}}
  \vspace{-0.38\textwidth}   
  \centerline{
    \hspace{0.0 \textwidth}  \color{black}{(c)}
    \hspace{0.435\textwidth}  \color{black}{(d)}
    \hfill}
  \vspace{0.35\textwidth}    
  \centerline{\hspace*{0.015\textwidth}
    \includegraphics[width=0.515\textwidth,clip=]{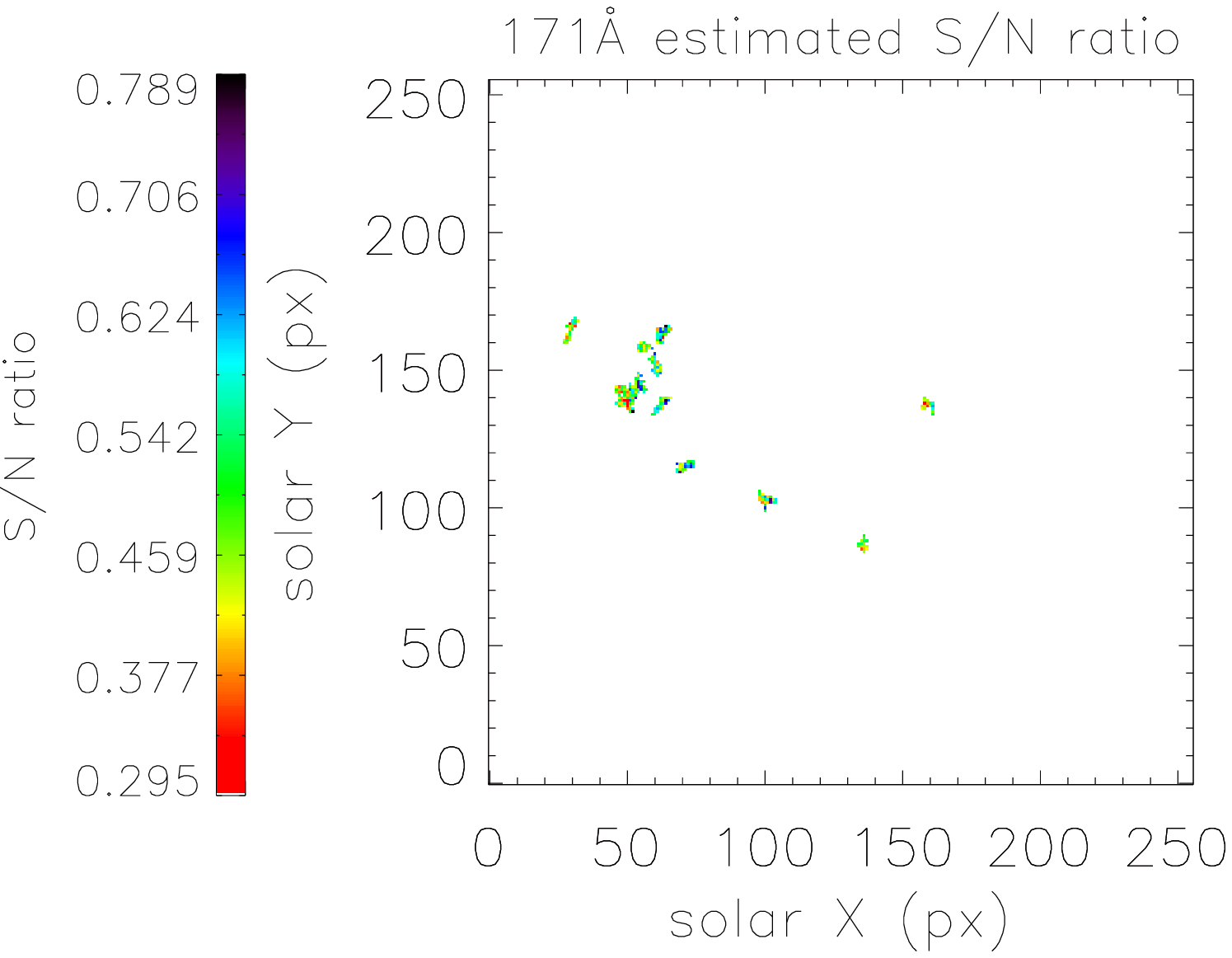}
   \hspace*{-0.03\textwidth}
    \includegraphics[width=0.515\textwidth,clip=]{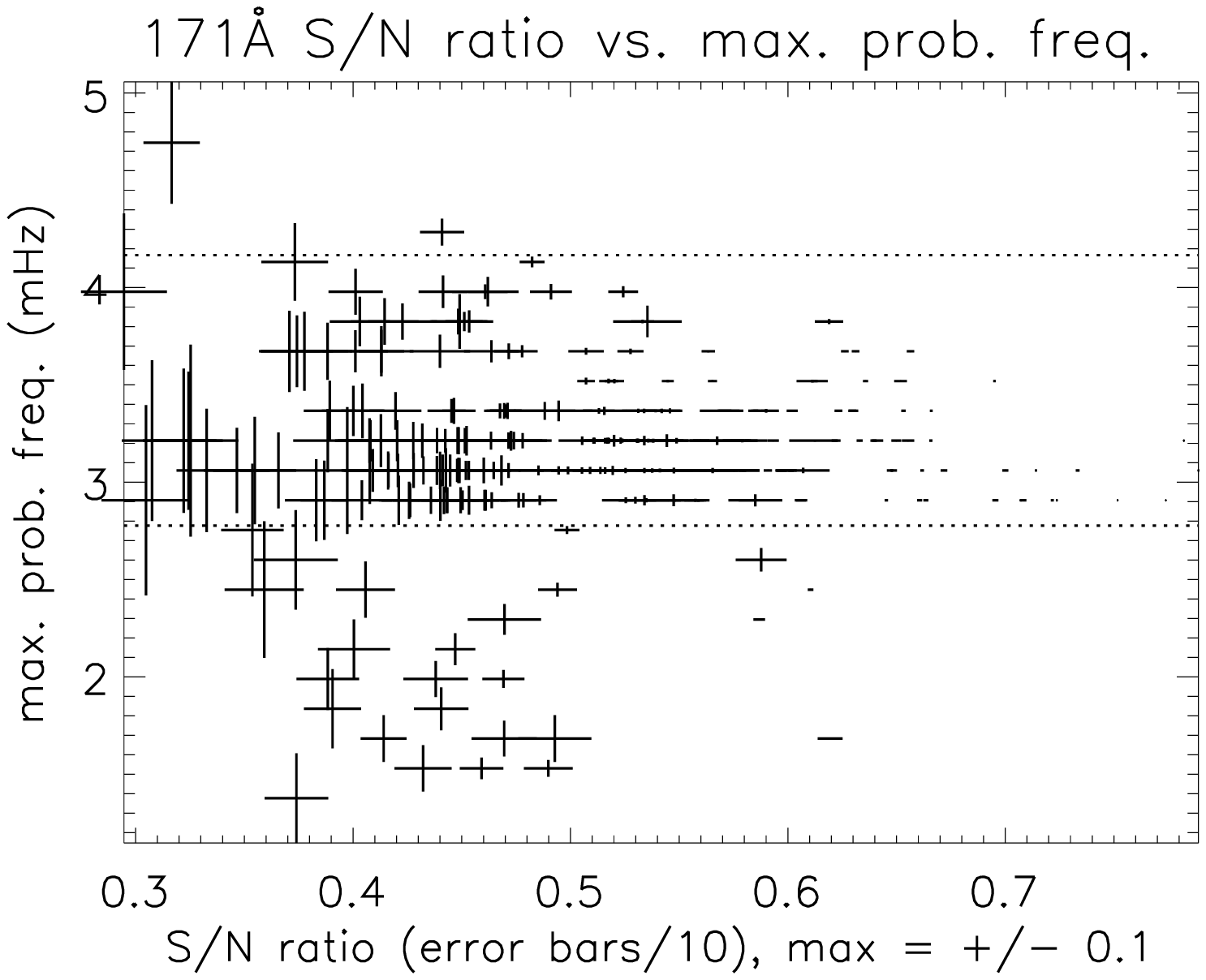}
   \vspace{0.0 \textwidth}}
  \vspace{-0.38\textwidth}   
  \centerline{
    \hspace{0.0 \textwidth}  \color{black}{(e)}
    \hspace{0.435\textwidth}  \color{black}{(f)}
    \hfill}
  \vspace{0.35\textwidth}    
  \caption{Example image data and results from the analysis algorithm
    for TRACE 171\AA\ data taken on 1 July 1998 (panel (a)). Panel (b)
    is the probability that a given pixel supports an oscillation
    within the frequency band indicated in panel (c), in this case, a
    five-minute oscillation frequency band.  Also indicated in panels
    (a\,--\,c) are the detected oscillation regions.  Panels (d\,--\,f) show
    similar maps to those shown and described in Figure 
    \protect\ref{fig:jul1_171_3min}.  Table \protect\ref{tab:jul1}(c)
    shows the values for the detected region for the parameters listed
    in Table \protect\ref{tab:measures}. }
  \label{fig:jul1_171_5min}
\end{figure}

\begin{figure}
  \centerline{\hspace*{0.015\textwidth}
    \includegraphics[width=0.515\textwidth,clip=]{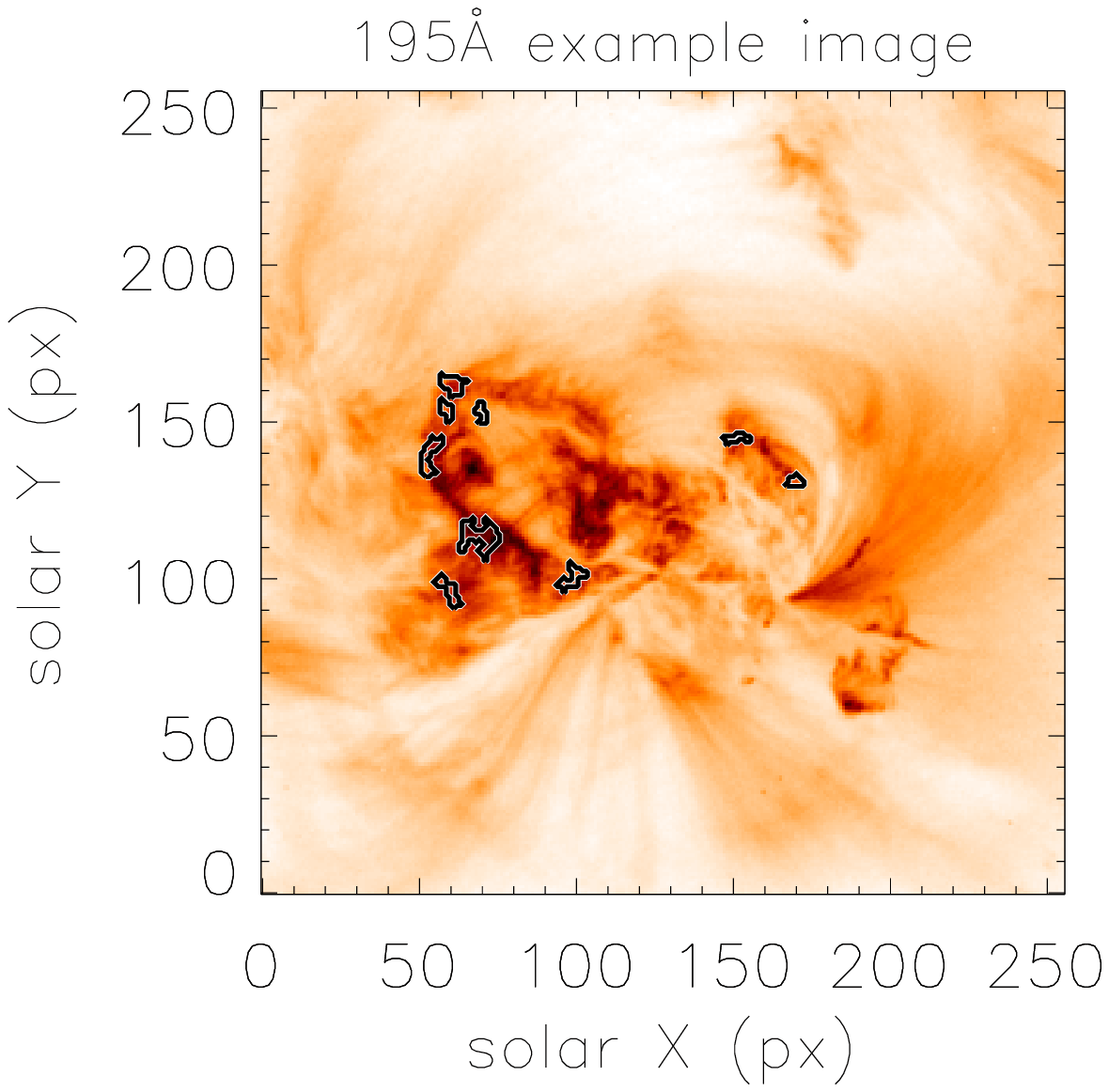}
   \hspace*{-0.03\textwidth}
    \includegraphics[width=0.515\textwidth,clip=]{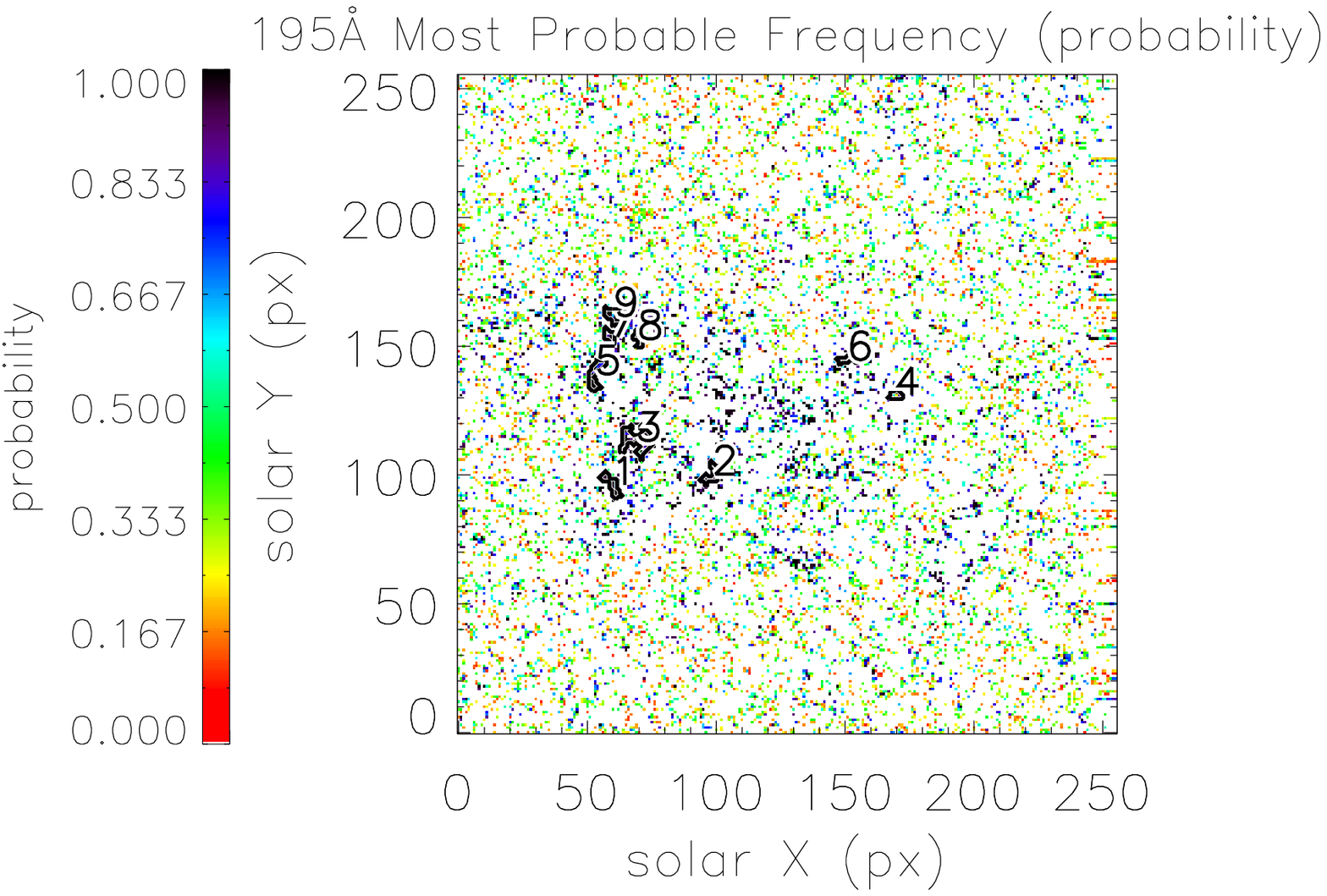}
   \vspace{0.0 \textwidth}}
  \vspace{-0.38\textwidth}   
  \centerline{
    \hspace{0.0 \textwidth}  \color{black}{(a)}
    \hspace{0.435\textwidth}  \color{black}{(b)}
    \hfill}
  \vspace{0.35\textwidth}    
  \centerline{\hspace*{0.015\textwidth}
    \includegraphics[width=0.515\textwidth,clip=]{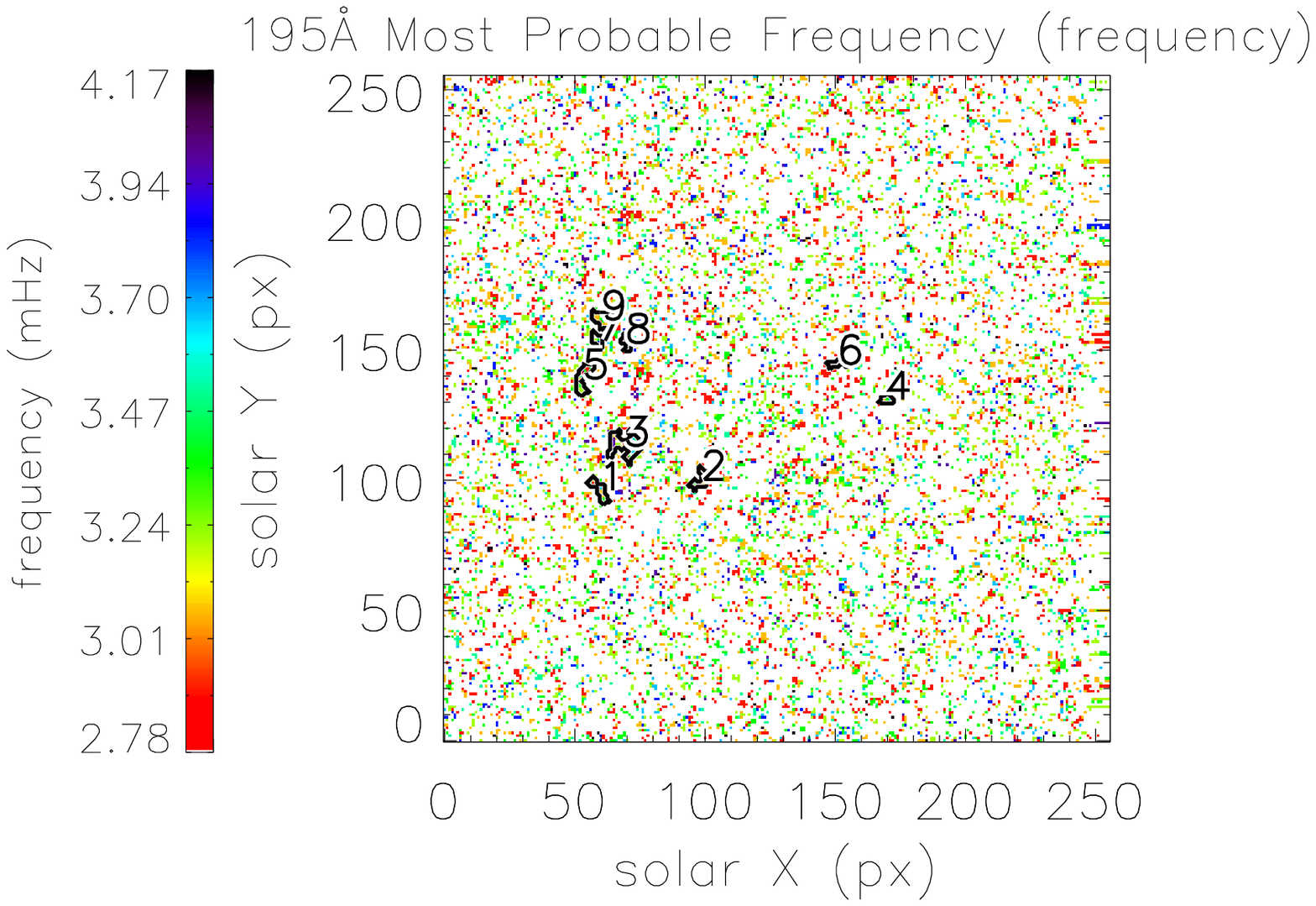}
   \hspace*{-0.03\textwidth}
    \includegraphics[width=0.515\textwidth,clip=]{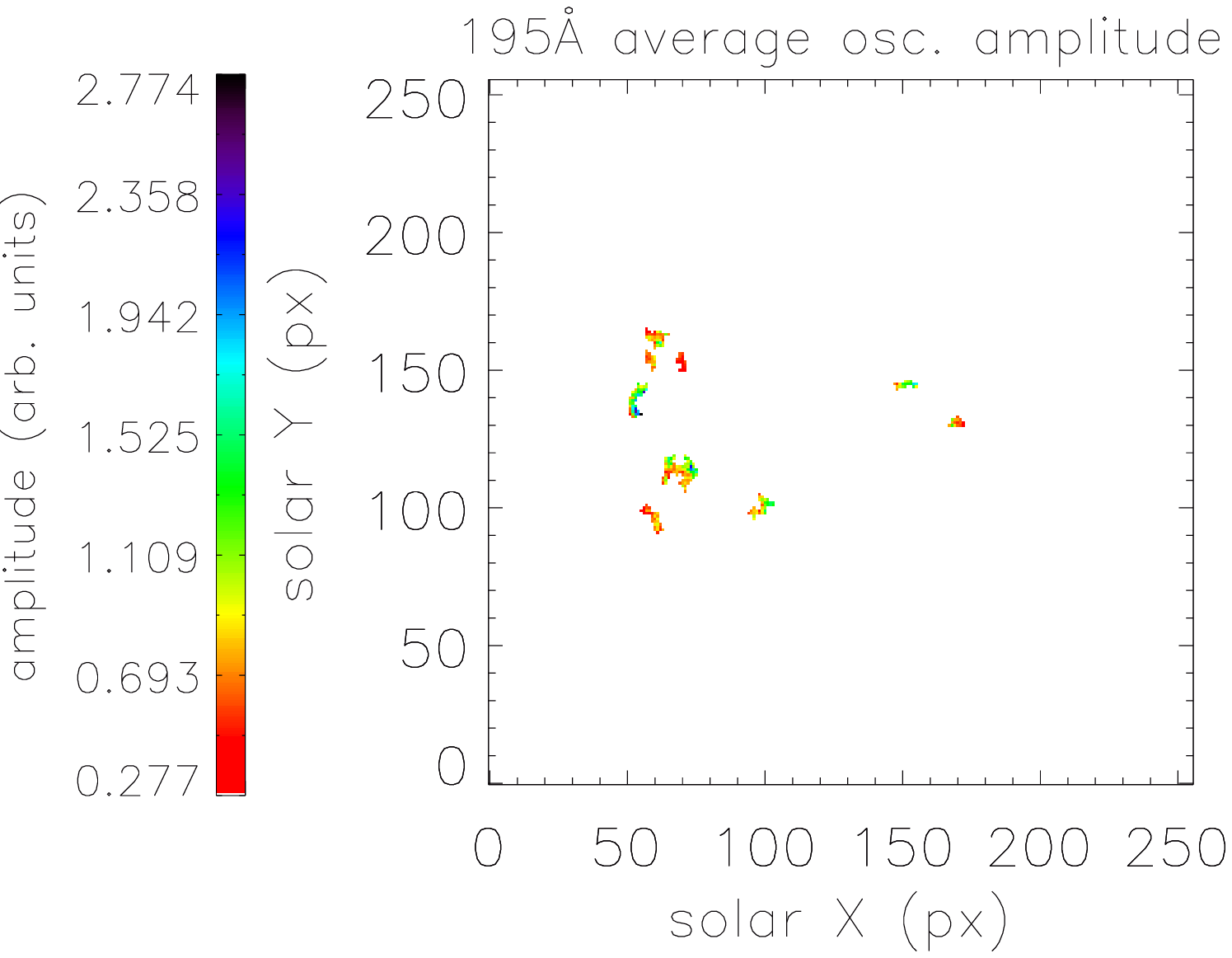}
   \vspace{0.0 \textwidth}}
  \vspace{-0.38\textwidth}   
  \centerline{
    \hspace{0.0 \textwidth}  \color{black}{(c)}
    \hspace{0.435\textwidth}  \color{black}{(d)}
    \hfill}
  \vspace{0.35\textwidth}    
  \centerline{\hspace*{0.015\textwidth}
    \includegraphics[width=0.515\textwidth,clip=]{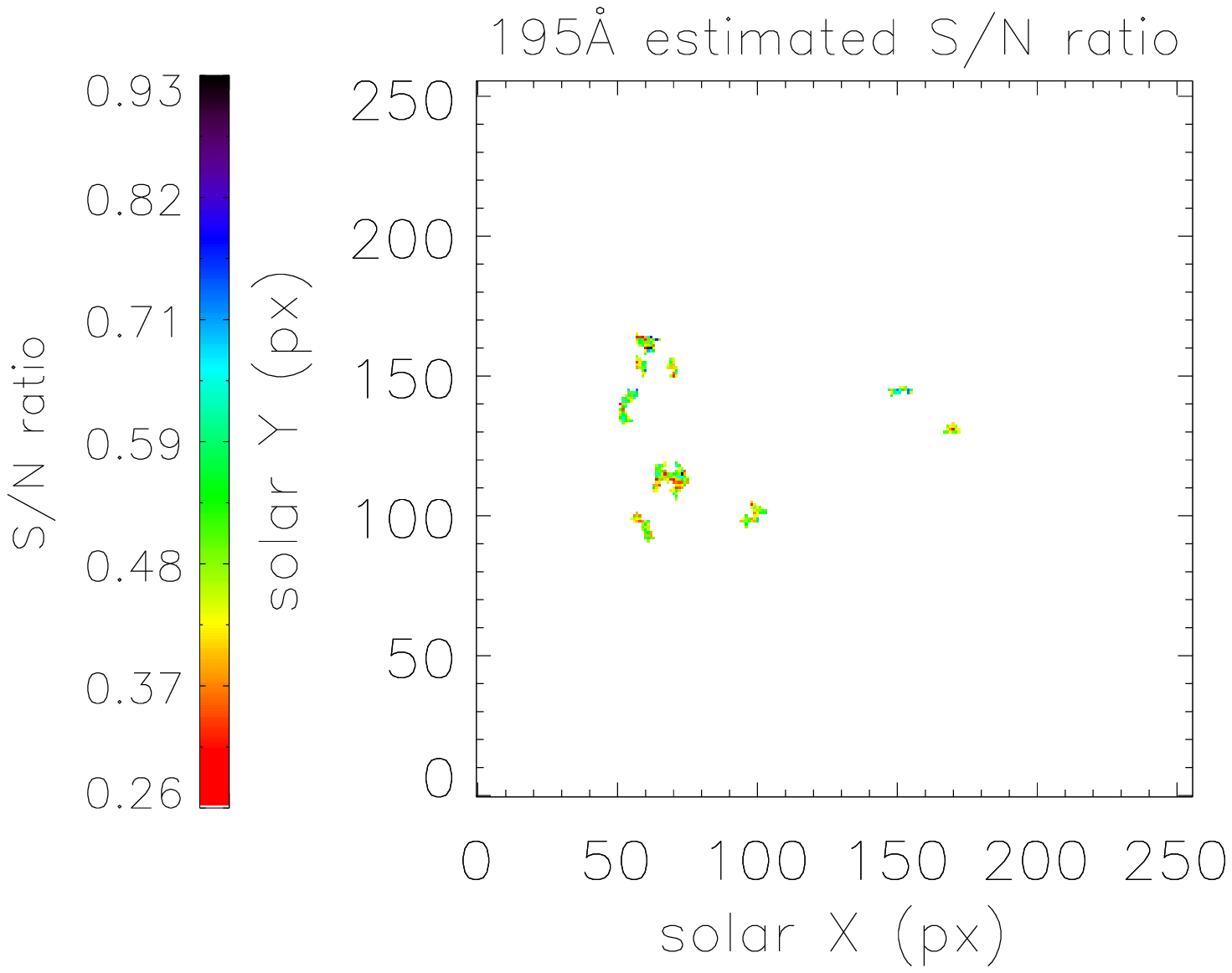}
   \hspace*{-0.03\textwidth}
    \includegraphics[width=0.515\textwidth,clip=]{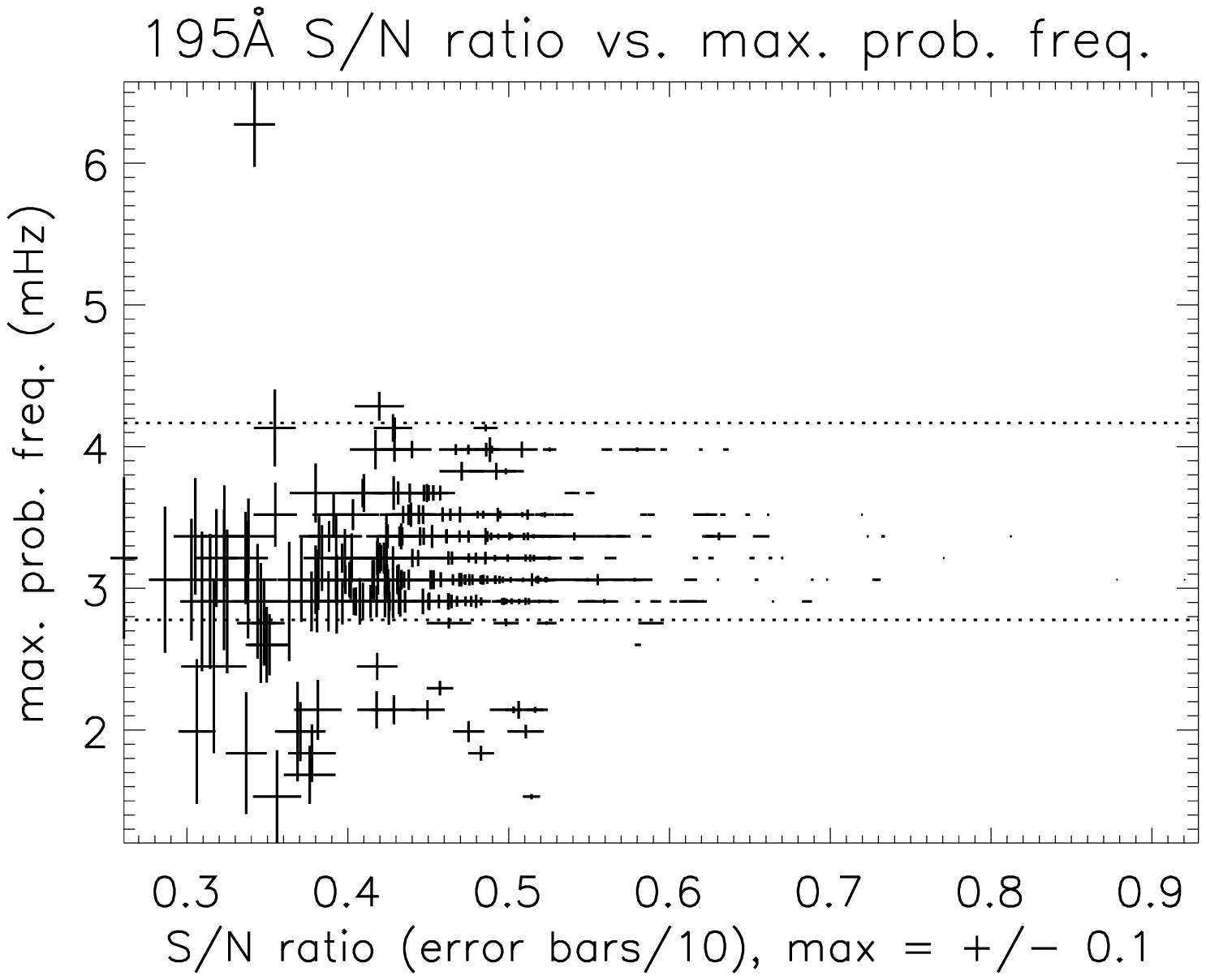}
   \vspace{0.0 \textwidth}}
  \vspace{-0.38\textwidth}   
  \centerline{
    \hspace{0.0 \textwidth}  \color{black}{(e)}
    \hspace{0.435\textwidth}  \color{black}{(f)}
    \hfill}
  \vspace{0.35\textwidth}    
  \caption{Example image data and results from the analysis algorithm
    for TRACE 195\AA\ data taken on 1 July 1998 (panel (a)) in the
    five-minute frequency band. Panels (b\,--\,f) show similar maps to those
    shown and described in Figure \protect\ref{fig:jul1_171_5min}.  Table
    \protect\ref{tab:jul1}(d) shows the values for the detected region
    for the parameters listed in Table \protect\ref{tab:measures}. }
  \label{fig:jul1_195_5min}
\end{figure}

The results of searching in the five-minute frequency band (240\,--\,360
seconds, or 2.78\,--\,4.17 mHz) are shown in Figures
\ref{fig:jul1_171_5min} (171 \AA) and \ref{fig:jul1_195_5min} (195
\AA) and in Table \ref{tab:jul1}(c,d).  The probability maps Figures
\ref{fig:jul1_171_3min}(b), \ref{fig:jul1_195_3min}(b) look very
different in this frequency range, with low-probability oscillations
scattered across the field of view.  It is noticeable, however, that
the higher-probability oscillations are concentrated in the core of
the active region, in the regions that appeared empty in Figures
\ref{fig:jul1_171_3min}(b), \ref{fig:jul1_195_3min}(b).  The algorithm
does qualify some pixel groups as supporting oscillations.
\inlinecite{king2003} do not claim any detections in these areas
(although it is not clear if they looked), and
\inlinecite{2008SoPh..252..321M} find only one coherent pixel group in
the same general region as those found here.  Some of the regions do
overlap, suggesting co-temporal and co-spatial propagation of
oscillations in two different layers of the atmosphere.  In the image,
the areas in question resemble TRACE moss
(\opencite{1999SoPh..190..409B};
\opencite{1999SoPh..190..419D};
\opencite{1999ApJ...520L.135F}). However the presence of groups of
oscillations at multiple temperatures argues that the algorithm has
found examples of leakage of five minute oscillations from lower down
in the atmosphere to the upper layers, as described by
\inlinecite{2004Natur.430..536D} and \inlinecite{2005ApJ...624L..61D}
and references therein.

Finally, the plotted points in Figures \ref{fig:jul1_171_5min}(f),
\ref{fig:jul1_195_5min}(f) appear to lie on horizontal lines across
the plot, which are simply the analysis frequencies.  The foregoing
analysis can be done with many more frequencies (see Section
\ref{sec:singfreq}), at the expense of using the FFT to perform the
analysis.  This would slow performance, but would lead to a more
precise knowledge of the frequencies present.  Since we are primarily
interested in detecting the presence of frequencies in a wide
frequency band, the precision of each frequency detected is not as
important as their detection in as little time as possible.

%
%
%
%

%
%

%
%
%
%
%
%
\begin{figure}
  \centerline{\hspace*{0.015\textwidth}
\includegraphics[width=0.515\textwidth,clip=]{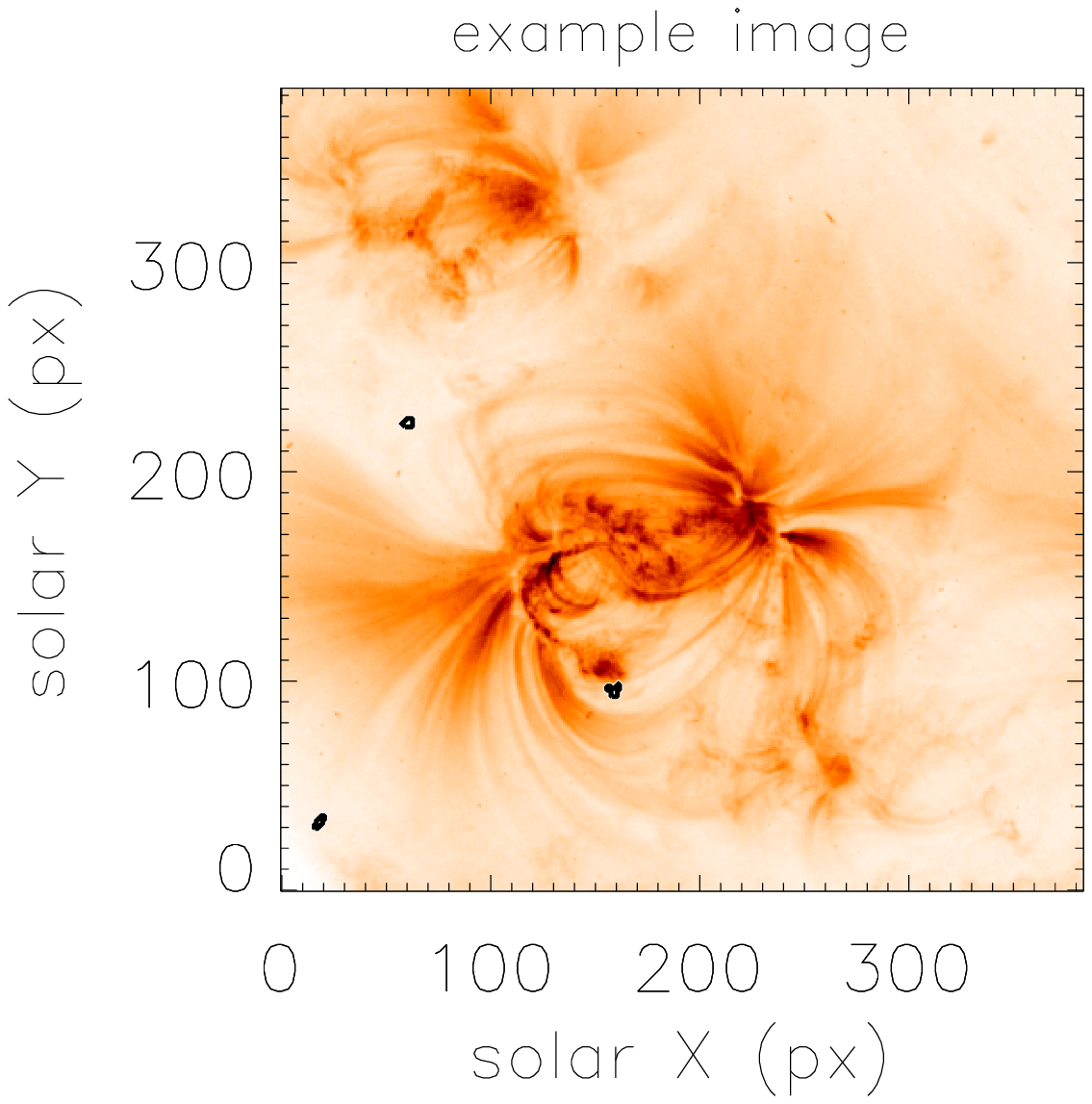}
    \hspace*{-0.03\textwidth}
\includegraphics[width=0.515\textwidth,clip=]{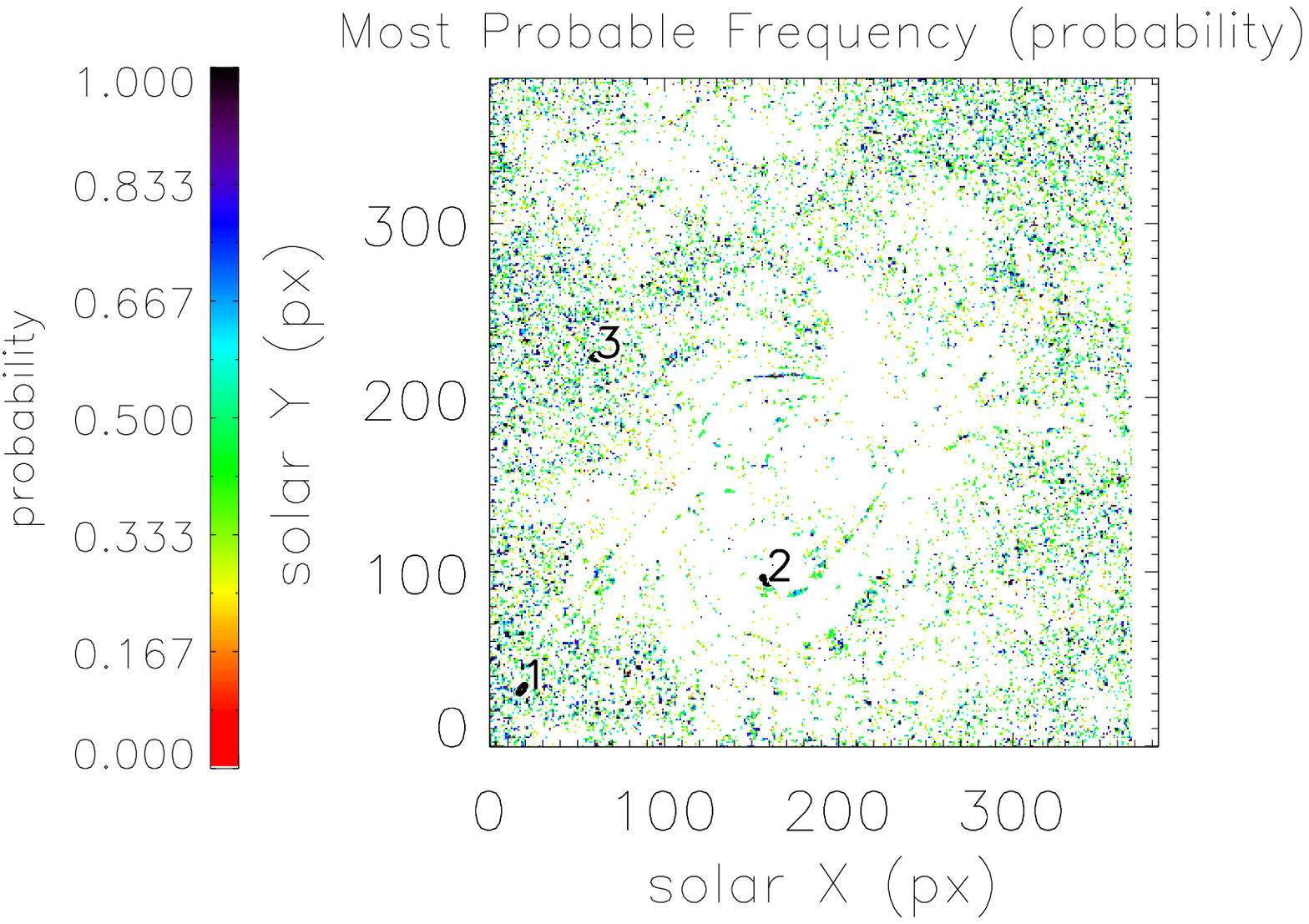}
    \vspace{0.0 \textwidth}}
  \vspace{-0.38\textwidth}   
  \centerline{
    \hspace{0.0 \textwidth}  \color{black}{(a)}
    \hspace{0.435\textwidth}  \color{black}{(b)}
    \hfill}
  \vspace{0.35\textwidth}    
  \centerline{\hspace*{0.015\textwidth}
\includegraphics[width=0.515\textwidth,clip=]{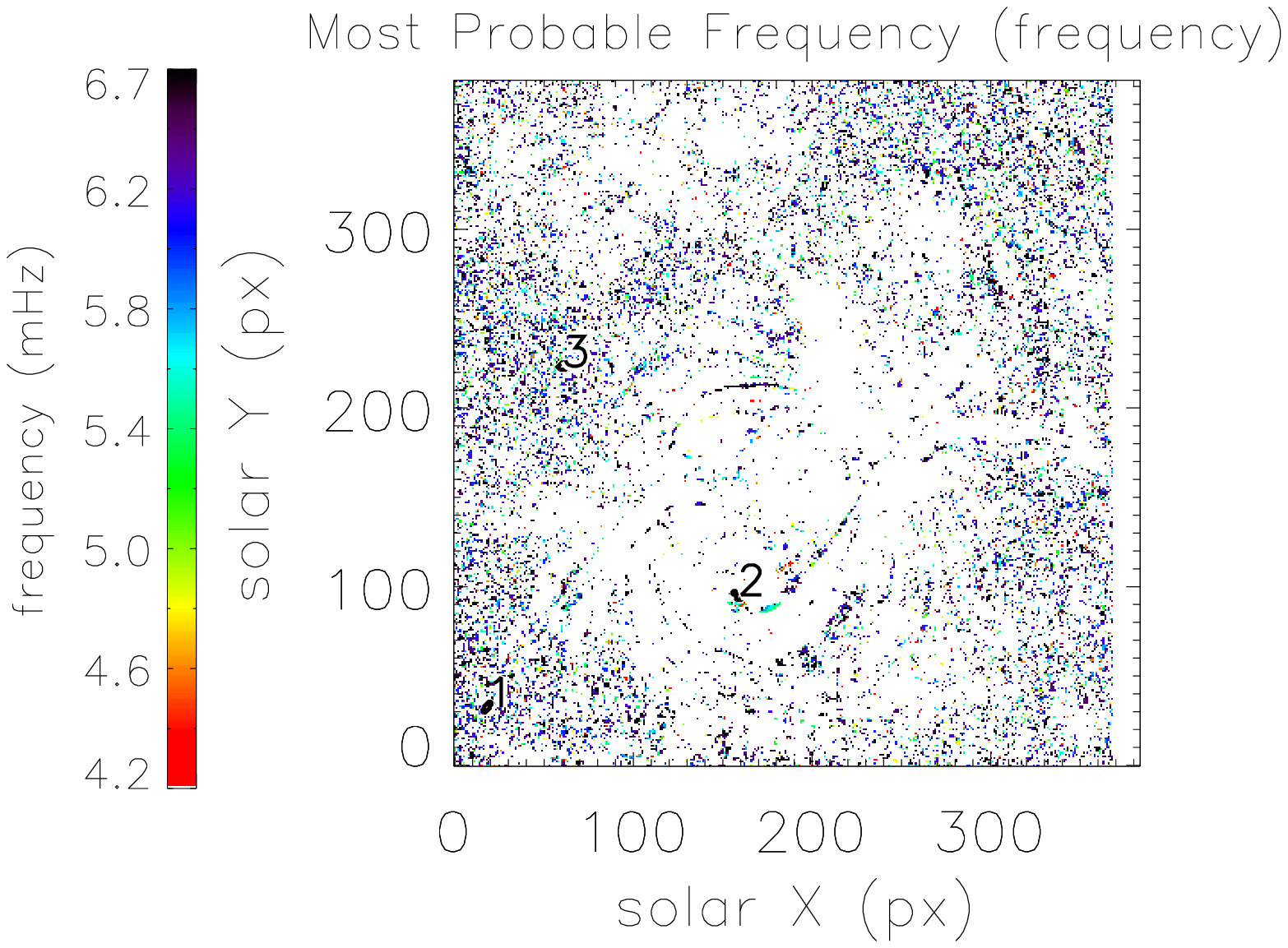}
    \hspace*{-0.03\textwidth}
\includegraphics[width=0.515\textwidth,clip=]{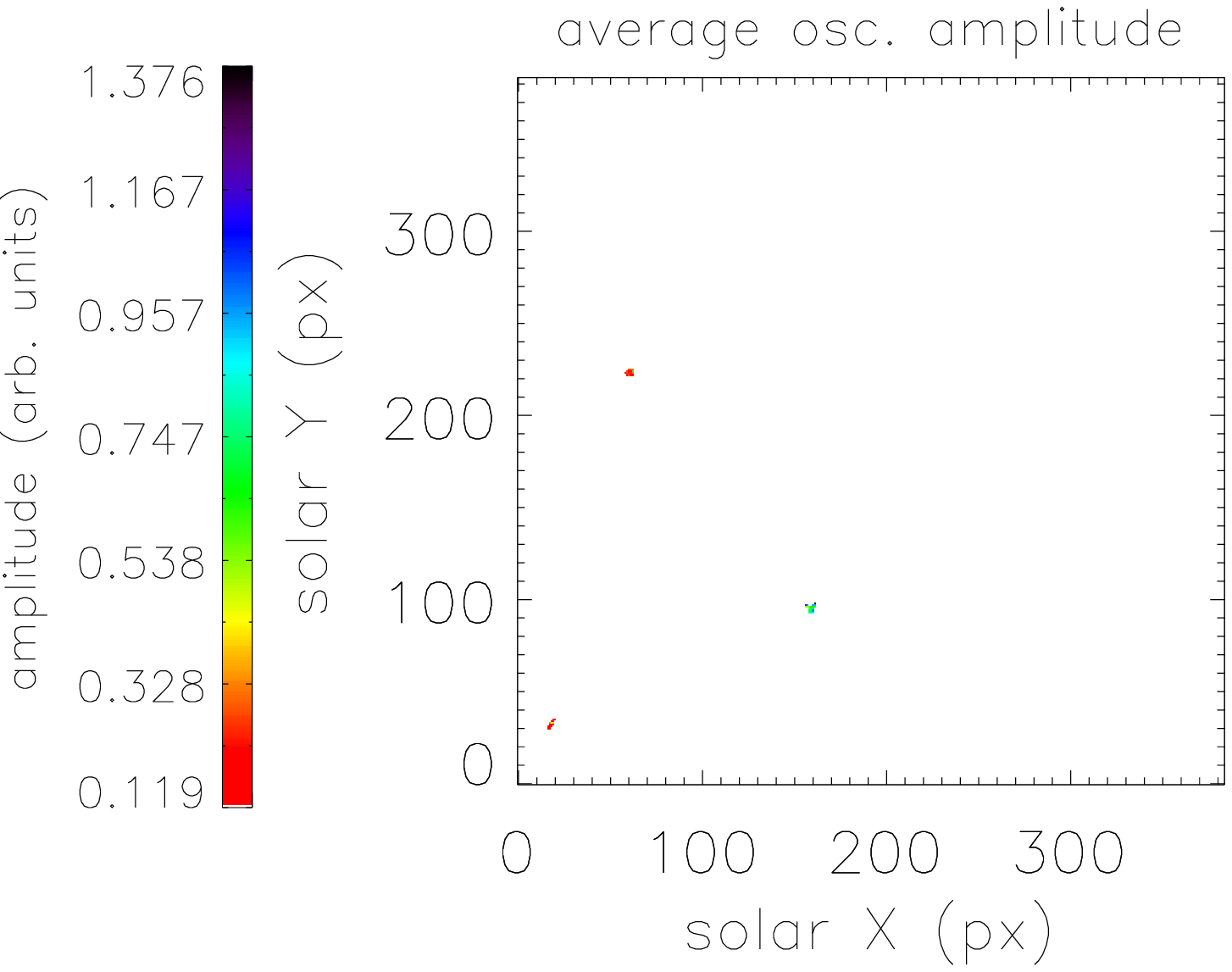}
    \vspace{0.0 \textwidth}}
  \vspace{-0.38\textwidth}   
  \centerline{
    \hspace{0.0 \textwidth}  \color{black}{(c)}
    \hspace{0.435\textwidth}  \color{black}{(d)}
    \hfill}
  \vspace{0.35\textwidth}    
  \centerline{\hspace*{0.015\textwidth}
\includegraphics[width=0.515\textwidth,clip=]{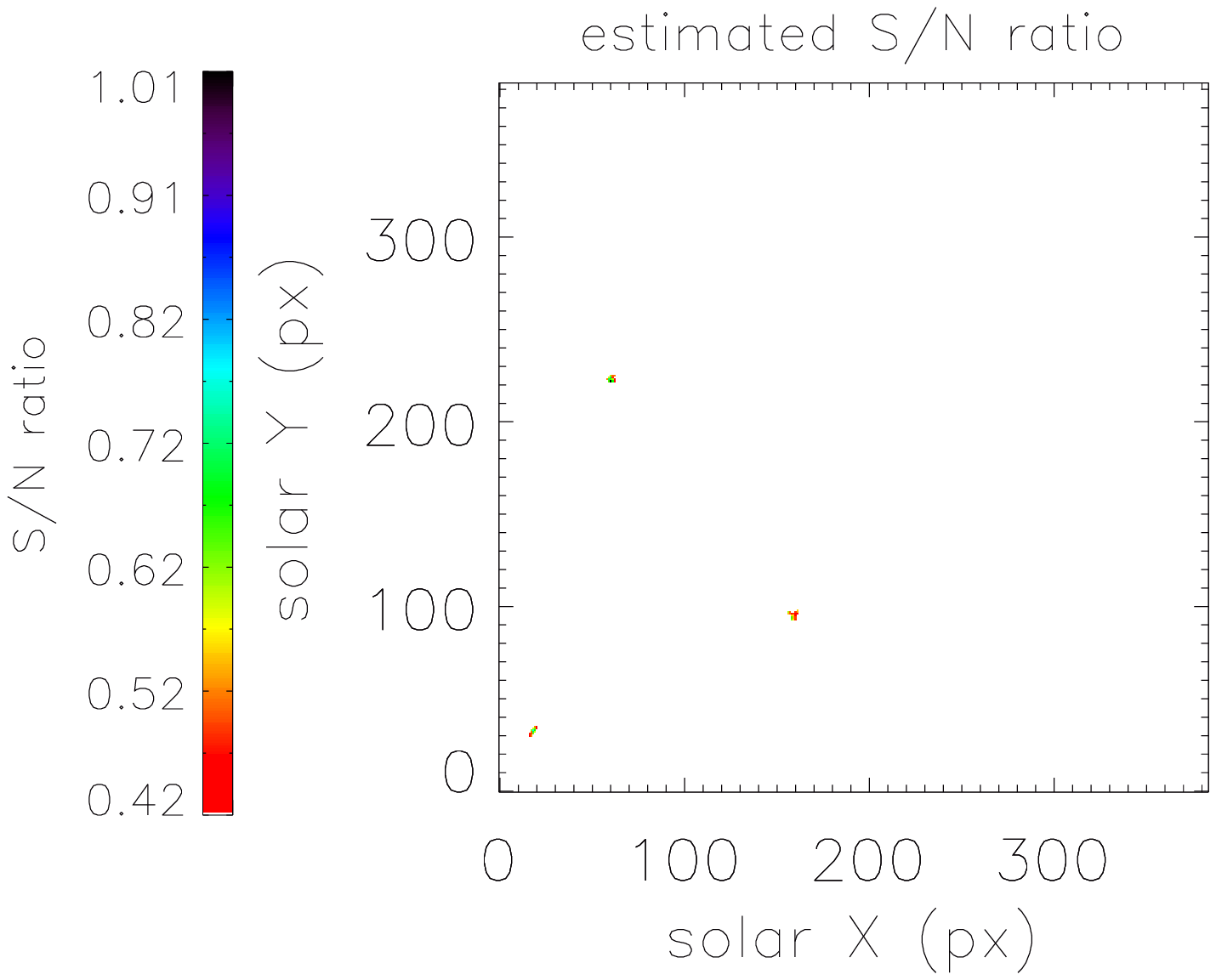}
    \hspace*{-0.03\textwidth}
    \includegraphics[width=0.515\textwidth,clip=]{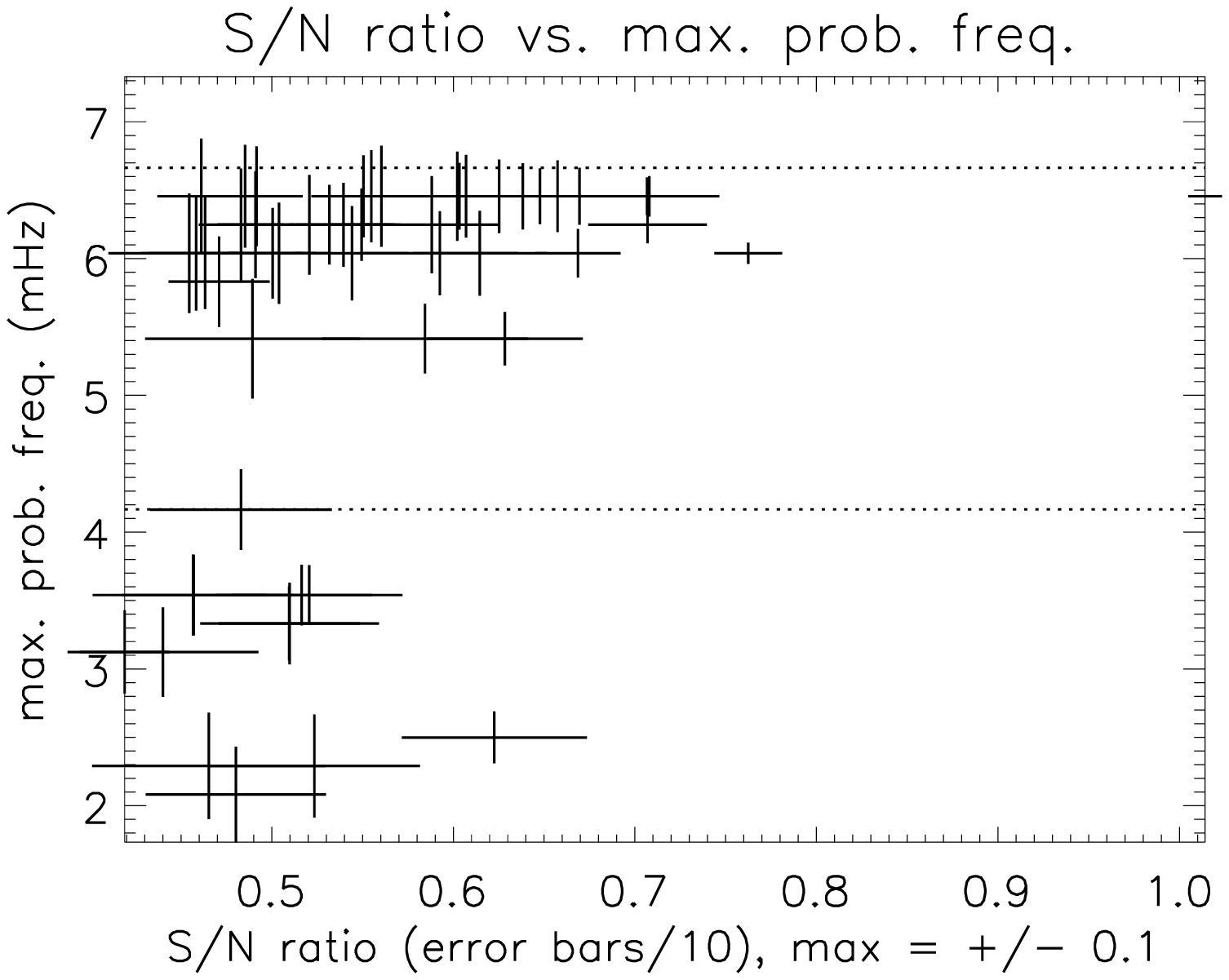}
   \vspace{0.0 \textwidth}}
  \vspace{-0.38\textwidth}   
  \centerline{
    \hspace{0.0 \textwidth}  \color{black}{(e)}
    \hspace{0.435\textwidth}  \color{black}{(f)}
    \hfill}
  \vspace{0.35\textwidth}    
  \caption{Example image data and results from the analysis algorithm
    for TRACE 171\AA\ data taken on 14 July 1998 (panel (a)) in the
    three-minute frequency band. Panels (b\,--\,f) show similar maps to
    those shown and described in Figure 
    \protect\ref{fig:jul1_171_5min}. Panels (b\,--\,f) show similar maps to
    those shown and described in \protect\ref{fig:jul1_171_5min}.
    Table \protect\ref{tab:jul14} shows the values for the detected
    region for the parameters listed in Table
    \protect\ref{tab:measures}.}
  \label{fig:jul14_3min}
\end{figure}

%
%
\begin{figure}
  \centerline{\hspace*{0.015\textwidth}
\includegraphics[width=0.515\textwidth,clip=]{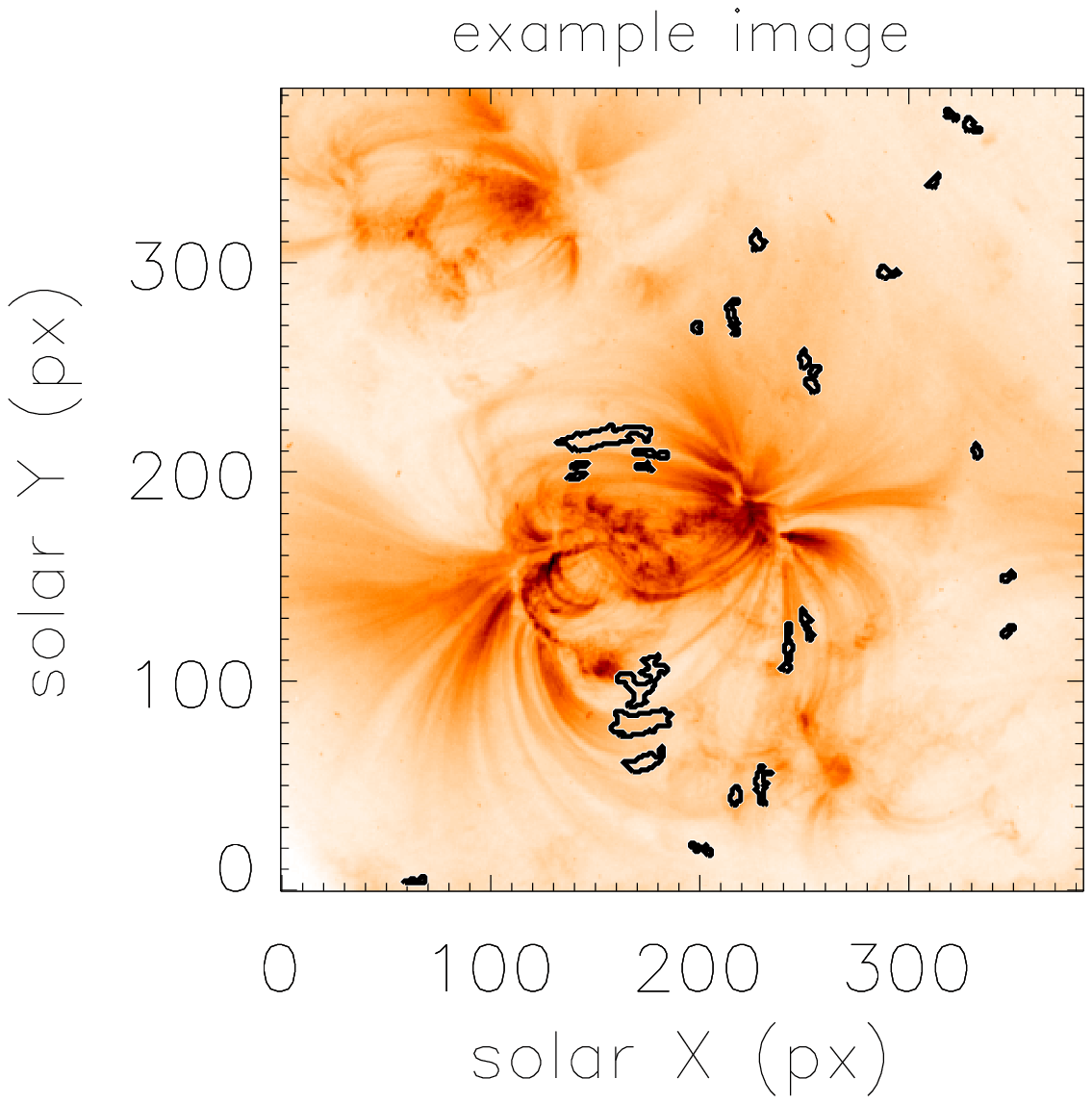}
    \hspace*{-0.03\textwidth}
\includegraphics[width=0.515\textwidth,clip=]{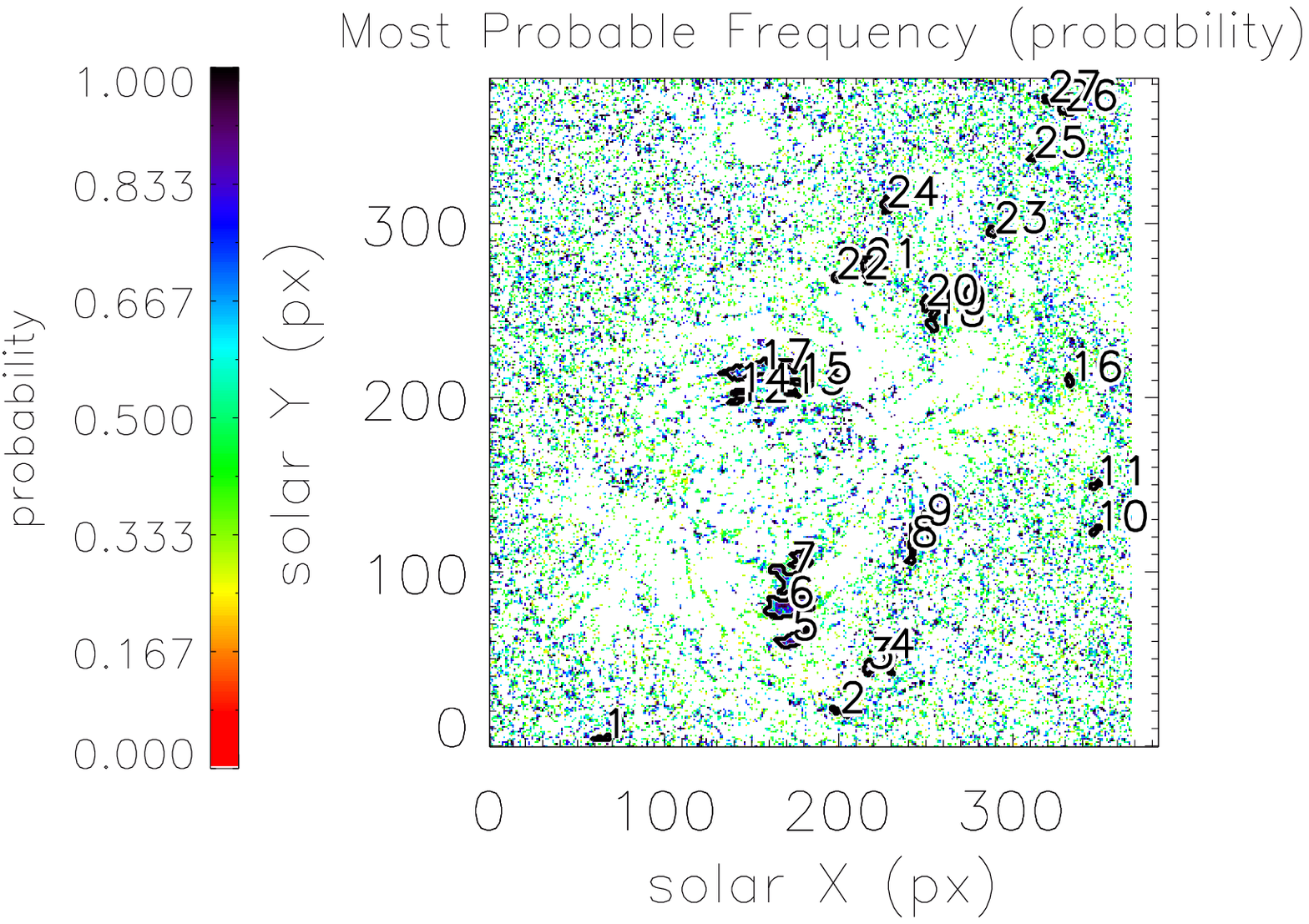}
    \vspace{0.0 \textwidth}}
  \vspace{-0.38\textwidth}   
  \centerline{
    \hspace{0.0 \textwidth}  \color{black}{(a)}
    \hspace{0.435\textwidth}  \color{black}{(b)}
    \hfill}
  \vspace{0.35\textwidth}    
  \centerline{\hspace*{0.015\textwidth}
\includegraphics[width=0.515\textwidth,clip=]{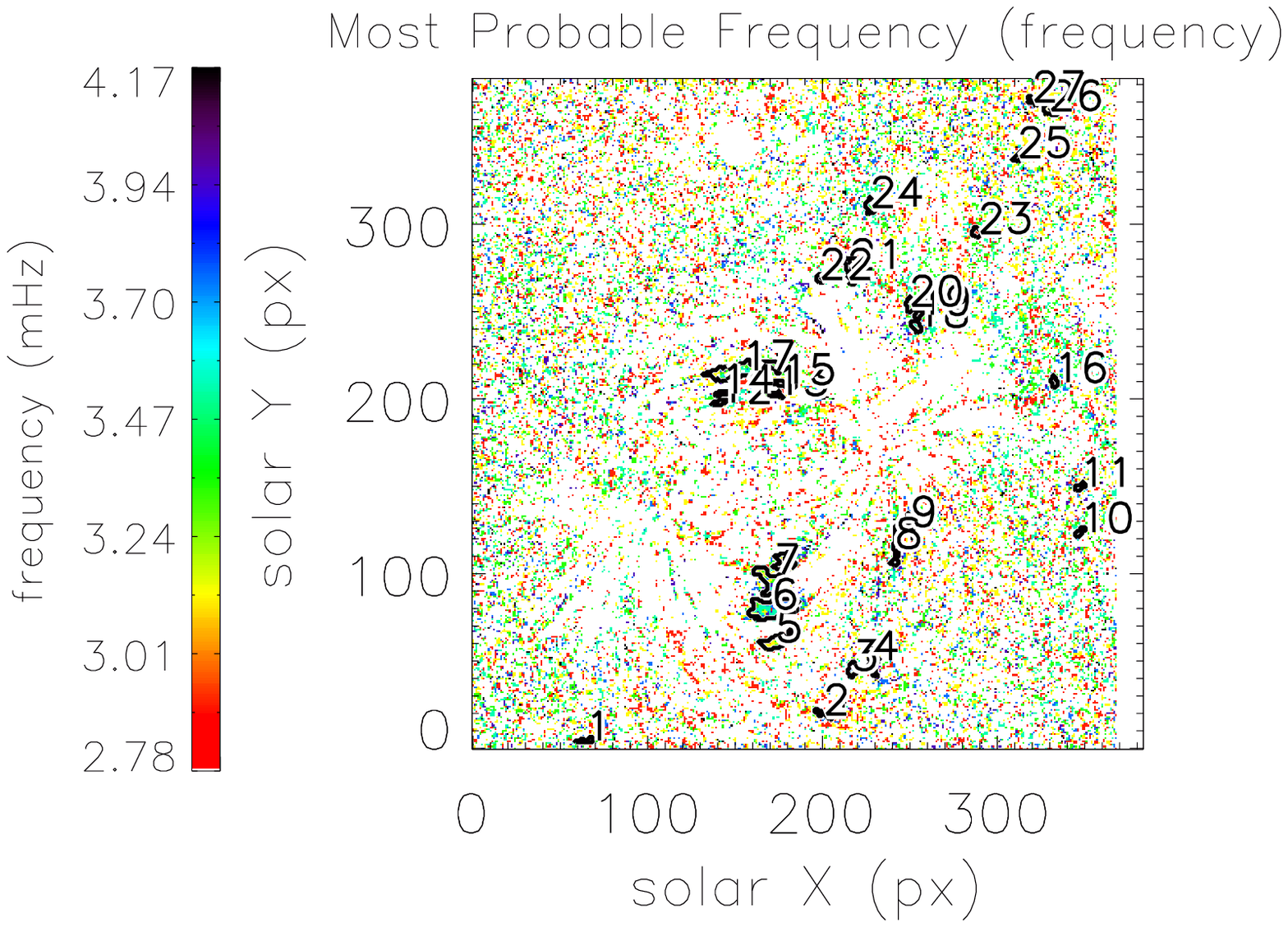}
    \hspace*{-0.03\textwidth}
\includegraphics[width=0.515\textwidth,clip=]{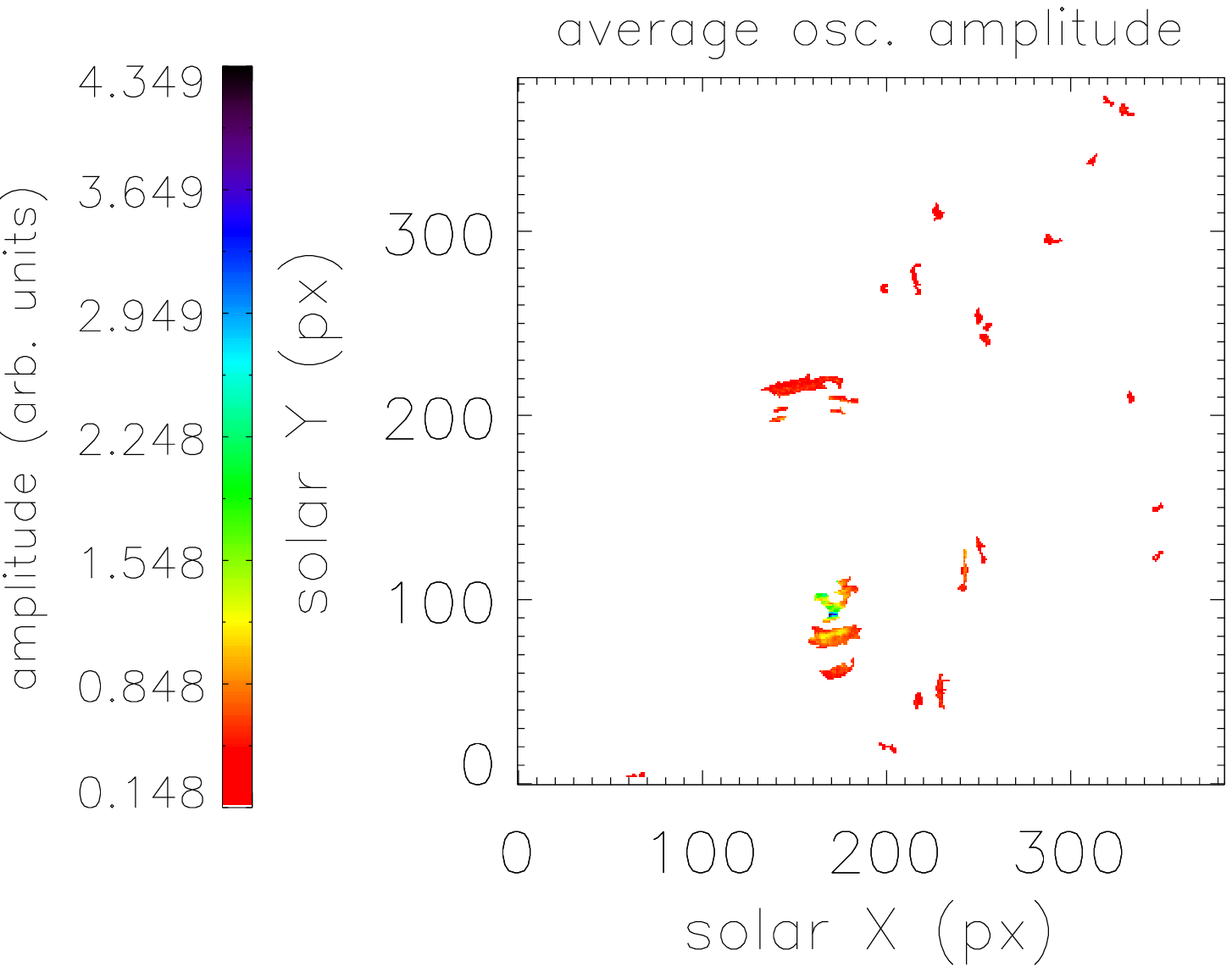}
    \vspace{0.0 \textwidth}}
  \vspace{-0.38\textwidth}   
  \centerline{
    \hspace{0.0 \textwidth}  \color{black}{(c)}
    \hspace{0.435\textwidth}  \color{black}{(d)}
    \hfill}
  \vspace{0.35\textwidth}    
  \centerline{\hspace*{0.015\textwidth}
\includegraphics[width=0.515\textwidth,clip=]{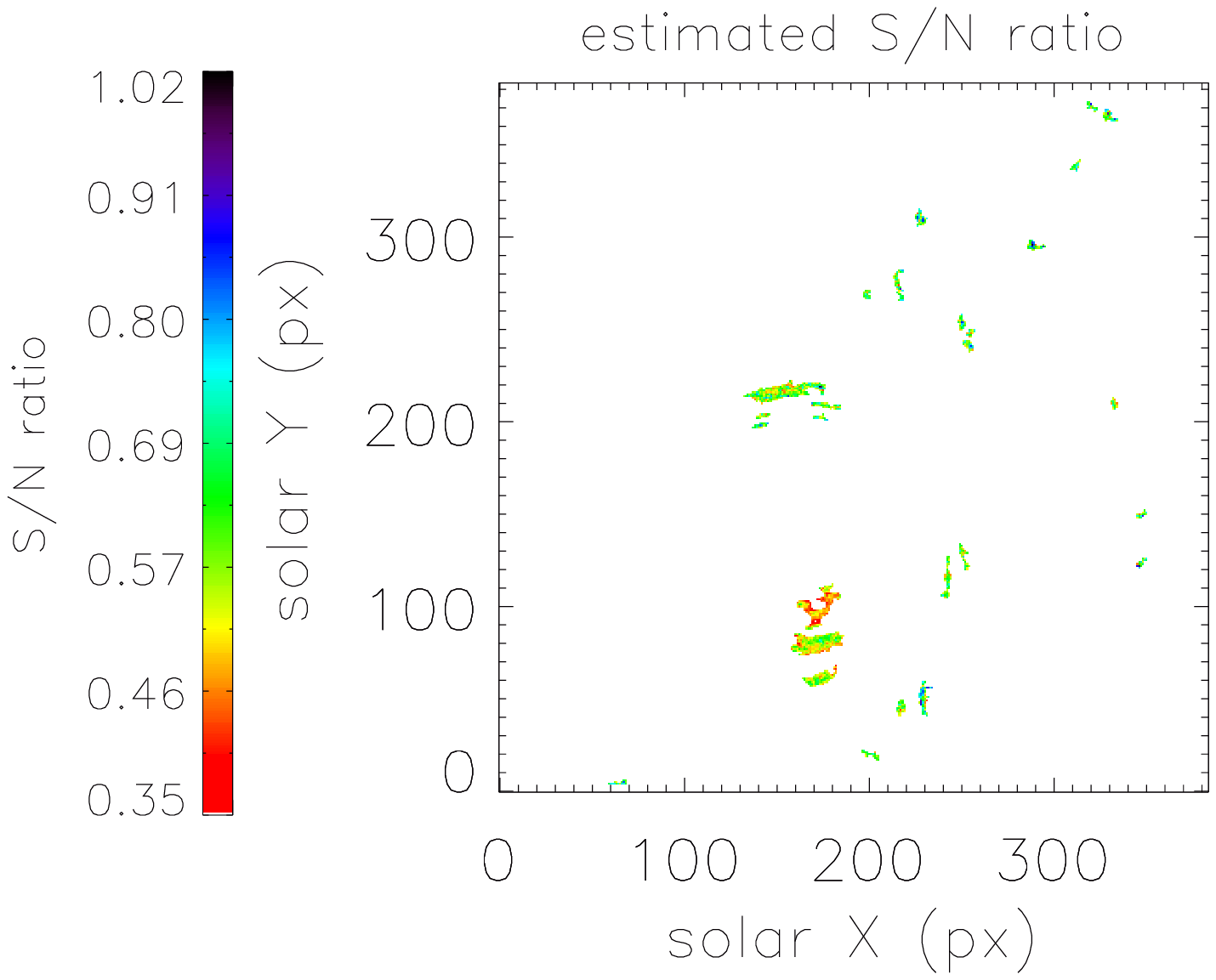}
    \hspace*{-0.03\textwidth}
    \includegraphics[width=0.515\textwidth,clip=]{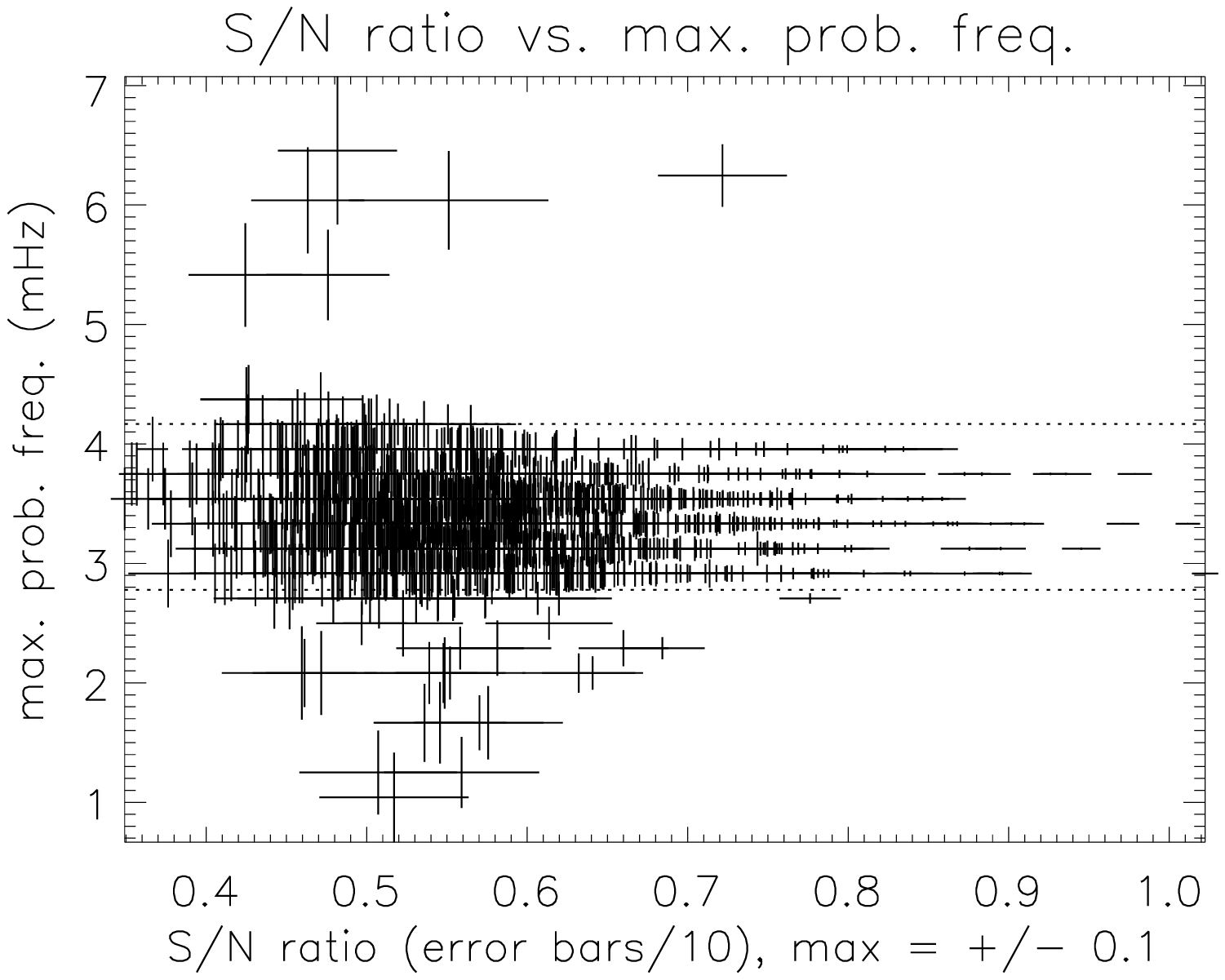}
   \vspace{0.0 \textwidth}}
  \vspace{-0.38\textwidth}   
  \centerline{
    \hspace{0.0 \textwidth}  \color{black}{(e)}
    \hspace{0.435\textwidth}  \color{black}{(f)}
    \hfill}
  \vspace{0.35\textwidth}    
  \caption{Example image data and results from the analysis algorithm
    for TRACE 171\AA\ data taken on 14 July 1998 (panel (a)) in the
    five-minute frequency band. Panels (b\,--\,f) show similar maps to
    those shown and described in Figure
    \protect\ref{fig:jul1_171_5min}. Panels (b\,--\,f) show similar
    maps to those shown and described in Figure
    \protect\ref{fig:jul1_171_5min}.  Table \protect\ref{tab:jul14}
    shows the values for the detected region for the parameters listed
    in Table \protect\ref{tab:measures}.}
  \label{fig:jul14_5min}
\end{figure}

\begin{table}
\begin{tabular}{ccrrccll}
data                        &  $\overline{f}\pm\sigma_{f}$& $A$ & $A_{0.95}$ & $F$ & $\overline{p}$ & $Q$ $[r(Q)]$ & $E$ $[r(E)]$ \\ \hline
3 min             &                            &      &           &     &                &            &            \\

(a) 171\AA & 6.04 $\pm$ 0.58 & 63 & 25 & 0.955 & 0.810 & 0.773 [1] & 48.691 [1] \\
(b) 195\AA & 6.00 $\pm$ 0.56 & 75 & 29 & 1.000 & 0.788 & 0.788 [1] & 59.127 [1] \\
                            &                            &      &           &     &                &            &            \\
 5 min                            &                            &      &           &     &                &            &            \\
 (c) 171\AA             &                            &      &           &     &                &            &            \\ \hline
$1$ & 3.112 $\pm$ 0.149 & 15 & 12 & 0.938 & 0.957 & 0.897 [3] & 13.457 [10] \\
$2$ & 3.276 $\pm$ 0.346 & 22 & 14 & 0.917 & 0.926 & 0.849 [6] & 18.675 [5] \\
$3$ & 3.405 $\pm$ 0.354 & 20 & 18 & 0.870 & 0.965 & 0.840 [8] & 16.791 [7] \\
$4$ & 3.134 $\pm$ 0.159 & 23 & 18 & 1.000 & 0.951 & 0.951 [1] & 21.877 [2] \\
$5$ & 3.340 $\pm$ 0.343 & 17 & 9 & 1.000 & 0.842 & 0.842 [7] & 14.306 [8] \\
$6$ & 3.268 $\pm$ 0.327 & 67 & 39 & 0.817 & 0.868 & 0.709 [10] & 47.520 [1] \\
$7$ & 3.297 $\pm$ 0.274 & 22 & 16 & 0.957 & 0.935 & 0.894 [4] & 19.676 [4] \\
$8$ & 3.195 $\pm$ 0.184 & 16 & 9 & 0.941 & 0.907 & 0.853 [5] & 13.652 [9] \\
$9$ & 3.067 $\pm$ 0.204 & 23 & 15 & 1.000 & 0.912 & 0.912 [2] & 20.982 [3] \\
$10$ & 3.228 $\pm$ 0.149 & 22 & 14 & 0.846 & 0.934 & 0.790 [9] & 17.379 [6] \\
                            &                            &      &           &     &                &            &            \\
(d) 195\AA             &                            &      &           &     &                &            &            \\ \hline

$1$ & 3.164 $\pm$ 0.214 & 31 & 15 & 0.912 & 0.863 & 0.787 [6] & 24.402 [5] \\
$2$ & 3.209 $\pm$ 0.334 & 31 & 18 & 0.886 & 0.904 & 0.801 [5] & 24.833 [4] \\
$3$ & 3.342 $\pm$ 0.395 & 75 & 42 & 0.852 & 0.878 & 0.748 [9] & 56.099 [1] \\
$4$ & 3.166 $\pm$ 0.254 & 16 & 6 & 1.000 & 0.840 & 0.840 [3] & 13.433 [8] \\
$5$ & 3.240 $\pm$ 0.286 & 41 & 25 & 0.911 & 0.895 & 0.816 [4] & 33.451 [2] \\
$6$ & 3.471 $\pm$ 0.289 & 19 & 11 & 1.000 & 0.911 & 0.911 [1] & 17.305 [6] \\
$7$ & 3.295 $\pm$ 0.277 & 17 & 12 & 0.850 & 0.914 & 0.777 [8] & 13.214 [9] \\
$8$ & 2.998 $\pm$ 0.133 & 17 & 11 & 0.944 & 0.910 & 0.859 [2] & 14.606 [7] \\
$9$ & 3.200 $\pm$ 0.174 & 34 & 21 & 0.895 & 0.870 & 0.778 [7] & 26.452 [3] \\

\end{tabular}
\caption{Quantities reported (see Table \protect\ref{tab:measures} for
  the automatically detected regions found in 1 July 1998 TRACE
  171\AA\ and 195\AA\ data -- see Figures \ref{fig:jul1_171_3min},
  \ref{fig:jul1_195_3min}, \ref{fig:jul1_171_5min}, and
  \ref{fig:jul1_195_5min} for maps and plots of the detected regions
  listed in (a), (b), (c) and (d) respectively above.  The quantities
  $Q$ and $E$ are also listed with their rank [$r$] when compared to all
  other regions found in the same dataset.}
\label{tab:jul1}
\end{table}

\begin{table}
\begin{tabular}{lcrrccr@{.}lr@{.}l}
region    &  $\overline{f}\pm\sigma_{f}$     & $A$ & $A_{0.95}$ & $F$& $\overline{p}$ & $Q$ & $[r(Q)]$ & $E$ &$[r(E)]$ \\ \hline
3 min                 &                            &      &           &     &                &            &            \\
$1$ & 6.303 $\pm$ 0.455 & 15 & 9 & 0.938 & 0.879 & 0&824 [2] & 12&354 [3] \\
$2$ & 6.383 $\pm$ 0.264 & 20 & 0 & 1.000 & 0.723 & 0&723 [3] & 14&466 [1] \\
$3$ & 6.547 $\pm$ 0.151 & 16 & 5 & 1.000 & 0.849 & 0&849 [1] & 13&581 [2] \\
                      &                            &      &           &     &                &            &            \\
5 min                 &                            &      &           &     &                &            &            \\ \hline
$1$ & 3.319 $\pm$ 0.052 & 16 & 6 & 0.941 & 0.815 & 0&767 [10] & 12&279 [22] \\
$2$ & 3.076 $\pm$ 0.127 & 22 & 1 & 0.917 & 0.782 & 0&717 [22] & 15&765 [16] \\
$3$ & 3.408 $\pm$ 0.348 & 30 & 3 & 0.938 & 0.781 & 0&732 [18] & 21&958 [12] \\
$4$ & 3.811 $\pm$ 0.185 & 53 & 26 & 0.930 & 0.862 & 0&802 [4] & 42&492 [6] \\
$5$ & 3.117 $\pm$ 0.164 & 93 & 3 & 0.979 & 0.778 & 0&762 [12] & 70&856 [4] \\
$6$ $[A3,8]$ & 3.616 $\pm$ 0.290 & 236 & 14 & 0.992 & 0.745 & 0&739 [16] & 174&427 [1] \\
$7$ $[A7]$ & 3.477 $\pm$ 0.358 & 197 & 1 & 0.985 & 0.644 & 0&634 [27] & 124&969 [3] \\
$8$ $[A4]$ & 3.556 $\pm$ 0.239 & 54 & 13 & 1.000 & 0.834 & 0&834 [1] & 45&025 [5] \\
$9$ $[A9]$ & 3.038 $\pm$ 0.116 & 34 & 2 & 1.000 & 0.745 & 0&745 [15] & 25&317 [9] \\
$10$ & 3.415 $\pm$ 0.323 & 15 & 6 & 0.882 & 0.856 & 0&756 [13] & 11&336 [26] \\
$11$ & 3.163 $\pm$ 0.231 & 16 & 4 & 0.941 & 0.832 & 0&783 [9] & 12&534 [21] \\
$12$ & 3.482 $\pm$ 0.275 & 18 & 7 & 1.000 & 0.823 & 0&823 [3] & 14&821 [17] \\
$13$ & 3.410 $\pm$ 0.213 & 16 & 5 & 1.000 & 0.823 & 0&823 [2] & 13&175 [20] \\
$14$ & 3.306 $\pm$ 0.262 & 16 & 1 & 1.000 & 0.716 & 0&716 [24] & 11&459 [25] \\
$15$ & 3.546 $\pm$ 0.145 & 36 & 5 & 0.947 & 0.757 & 0&718 [20] & 25&835 [8] \\
$16$ & 3.430 $\pm$ 0.182 & 17 & 1 & 0.944 & 0.693 & 0&654 [26] & 11&126 [27] \\
$17$ $[A1,2,6]$ & 3.359 $\pm$ 0.222 & 247 & 35 & 0.939 & 0.749 & 0&703 [25] & 173&700 [2]\\
$18$ & 3.380 $\pm$ 0.272 & 26 & 8 & 0.963 & 0.764 & 0&736 [17] & 19&134 [15] \\
$19$ & 3.332 $\pm$ 0.108 & 16 & 3 & 0.941 & 0.761 & 0&716 [23] & 11&463 [24] \\
$20$ & 3.455 $\pm$ 0.132 & 27 & 7 & 1.000 & 0.788 & 0&788 [7] & 21&274 [14] \\
$21$ & 3.270 $\pm$ 0.248 & 44 & 7 & 0.936 & 0.773 & 0&724 [19] & 31&839 [7] \\
$22$ & 3.436 $\pm$ 0.215 & 16 & 3 & 0.941 & 0.762 & 0&717 [21] & 11&475 [23] \\
$23$ & 3.047 $\pm$ 0.235 & 30 & 8 & 0.968 & 0.819 & 0&792 [6] & 23&767 [11] \\
$24$ & 3.376 $\pm$ 0.186 & 33 & 12 & 0.917 & 0.833 & 0&764 [11] & 25&203 [10] \\
$25$ & 3.185 $\pm$ 0.263 & 17 & 4 & 1.000 & 0.787 & 0&787 [8] & 13&377 [19] \\
$26$ & 3.124 $\pm$ 0.176 & 29 & 8 & 0.935 & 0.799 & 0&748 [14] & 21&684 [13] \\
$27$ & 3.173 $\pm$ 0.202 & 17 & 2 & 1.000 & 0.799 & 0&799 [5] & 13&583 [18] \\
\end{tabular}
\caption{Quantities reported (see Table \protect\ref{tab:measures}) for
  the automatically detected regions found in 14 July 1998 TRACE
  171\AA\ -- see Figure \protect\ref{fig:jul14_5min}, for maps and
  plots of the detected regions. The quantities $Q$ and $E$ are also
  listed with their rank $r$ when compared to all other regions found
  in the same dataset.  Region 4 is the same region as that studied by
  \protect\inlinecite{nak1999} and \protect\inlinecite{irelanddem02},
  and region 8 is the base of the same loop.  The numbers $Ax$ in the
  region column refer to the oscillations found manually in Figure 1
  of \protect\inlinecite{1999ApJ...520..880A}.}
\label{tab:jul14}
\end{table}

\subsection{14 July 1998}
\label{sec:results:jul14}
The original data was taken on 14 July 1998 12:45:19 -- 13:42:44 UT
with an average cadence of 73 seconds (sixty samples) and an image
size of $512\times 512$ pixels of equivalent size $0.5''$.  Data are
also $2\times 2$ summed in space to increase signal-to-noise ratio.
TRACE was observing NOAA AR 8270 when a GOES class M4.6 flare 
occurred at around 12:55 UT.

A movie of the analyzed data cube shows that the first eight samples
show no sign of any flaring.  After that, the flare continues for
about another 15\,--18\, frames.  Finally, the flare dissipates and
post-flare loops are observable (20 frames).  The flare event
transfers momentum to the surrounding medium, causing loops to
oscillate.  In addition, the whole region evacuates -- the flare
appears to blast material away, or at least change its temperature
enough to put the material out of the TRACE 171\AA\ passband.  The
physical phenomena captured by these observations imply that the
resultant time-series have a significant background trend which varies
from location to location.  The detrending timescale ($500$ seconds)
removes the secular background intensity variations such as flaring
and dimming.

Results for the analysis in a three (120-240 seconds) and five minute
frequency band (240-360 seconds) are given in Figure
\ref{fig:jul14_3min} and Figure \ref{fig:jul14_5min} repestively, and
Table \ref{tab:jul14}.  Only three three-minute frequency band regions
are found.  Two of them (regions 1 and 3) are distant from the central
active region over relatively dark pieces of corona where the signal
is weak.  Region 2 (the largest of the three) overlies a portion of
the active region where many oscillations are detected in the
five-minute frequency band; the detection in the three-minute
frequency band may be due to multiple loops each oscillating in the
five-minute frequency band that coincidentally gives the appearance of
a three-minute oscillation.

\begin{figure}
  \centerline{
    \includegraphics[width=1.0\textwidth,clip=]{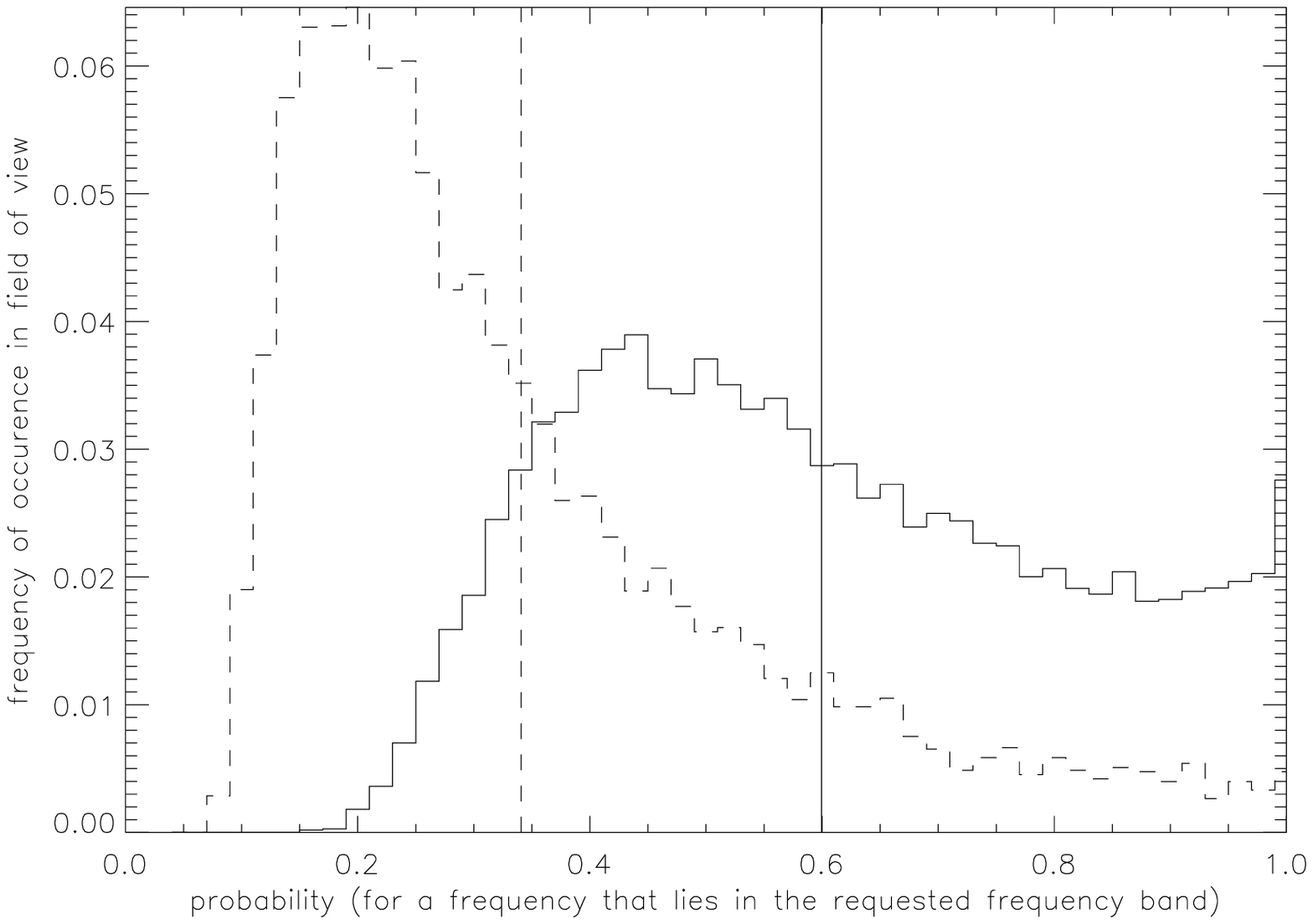}}
  \caption{Distribution of non-zero probabilities for the 1 July and
    14 July 171\AA\ data.  The dashed line refers to the 1 July
    171\AA\ data analyzed in the three-minute frequency band
    (120\,--\,240 seconds), as shown in Figure
    \protect\ref{fig:probex}(b).  The solid line refers to the 14 July
    171\AA\ data analyzed in the five-minute frequency band
    (240\,--\,360 seconds), as shown in Figure
    \protect\ref{fig:probex}(e).  Average values to each frequency
    distribution are indicated by the vertical lines of the same line
    style.  See Section \protect\ref{sec:results:jul14} for more
    detail.}\label{fig:probcompare}
\end{figure}

The detection algorithm finds eight of the nine transversely
oscillating as described in \inlinecite{1999ApJ...520..880A} (their
Figure 1, oscillations 1 through 9 excepting 5). Many of the other
claimed detections of Figure \ref{fig:jul14_5min} are in small regions
distant from the flare site.  These locations also typically show an
oscillation which is more of a gentle sway, that is, the oscillation
decays very quickly.  It is instructive to consider Figure
\ref{fig:probex}(e) in comparison to Figure \ref{fig:probex}(b).  The
spatial probability distribution is much denser at all lengthscales in
Figure \ref{fig:probex}(e) than in Figure \ref{fig:probex}(b).  Figure
\ref{fig:probcompare} shows that the material of Figure
\ref{fig:probex}(e) has a greater proportion of higher-probability
oscillations than the material in Figure \ref{fig:probex}(b).  Given
the differing spatial distribution, it is clear that everywhere in
TRACE field of view on 14 July 1998 was much more likely to oscillate
in the five-minute frequency band.  Since the probability map is
filtered for regions that show a locally high probability, many more
regions are found in this dataset than in the previous dataset.

\section{Detecting Other Oscillatory Signals with this Algorithm}
\label{sec:discuss}
This paper is primarily concerned with detecting areas in the solar
atmosphere that oscillate with a single frequency, described by
Equation (\ref{eqn:model}).  Other oscillations have been found in the
solar atmosphere that are not perfectly described by Equation
(\ref{eqn:model}). Bretthorst (1988, ch. 6.1.4) notes that Equation
(\ref{eqn:model}) still gives strongly peaked probability
distributions close to the true frequencies even when the actual
signal exhibits features not present in the model, such as periodic
but non-harmonic oscillations and non-stationary and non-Gaussian
noise.  In the sections below we discuss some commonly occuring
periodic signals in solar atmosphere, that contain features not
modeled by Equation (\ref{eqn:model}), and their effect on the
probability distributions that are at the center of the proposed
detection algorithm.
\begin{figure}
  \centerline{
    \includegraphics[width=1.0\textwidth,clip=]{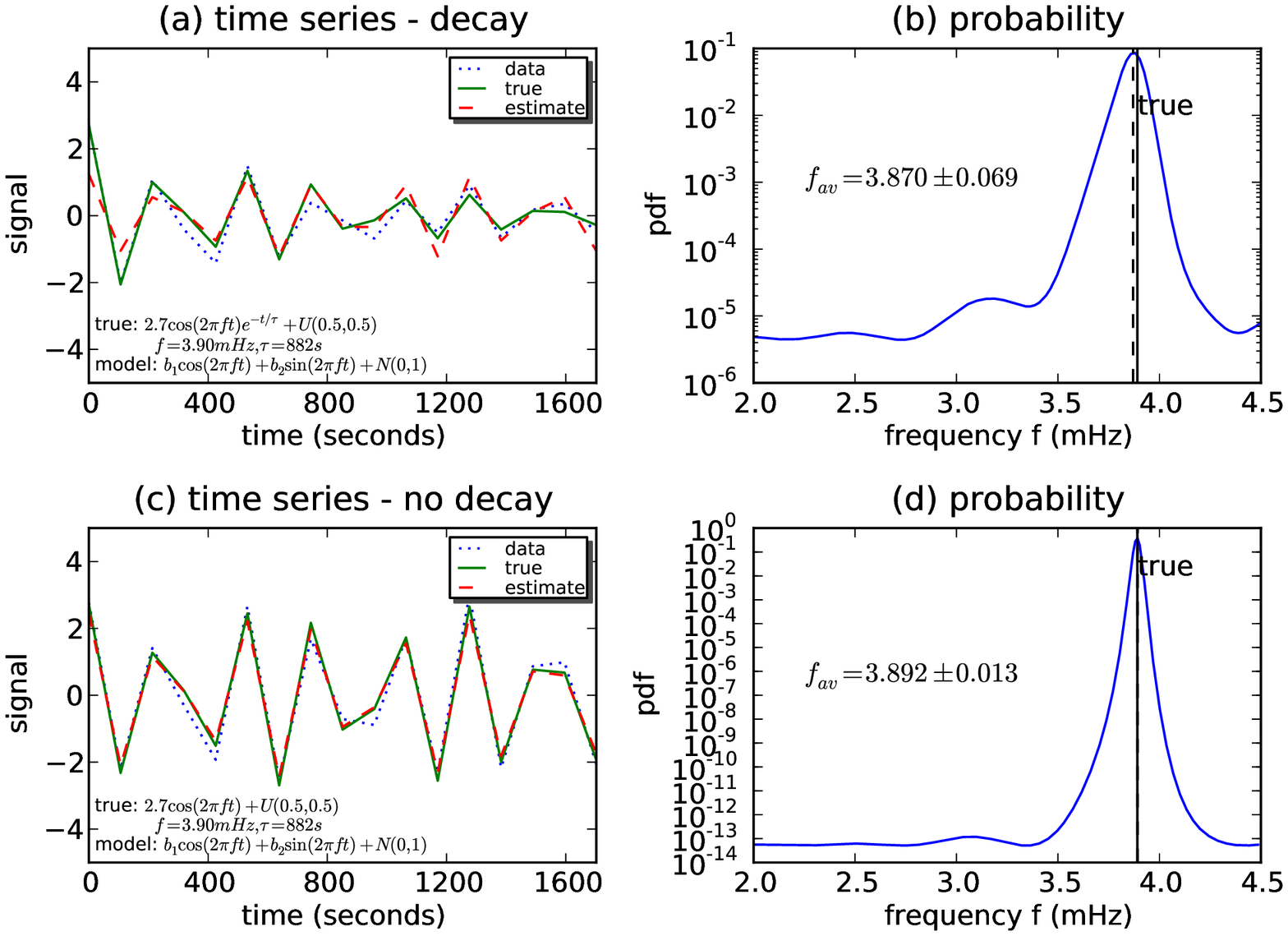}}
  \caption{Effect of an exponential decay in the observed signal on
    the detection of an oscillation.  Panel (a) shows the observed
    noisy time series signal (dotted line), the true (non-noisy)
    signal (solid line) and the estimated signal (dashed line).  Panel
    (b) shows the probability-density function for the observed signal
    as defined by Equation (\protect\ref{eqn:prob}).  Panels (c) and
    (d) are the same as (a) except the exponential decay is no longer
    present. The probability-density function in Panel (d) is much
    narrower than in Panel (b).  The effect of the exponential decay
    is to make the location of the average frequency more
    uncertain. This is because the signal-to-noise ratio decreases due
    to the exponential decay and therefore there is less information
    to determine the frequency when compared to the observed signal of
    panel (c).  See Section \ref{sec:decay} for more
    detail.}\label{fig:decaycompare}
\end{figure}
\subsection{Decaying Oscillations}\label{sec:decay}
Figure \ref{fig:decaycompare} applies the single-frequency model to an
example dataset based on the transverse loop oscillation described by
\inlinecite{nak1999}.  The data (shown in Figure
\ref{fig:decaycompare}(a)) exhibits approximately the same properties
as the true observation.  Figure \ref{fig:decaycompare}(b) shows the
measured frequency to be close to the true frequency of 3.9\,mHz.  For
comparison, Figure \ref{fig:decaycompare}(c) shows the same
time-series as Figure \ref{fig:decaycompare}(a) except with no decay,
along with the probability distribution for the frequency in Figure
\ref{fig:decaycompare}(d).  The only difference between the two is the
error in the determination of the frequency.  This is because the
decay in the first time-series decreases the signal to noise ratio,
and so later portions of the time-series contribute less information
to the determination of the frequency, and so the error increases.
This shows that for decaying oscillations of the type already
observed, the probability based methods of Section 2 yield good
results.  Further, it also demonstrates that even although we used an
inappropriate model, the model of Section 2 still gives a sufficiently
peaked distribution at the right frequency to enable detection.

\subsection{Non-Stationary Frequency Oscillations}\label{sec:nonstat}
\begin{figure}
  \centerline{
    \includegraphics[width=1.0\textwidth,clip=]{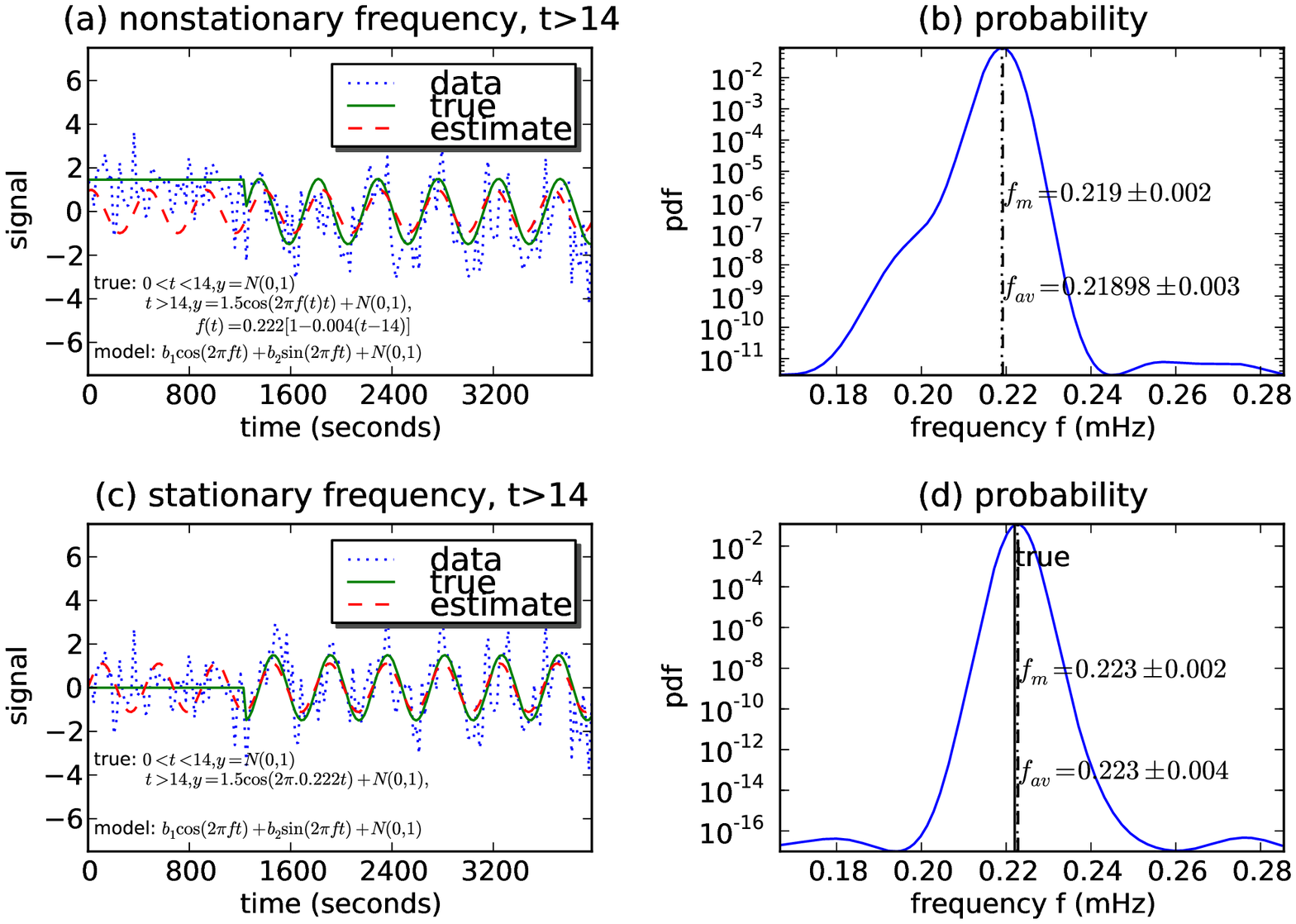}}
  \caption{Effect of a non-stationary frequency in the observed signal
    on the detection of an oscillation.  Panel (a) shows the observed
    noisy time series signal (dotted line), the true (non-noisy)
    signal (solid line) and the estimated signal (dashed line).  The
    true signal consists of noise at $0<t<14$ and a time-varying
    frequency at $t>14$.  Panel (b) shows the probability-density
    function for the observed signal as defined by Equation
    (\protect\ref{eqn:prob}).  Panels (c) and (d) show the same
    quantities as (a) and (b) respectively except that the frequency
    in the observed and true signals are constant at $t>14$.  The peak
    of the probability-density in panel (b) is at a lower frequency
    than in panel (d) because the observed signal contains more low
    frequencies.  Note that in both cases the probability-density
    function is strongly peaked and so most of the probability lies in
    a narrow band around the peak.  See Section
    \protect\ref{sec:nonstat} for more detail.}\label{fig:varying}
\end{figure}
Figure \ref{fig:varying} applies the single-frequency model to an
example dataset based on the ``tadpole'' signature of
\inlinecite{2004MNRAS.349..705N}. The example data-set reproduces the
extent of the signal as mentioned in \inlinecite{2004MNRAS.349..705N},
who claim that there is a time-varying (i.e., non-stationary)
frequency present in a signal found by \inlinecite{2003AA...406..709K}
(see also \opencite{2001MNRAS.326..428W},
\opencite{2002MNRAS.336..747W}) in the Solar Eclipse Imaging
System (SECIS) of Queens University, Northern Ireland eclispe data
taken on 11 August 1999.  Their observation is modeled as having a
signal to noise ratio of 1.5, and a non-stationary frequency that
lowers by 5\% over the range (Figure \ref{fig:varying}(a)).  The
result of an analysis using Equation (\ref{eqn:ruan}) with Equation
(\ref{eqn:model}) (note that Equation (\ref{eqn:prob}), as used in the
automated detection algorithm, is essentially the same) is shown in
Figure \ref{fig:varying}(b).  Figure \ref{fig:varying}(c) shows the
same extent of signal except that now the oscillation is stationary.
Note that the algorithm will not detect the oscillation as being
non-stationary or as not extending across the observation range, as
neither of these effects are included in the model oscillation
Equation (\ref{eqn:model}).  However, the probability distribution
functions Figures \ref{fig:varying}(b,d) are strongly and singly
peaked, and they would therefore be detected given a wide enough
detection window $\omega_{1},\omega_{2}$ (see Equation \ref{eqn:int}).

\subsection{Multiple discrete frequencies}\label{sec:multiple}
\begin{figure}
  \centerline{
    \includegraphics[width=1.0\textwidth,clip=]{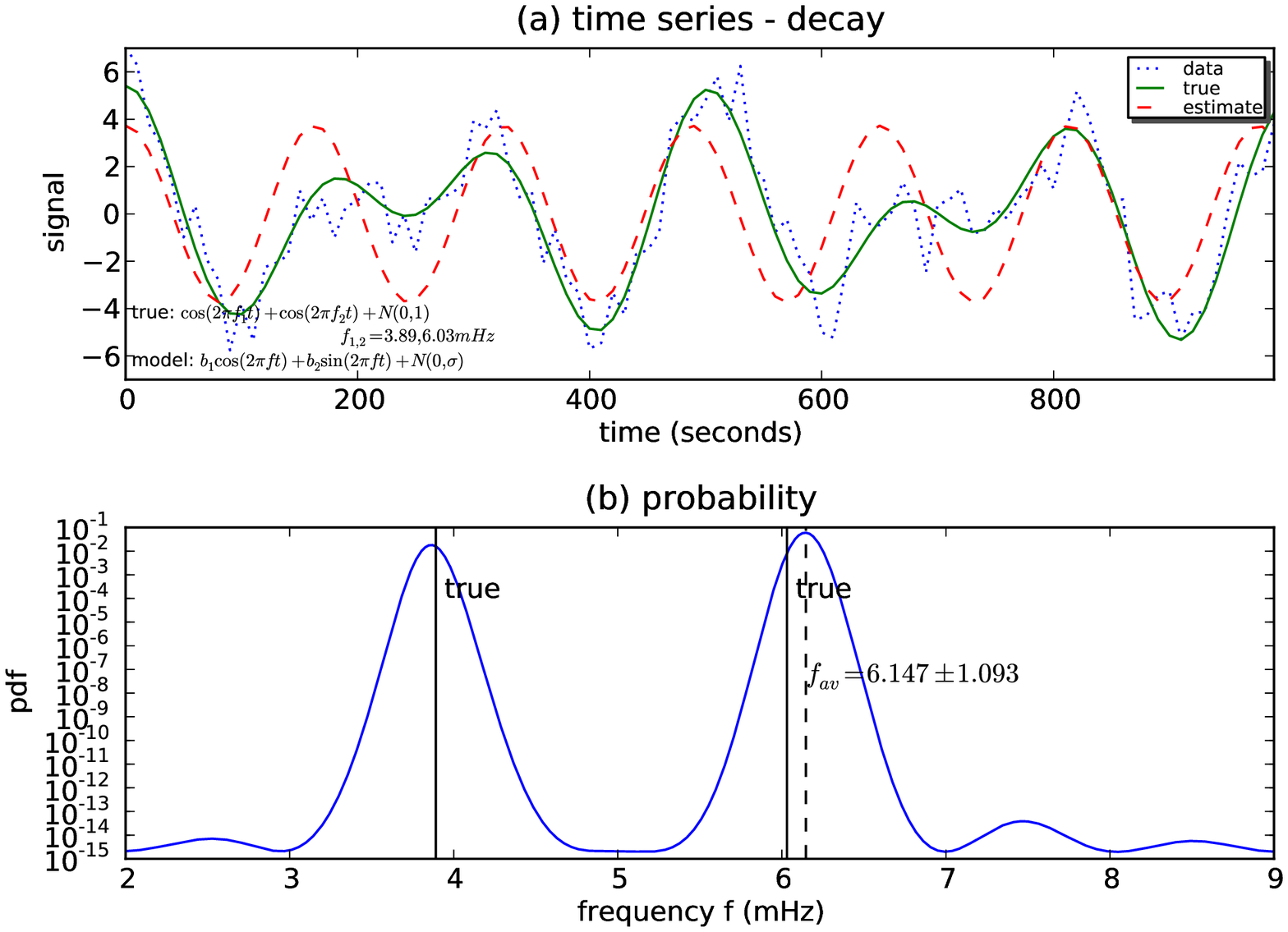}}
  \caption{Analysis of a time-series that has two widely separated
    present.  In panel (a), the observed signal (dotted line) is a
    noisy observation of the true signal (solid).  The estimated
    signal (dashed) is found using Equation (\ref{eqn:prob}), a single
    frequency model.  Panel (b) shows the probability-density function
    (Equation (\ref{eqn:prob})) of the observed signal. In this case,
    the maximum probability density is close to the frequency of the
    higher frequency signal because the signal contains more
    information about this frequency than the lower frequency.  See
    Section \protect\ref{sec:multiple} for more
    detail.}\label{fig:twofreq}
\end{figure}
The results stated in this paper are derived with regard to a single
frequency model, Equation (\ref{eqn:model}).  Initial identification of a
pixel as containing a significant oscillation relies on the
integrated probability over a user-defined frequency band exceeding a
high user-defined limit (in this case, 0.95, see
Section\ref{sec:algorithm}).  If there are multiple frequencies inside
the user-defined frequency band and the integrated probability exceeds
0.95, then the algorithm will report a detection at that pixel and
report the average frequency (and error) within that frequency band.  It
cannot do any more, since the model assumes the presence of only a
single frequency.

Consider now the case of two well-separated frequencies such that one
frequency lies inside the user-defined frequency band
($\omega_{1},\omega_{2}$, Equation (\ref{eqn:int})) and the other lies
outside the user-defined frequency band (Figure \ref{fig:twofreq}a).
Both frequencies are present in the observation, and both have the
same amplitude and signal-to-noise ratio.  The model however, supposes
the presence of one frequency.  Hence the probability will be split
between these two frequencies (Figure \ref{fig:twofreq}(b)).  Which
frequency has the highest probably depends on the evidence for each in
the time-series, that is, the number of oscillations in the time
series and the signal-to-noise ratio for each frequency present.  In
the case above, the most probable frequency is in the three-minute
wave band.  If one were looking in this frequency band, then a peak
would be detected and the algorithm would decide if there were enough
probability in the user-defined frequency band
$[\omega_{1},\omega_{2}]$ to claim a detection.  However if the
user-defined frequency band $[\omega_{1},\omega_{2}]$ was in the
five-minute frequency band, then since the majority of the probability
lies outside this range, no detection would be claimed.  However,
we note that observations of multiple oscillations in single
structures in the corona show one dominant oscillation and a second,
much weaker oscillation (\inlinecite{2004SoPh..223...77V}).
Therefore, the situation of Figure \ref{fig:twofreq}(a) -- two
oscillations of equal amplitude but different frequency -- is
observationally unlikely in the corona.  However, one may use the
approach here to first identify pixels that have a dominant
oscillation present, and use other methods to examine for the presence
of secondary, smaller amplitude oscillations.

Deciding how many frequencies a time-series supports is specifically
excluded by the model choice at the start, that is, a single
oscillation plus noise.  To properly decide how many frequencies are
present in a time-series, one must define a model for each case to be
considered (either no frequencies, one frequency, two frequencies,
three frequencies, etcetera), calculate the probability of each
model, and decide which model is the most probable.  Bayesian
probability tests hypotheses (given the data), and those hypotheses
must be explicitly stated.  Introducing other hypotheses, such as
multiple frequency models, is asking a different question of the data.
For example, multiple frequencies have long been observed in the type
of sunspot observations examined by \inlinecite{2008ApJ...681..672M},
and therefore a single-oscillation model is inappropriate to start
with; one simply would not use this model in this case.  In addition,
the approach taken by \inlinecite{2008ApJ...681..672M} is
prohibitively expensive computationally at present, and so is not
suitable for the automated detection of oscillations in large
datasets.

\section{Conclusion}
\label{sec:conc}
Oscillations in the solar atmosphere have already demonstrated their
worth as probes of the physical conditions present.  SDO will be a
major contributor to the study of solar atmospheric oscillations, and
automated detection algorithms will be necessary in order to maximize
the scientific potential here.  The algorithm parameters in the
filtering procedure (see Section \ref{sec:algorithm}) will have to be
tailored to SDO data; other algorithms will also probably have to
undergo a ``tweaking'' phase to operate optimally.

The algorithm described here is a first attempt at implementing an
automated coronal oscillation detector based on a Bayesian
understanding of probability. It shows promise in being able to find
areas of the solar atmosphere that are highly likely to support an
oscillation.  Section \ref{sec:results} shows that it is able to find
both longitudinal and transversely oscillating loops at low estimated
signal-to-noise ratio over a complex background scene (the
non-oscillating or slowly varying background, where slowly varying is
understood as evolution longer than the characteristic timescale of
the background trend subtraction).  We note that even by eye, these
events appear to be at low signal-to-noise ratio, and the estimates
calculated here (Figures
\ref{fig:jul1_171_3min}\,--\,\ref{fig:jul14_5min}) agree with this
observation.

We also examined the same quiet-Sun TRACE 1600 and 1700\AA\ data as
\inlinecite{2004ApJ...609L..95M}. In that paper, the authors calculate
the travel time of waves in two different UV wave bands, with the
understanding that the entire FOV contains oscillations. We find that
almost the entire FOV contains oscillations. The algorithm returns
positive detections over almost all of the FOV for other TRACE UV data
(\opencite{2004ApJ...602..436M}) . In both cases the detection
itself takes only a few seconds: however, the other parameters such as
signal-to-noise ratio takes many multiples of the detection time. This
suggests that a line production version which generates data products
similar to Figures \ref{fig:jul1_171_3min}, \ref{fig:jul14_5min}, or
\ref{fig:jul1_3min_zoomin} be restricted to data where we know that
only a small percentage of the FOV contains a signal, such as the
higher temperature corona. An alternate mode of operation is that the
reported quantities are restricted to detections only, with further
quantities (such as signal to noise ratios) calculated offline by the
interested user.

The model oscillation that Equation (\ref{eqn:model}) describes, for a
single pixel, a sinusoidol intensity oscillation existing for the
entire duration of the time-series, in the presence of Gaussian noise.
Although this seems quite a restrictive model, it is clear from the
results of Section \ref{sec:results:jul14} that it is sufficient to
enable the detection of decaying transverse oscillations. Indeed,
\inlinecite{irelanddem02} showed that the oscillation studied by
\inlinecite{nak1999} (region 4 in Table \ref{tab:jul14}) displays
non-Gaussian noise and a linear change in period over the measurable
duration of the oscillatory motion, complications not contained in the
model choice, Equation (\ref{eqn:model}).  A transverse oscillation on a
single pixel appears as a periodic non-sinusoidal change in intensity
as the loop swipes across the field of view.  Despite the great
difference between the model choice and the appearance of such an
oscillation in this analysis, the resemblance between the two is
sufficient for our algorithm to be able to detect transversely
oscillating material.  This is an example of the sufficiency of the
model, (Equation (\ref{eqn:model})), to adequately describe a wide range
of periodic behavior, as previously noted by
\inlinecite{bretthorst1988book}.

The algorithm runs quickly on the data (less than a minute, reported
in Table \ref{tab:perform}) on an Apple MacBook (dual core, 1.6 Ghz
processor), without any special efforts at algorithmic optimization.
Analysis of binned (prepared) full-disk SDO data (see Table
\ref{tab:perform}) is possible with existing computers.  Note,
however, that the analysis times quoted do not take into account the
time required to prepare the data for automated oscillation analysis.
This overhead will presumably be the same for all detection
algorithms, and must be considered in design of a full-scale
operational SDO oscillation-detection algorithm in order for it to run
faster than the time taken to acquire the data.

This paper introduces two new features to the discussion of automated
detection of oscillations in the solar atmosphere.  Firstly, the
algorithm is implemented using a Bayesian interpretation of
probability as a ``degree of belief'' as opposed to the standard
interpretation as ``frequency of occurence'' (leading to powerful and
convenient formulae such as Equations \ref{eqn:ruan} and
\ref{eqn:prob}).  At the core of the algorithm lies the probability
that solar atmospheric time-series can be described as a single
sinusoidal oscillation at a fixed frequency, subject to distortion
Gaussian noise.  As \inlinecite{bretthorst1988book} notes, this is a
good approximation for many purposes.  This algorithm does not say
anything about the presence of two or more sinusoidal signals in a
single time-series.  However, it is certainly possible to develop an
analysis algorithm to assign probabilities to each of the three
hypotheses that the time time-series is either noise, contains a
single frequency or contains two frequencies.  Most of the required
mathematics is quoted or referred to in this work; such an algorithm
will be the described in future developments.

Secondly, we have introduced ``quality measures'' in an attempt to
grade the regions that survive the region finding and filtering
process. This appears to be necessary given the large number of
oscillating regions that can occur in a given dataset, for example the
14 July 1998 TRACE 171\AA\ data. Further criteria may be set by
individual users. For example high values of the ratio $A_{0.95}/A$
indicate that the region has a high proportion of pixels very probably
supporting a frequency; a threshold could be set to filter only those
regions that have a high proportion of very certain pixels. In
addition, the current VOEvent standard makes for the provision of
extra algorithm information such as arbitrary parameters ($R$,
$m$) to be carried along with any results. This means the user will be
fully informed of all the parameter values used to obtain the results.

Future algorithm development will concentrate on improving the
probability map filtering (step 6 of Figure \ref{fig:algorithm}, and
also Section \ref{sec:algorithm}), extending the analysis to assign
probabilities to multi-frequency models, and to distinguishing between
longitudinal and transverse oscillations (it may be possible to assign
a probability that a given wave-mode has been observed).
Understanding the wave mode demands an understanding of the structure
on which it is supported, which naturally leads to the automated
detection and characterization of loop structures, which is a complex
topic by itself (see \inlinecite{2008SoPh..248..359A} for a
review).  We hope that the algorithm presented here is a first step
towards automated coronal seismology.

\begin{table}
\begin{tabular}{ccc}
data                & data size   & algorithm time [observation duration] \\ \hline
1 July            & $256 \times\ 256$    &   30 [6231]   \\
                    & 201 samples  &               \\
                    & 31 s cadence &               \\
14 July           &$ 368 \times\ 368$    &   25 [3431]   \\
                    & 47 samples   &               \\
                    & 31 s cadence &               \\
raw SDO             & $4096 \times\ 4096$  &   7641 [2000] \\
                    & 200 samples  &               \\
                    & 10 s cadence &               \\
$2\times\ 2\times 2$ rebinned      & $2048 \times\ 2048$  &   955 [2000]  \\
   SDO              & 200 samples  &               \\
                    & 20 s cadence &               \\
\end{tabular}
\caption{Performance of the algorithm on the analyzed data compared to
  data acquisition time.  Also shown is the projected performance on
  SDO data.}
\label{tab:perform}
\end{table}

\begin{acks}
  This work was supported by the NASA SESDA contract and a NASA
  Heliophysics Guest Investigator award (NNG08EL33C). SOHO
  is a joint project of international co-operation by ESA and NASA.
\end{acks}
\bibliographystyle{spr-mp-sola}
\bibliography{tsbib}  

\appendix
\section{The General Linear Model}
\label{sec:genlin}

A time-series is a special case of a more general linear model
description.  In this section we briefly recap the argument of
\inlinecite{ruan} in deriving Equation (\ref{eqn:prob}). In this
description, a signal $d(i)$ observed at times $t_{i}$ $(1\le i \le
N)$, is modeled as
\begin{equation}\label{eqn:genlin}
d(i) = \sum_{k=1}^{M}b_{k}g_{k}(i) + x(i),
\end{equation}
with $M$ basis functions $g_{k}(i)$, each of amplitude $b_{k}$,
evaluated at time $t_{i}$ (parameterized by $\ww$) and Gaussian
distributed noise $x(i)$ of mean 0 and standard deviation $\sigma$.
In the context of Section \ref{sec:btsa}, the observation $D$ is
equivalent to the signal $d(i)$, and the hypothesis $H$ is that the
data can be described by the right hand side of Equation
(\ref{eqn:genlin}), including the noise.  In matrix form, the above
equation may be written as
\begin{equation}\label{eqn:matgenlin}
\md = \mG \mb + \mx,
\end{equation}
where $\md$ is a $N\times1$ matrix of the observed data, and $\mx$ is
a $N\times1$ matrix of identically distributed and independent
Gaussian noise samples. The matrix $\mG$ is size $N\times M$; each
column of $\mG$ is one of the basis functions evaluated at all
$t_{i}$.  The matrix $\mb$ is a $M\times 1$ matrix, the linear
coefficients of each of the (column) basis functions in $\mG$.  The
likelihood function of the observed data is
\begin{eqnarray}\label{eqn:like}
p( \md | \ww, \sigma,\mb,I)  & = &
\frac{1}
{\left(2\pi\sigma^{2}\right)^{N/2}}
\exp
\left[-
\frac{\mx^{T}\mx}{2\sigma^{2}}
\right] 
 \nonumber \\
& = &
\frac{1}
{\left(2\pi\sigma^{2}\right)^{N/2}}
\exp
\left[-
\frac{(\md - \mG \mb)^{T}(\md - \mG \mb)}{2\sigma^{2}}
\right]
\end{eqnarray}
where $\ww$ parameterize the basis functions $g_k$ and hence $\mG$.
This equation is derived by multiplying together the probability
distributions of the noise $x_{i}$ at each time $i$.  The exponent
shows that maximizing the probability is equivalent to minimizing the
difference between the data and the basis functions (weighted by their
amplitudes); indeed, this equation forms the basis of least-squares
fitting in the presence of Gaussian noise.

In analysis, one is primarily interested in the values of the
parameters $\ww$, and secondarily interested in the other values such
as $\mb$ and $\sigma$.  \inlinecite{ruan} and
\inlinecite{bretthorst1988book} describe the process by which the
``nuisance parameters'' $\mb$ and $\sigma$ are removed from further
consideration by {\it marginalization}.  The statement of Bayes'
theorem for the general linear model, using Equations (\ref{eqn:bayes})
and (\ref{eqn:like}) is
\[
p(\ww,\mb,\sigma|\md,I) = \frac{p(\ww,\mb,\sigma|I)p(\md| \ww, \sigma,\mb,I)}{p(\md|I)}.
\]
Integrating over $\mb$ and $\sigma$ using a prior
$p(\ww,\mb,\sigma|I)$, removes these variables from further explicit
consideration, and is an example of Bayesian marginalization.  On
integration, this obtains the marginal posterior distribution for the
parameters $\ww$:
\begin{equation}\label{eqn:post}
p(\ww|\md,I) = \int_{\mb}\int_{\sigma} p(\ww,\mb,\sigma|\md,I) \mbox{d}\sigma \mbox{d}\mb.
\end{equation}
\inlinecite{ruan} and \inlinecite{bretthorst1988book} use uniform
priors for the amplitude parameters $\mb$ ($p(\mb,I)$ $ = constant$)
and the Jeffreys prior ($p(\sigma|I)\propto 1/\sigma$) for $\sigma$.
On integration,
\begin{equation}\label{eqn:ruan}
p(\ww|,\md, I) \propto
\frac{
\left[
\md^{T}\md - \md^{T}\mG\left(\mG^{T}\mG\right)^{-1}\mG^{T}\md
\right]^{(M-N)/2}
}
{
\sqrt{\det\left(\mG^{T}\mG\right)}
}
\end{equation}
Equation (\ref{eqn:ruan}) is a function of $\ww$ only; the standard
deviation (i.e., the noise level of the time-series) or the amplitude
of the basis functions (the values $\mb$) need not be known in order
for estimates of $\ww$ to be found.  This is a very powerful equation,
with clear application to solar time-series analysis, where estimates
of the noise level and oscillation amplitude in the data are often
difficult to obtain by direct fitting of model functions to the
observed time-series.  It should also be noted that Equation
(\ref{eqn:ruan}) arises from Equations (\ref{eqn:matgenlin}) and
(\ref{eqn:like}), which is the basis of a general least squares fit to
the data given Gaussian distributed noise.

\section{Estimating Basis Function Amplitudes and Variance}
\label{sec:estimate}
Maximizing Equation (\ref{eqn:like}) with respect to $\mb$ (on
substitution of Equation (\ref{eqn:matgenlin})) leads to an amplitude estimate:
\begin{equation}\label{eqn:amp}
\hat{\mb} = \left(\mG^{T}\mG\right)^{-1}\mG^{T}\md .
\end{equation}
This maximization is identical to the ``least squares'' fit to the
data for a given value of $\ww$.  Given these amplitudes, the model
fit is then
\begin{equation}\label{eqn:f}
\hat{\mf} = \mG\hat{\mb}.
\end{equation}
In addition, \inlinecite{ruan} show that an estimate (found by
maximizing the posterior after marginalizing the amplitudes) to the
Gaussian variance is
\begin{equation}\label{eqn:sigma}
{\hat{\sigma}}^{2} = \frac{1}{N-M}\left[\md^{T}\md - \mf^{T}\mf\right]
\end{equation}
This estimated variance is the data energy minus the estimated signal
energy, divided by the number of degrees of freedom.  Note that $\mb$
and ${\hat{\sigma}}^{2}$ are functions of the analyzing frequency
[$\omega$].  Equations (\ref{eqn:amp}) and (\ref{eqn:sigma}) are calculated
for the oscillating regions detected via Section \ref{sec:detectalg}
and returned as part of the results in Section \ref{sec:results} (see
Figures \ref{fig:jul1_171_3min}, \ref{fig:jul1_195_3min},
\ref{fig:jul1_171_5min}, and \ref{fig:jul1_195_5min}).

\end{article} 
\end{document}